\documentstyle[12pt,epsfig]{article}
\begin{document}


\def\gtrsim{\mathrel{\raise.3ex\hbox{$>$}\mkern-14mu
             \lower0.6ex\hbox{$\sim$}}}
\def\lesssim{\mathrel{\raise.3ex\hbox{$<$}\mkern-14mu
             \lower0.6ex\hbox{$\sim$}}}

\parindent=0pt
\title{\bf Shell Model Monte Carlo Methods}
\bigskip
\author{S.E.~Koonin, D.J.~Dean${}^\dagger$ and K.~Langanke \\
{\it W.K. Kellogg Radiation Laboratory, California Institute of
Technology},\\
{\it Pasadena CA 91125 USA}}
\date{}
\maketitle

\bigskip
\begin{center}
{\it Abstract}
\end{center}

We review quantum Monte Carlo methods for dealing with large shell model
problems.  These methods reduce the imaginary-time many-body evolution
operator to a coherent superposition of one-body evolutions in fluctuating
one-body fields; the resultant path integral is evaluated stochastically.  We
first discuss the motivation, formalism, and implementation of such Shell
Model Monte Carlo (SMMC) methods. There then follows a sampler of results and
insights obtained from a number of
applications.  These include the ground state and thermal properties of {\it
pf}-shell nuclei,
the thermal and rotational behavior of rare-earth and
$\gamma$-soft nuclei, and the calculation of double beta-decay matrix elements.
Finally, prospects
for further progress in such calculations are discussed.
\par

\vfill
${}^\dagger$ Present address: Physics division, Oak Ridge National
Laboratory, P.O. Box 2008, Oak Ridge, Tennessee 37831, U.S.A.

\newpage



\vfill\eject
\newcounter{subeqn}
\makeatletter
\newcommand\timnoletter
{
\makeatletter\renewcommand\theequation{\thesection.\arabic{equation}}
\addtocounter{equation}{1}\@addtoreset{equation}{section}\makeatother }
\makeatother
\makeatletter
\newcommand\timletter
{
\setcounter{subeqn}{1}\makeatletter\renewcommand
\theequation{\thesection.\arabic{equation}\alph{subeqn}
\addtocounter{subeqn}{1}\addtocounter{equation}{-1}}
\@addtoreset{equation}{section}\makeatother }
\makeatother

\newenvironment{timeqnarray}{\timletter\begin{eqnarray}}
{\end{eqnarray}\timnoletter}

\parindent=10pt

\noindent
\section{Introduction}

The description of nuclear structure began more than 60 years ago with the
discovery of the neutron. Major
milestones were the discovery of single-particle shells \cite{Haxel},
collective motion \cite{Bohr}, and their
reconciliation \cite{Ring}.
However, a number of recent developments impose new and
more stringent tests of
our ability to describe nuclear structure.  Heavy-ion induced reactions allow
the study of nuclear behavior at
extremes of temperature, angular momentum, or isospin \cite{Mosel}.
Increasingly
precise experiments with
electron \cite{Musolf}, pion \cite{Weise}, kaon \cite{Frekers,Axen}, and
nucleon \cite{Schwandt,Osterfeld} beams probe new modes
of excitation.  As our
understanding of supernovae \cite{Bethe90} and nucleosynthesis \cite{Fowler83}
is refined
there is a
corresponding need to know more
precisely the relevant nuclear properties. And new pictures of nuclear
structure such as the Interacting Boson
Model \cite{Arima} make implicit or explicit assumptions
about the solutions of the
underlying fermion problem
that demand verification.

The range and diversity of nuclear behavior (perhaps the greatest of any
quantal many-body system) have
naturally engendered a host of models.   Short of a complete solution to the
many-nucleon problem \cite{Vijay}, the
interacting shell model is widely regarded as the most broadly capable
description of low-energy nuclear
structure, and the one most directly traceable to the fundamental many-body
problem.  Many studies have
demonstrated that exact diagonalizations of shell-model Hamiltonians can
accurately and consistently
describe a wide range of nuclear properties, {\it if} the many-body basis is
sufficiently large.  Pioneering
papers include work in the $0p$ \cite{Kurath}, $0s$-$1d$ \cite{Wild84} and
$0f_{7/2}$ \cite{McGrory} shells;
more recent examples are \cite{Monster,Zuker,Vallieres}.  Unfortunately,
the combinatorial scaling of the
many-body space with the size of the single particle basis or the number of
valence nucleons restricts such
exact diagonalizations to either light nuclei or to heavier nuclei with only
a few valence particles \cite{Wild88}.

The Shell Model Monte Carlo (SMMC) methods that have been developed in the
past few years
circumvent some of these difficulties while retaining the rigor, flexibility,
and predictive power of
traditional shell model calculations.  These methods are based on a Monte
Carlo evaluation of the path
integral obtained by a Hubbard-Stratonovich (HS) transformation
\cite{Hubbard} of the
imaginary-time evolution
operator.
The many-body problem is thus reduced to a set of one-body
problems in fluctuating
auxiliary fields \cite{Sugiyama}.
The method enforces the Pauli principle exactly, and the
storage and computation time
scale gently with the single-particle basis or the number of particles.
Auxiliary field methods have been
applied to condensed matter systems such as the Hubbard Model
\cite{Linden,Hirsch},
yielding
important information
about electron correlations and magnetic properties.

A series of papers \cite{Johnson,Lang,Ormand,Alhassid}
have described the development of SMMC methods.  The
purpose of the present
exposition is to summarize, in one document, the philosophy, formalism,
numerical methods, and
computational implementation of these techniques.  A coherent presentation is
particularly useful, as there
is a growing body of experience in such calculations and there have been a
few missteps along the way. We
also offer a sampler of ``first-of'' calculations that illustrate the power
and limitations of the SMMC
approach, as well as some interesting physics; some of the results presented
have not been published
previously.

Our presentation is organized as follows. In section~2 we review the
rationale and definition of the
interacting shell model, some of the important characteristics of the
two-body interaction, and discuss some
of the successes that direct diagonalization calculations have had in
describing a broad range of nuclear
properties. In section~3, we give an overview of SMMC methods, showing how the
HS transformation can be
used to express physically interesting observables as ratios of
high-dimensional integrals. In
section~4, we discuss the various ways in which a realistic shell model
hamiltonian can be cast in a
form suitable  for the HS transform. Section~5 is devoted to a summary of the
relevant aspects of MC
quadrature. In section~6, we discuss various details of implementation:
subtleties of the algorithm,
numerical issues, and computational implementation. In section~7
we discuss the notorious
sign problems and a practical method for their solution as well as the
validation of the overall method.
Section~8 is a sampler
of various types of SMMC
calculations (virtually all intractable by other methods), and in section~9
we offer a perspective on future
work. A brief appendix is devoted to the relevant properties of
independent-fermion quantum mechanics.
Our review covers work through December, 1995.

\noindent
\section{Review of the shell model}

In this section, we present a brief definition and overview of the nuclear
shell model.  Our
goals are to establish conventions and notation and to give the non-expert
some
appreciation for the special features of the nuclear problem relative to
other quantal
many-body systems.  More detailed discussions can be found in several texts
\cite{Ring,Lawson,Heyde,Talmi,Brussard}.

The notion of independent particles moving in a common one-body potential is
central to
our description of atoms, metals, and hadrons. It is also realized in nuclei,
and the shell
structure associated with the magic numbers was first put on a firm basis in
1949, when
the magic numbers were explained by an harmonic oscillator spectrum with a
strong,
inverted (with respect to the atomic case) spin-orbit potential \cite{Haxel}.

But nuclei differ from the other quantal systems cited above in that the
residual
interaction between the valence fermions is strong and so severely perturbs
the naive
single-particle picture.  This interaction mixes together many different
configurations to
produce the true eigenstates and, because of its coherence, there emerge
phenomena such
as pairing, modification of sum rules, deformation, and collective rotations
and vibrations.
An accurate treatment of the residual interaction is therefore essential to
properly
describe nuclei.

The nuclear shell model is defined by a set of spin-orbit coupled
single-particle states
with quantum numbers $ljm$ denoting the orbital angular momentum ($l$) and
the total
angular momenta ($j$) and its $z$-component, $m$.  Although non-spherical
one-body
potentials are a common efficiency used in describing deformed nuclei, for
the rotationally
invariant hamiltonians used in SMMC so far, these states have energies
$\varepsilon_{lj}$
that are
independent of $m$.  The single-particle states and energies may be different
for neutrons
and protons, in which case it is convenient to include also the isospin
component
$t_3=\pm 1/2$
in the state description.  We will use the label $\alpha$ for the
set of quantum
numbers $ljm$ or $ljmt_3$, as appropriate.

The particular single-particle states included in a given calculation depend
upon the
physics being addressed, but at least one major shell is believed to be
necessary to
adequately describe low-lying states of a given nucleus.  We will use $N_s$
to denote the
number of such states.  Thus, for example, $N_s$ is (12, 20, 32, 44) for
either neutrons or
protons in the ($1s0d$, $1p0f$, $2s1dg_{7/2}h_{11/2}$, $2p1fh_{9/2}i_{13/2}$)
shells.

The totality of Pauli-allowed configurations of the valence nucleons in the
single particle
states defines the model space in which the many-body Hamiltonian acts.
Computing the
dimension of this space (number of such
many-body states) is a straightforward
combinatorial
exercise.  As noted in the Introduction, the dimension increases strongly
with either
$N_s$ or the number of valence nucleons, and can vastly exceed $10^8$ for
realistic
applications of current interest.
The size of the Hamiltonian matrix to be considered can be reduced
somewhat
by exploiting rotational and isospin invariance properties.  Even so,
constructing such a
matrix and finding its lowest eigenvalues and eigenstates is difficult in
some cases of
interest and impossible in most, and thermal properties are completely
inaccessible
without {\it ad hoc} assumptions.

The shell model Hamiltonian can be written in the form
$\hat{H}=\hat{H}_1 +\hat{H}_2$ where
\timletter
\begin{eqnarray}
\hat{H}_1&= &\sum_\alpha \varepsilon_\alpha a^{\dagger}_\alpha a_\alpha \; ,
\\
\hat{H}_2&= &{{1}\over{2}}
\sum_{\alpha\beta\gamma\delta} V_{\alpha\beta\gamma\delta}
a^\dagger_\alpha a^{\dagger}_\beta a_\delta a_\gamma \; .
\end{eqnarray}
\timnoletter
Here, $a^{\dagger}$ and $a$ are fermion creation and annihilation operators,
and the $V$ are the uncoupled matrix elements of the two-body interaction.
These
latter must respect rotational and time-reversal invariance, and
parity conservation.
To make explicit the rotational invariance and shell structure, we can
rewrite the two-body Hamiltonian as
\begin{eqnarray}
\hat{H}_2 & = &{1 \over 2} \sum_{abcd}\sum_J
V_{J}(ab,cd)\sum_{M} \hat{A}_{JM}^\dagger
(ab)\hat{A}_{JM}(cd) \nonumber\\
& = & {1 \over 4} \sum_{abcd}\sum_J
\left[ (1+\delta_{ab})(1+\delta_{cd}) \right]^{1/2}
V_{J}^{A}(ab,cd)\sum_{M} \hat{A}_{JM}^\dagger
(ab)\hat{A}_{JM}(cd) \; ,
\end{eqnarray}
where the sum is taken over all proton and neutron single-particle orbits
(denoted by
$a,b,c,d$) and the pair creation and annihilation operators are given by
\timletter
\begin{eqnarray}
\hat{A}_{JM}^\dagger (a b) & = &
\sum_{m_{a} m_{b}}(j_a m_a j_b m_b
| J M) a_{j_b m_b}^\dagger
a_{j_a m_a}^\dagger = -[ a_{j_a}^\dagger \times
a_{j_b}^\dagger ]^{JM},
\\
\hat{A}_{JM}(ab) & = &
\sum_{m_{a}m_{b}}(j_a m_a j_b m_b| J M) a_{j_a m_a}
a_{j_b m_b} = [ a_{j_a} \times a_{j_b} ]^{JM}\;.
\end{eqnarray}
\timnoletter

The $V_J(ab,cd)$ are the angular-momentum coupled two-body matrix elements
(TBME) of a
scalar potential $V(\vec{r}_1,\vec{r}_2)$ defined as
\begin{equation}
V_J(ab,cd) = \langle [\psi_{j_a}(\vec{r}_1)\times
\psi_{j_b}(\vec{r}_2)]^{JM} |V(\vec{r}_1,\vec{r}_2) |
[\psi_{j_c}(\vec{r}_1)\times \psi_{j_d}(\vec{r}_2)]^{JM} \rangle\;,
\end{equation}
(independent of $M$) while the anti-symmetrized two-body matrix elements
$V_J^A(ab,cd)$ are given by
\begin{equation}
V_J^A(ab,cd)= \left[ (1+\delta_{ab})
(1+\delta_{cd})\right]^{-1/2}
\left[ V_J(ab,cd) -(-1)^{j_c+j_d-J}V_J(ab,dc)\right]\;.
\end{equation}

There are a number of methods for deriving the residual interaction $V$
(which acts in the
model space) from the underlying "bare" internucleon interaction
\cite{Kuo,Barrett,Machleidt}.
Although a complete specification of $V$ requires many two-body matrix
elements \cite{BW,Richter} (e.g.,
63 in the complete $sd$-shell and 195 in the complete $pf$-shell), successful
interactions that
reproduce a large body of experimental data show a few simple features (e.g.,
isoscalar
pairing, attractive quadrupole, repulsive dipole, $\ldots$), with the rest of
the TBMEs
being small and random \cite{Richter,Zuker94}.
A venerable approximation is to truncate the
interaction to
just the isoscalar pairing and the quadrupole interaction \cite{Baranger}.

Various aspects of the eigenstates of ${\hat H}$ can be understood through such
approximations as Hartree-Fock \cite{Flocard,Goeke} or Random Phase
\cite{Bertsch}.
However, such
approaches are unsuitable for precise work. The preferred method is to expand
the
wavefunction in a many-body basis (truncated by some criteria, if necessary)
and then
construct and diagonalize the resulting hamiltonian matrix.
Several different approaches to these tasks
have been followed over the years. In the {\it jj-coupling scheme},
developed by the Rochester-Oak Ridge group \cite{McGrory},
the antisymmetric $N$-particle states in each
$j$-shell are first built recursively,
before the multishell model space is constructed from these
single-shell states. The most time-consuming part of this method
is the calculation of the single-particle coefficients of fractional
parentage to form antisymmetric $N$-particle states and it has been found
that this scheme has severe limitations when the shell dimensions
are large. In the {\it m-scheme} approach \cite{Livermore,Ohio}
each $N$-particle Slater determinant is represented by a computer bit
using the binaries 0 and 1 to represent unoccupied and occupied
levels. An advantage is that the $m$-scheme reduces the calculation
of matrix elements (in second quantization) to very efficient logical
operations. However, it becomes impractical when the
dimensions involved get very large (i.e., with increasing numbers of particles
and/or orbitals). The {\it OXBASH} code combines $jj$-coupling and
$m$-scheme \cite{Oxbash1,Oxbash2} by constructing the basis states in the
$m$-scheme
approach, but following the $jj$-coupling scheme in the diagonalization
of the Hamilton matrix. Recently, the Drexel group has developed an
alternative approach based on a pair-truncated fermion model space
and permutation group ideas \cite{Vallieres}.
A more detailed description of the various
shell model concepts and codes can be found in \cite{Vall91}.

In conventional shell model applications the dimension of the model space
makes a complete diagonalization of the Hamiltonian impractical.
As one is usually interested only in the nuclear spectrum at low
energies, the diagonalization is often performed using the Lanczos
algorithm, which is an efficient way to find the few lowest (or highest)
eigenvalues and eigenvectors of a large matrix \cite{Lanczos,Koonin}.

Physical applications of the shell model were pioneered by
Lawson and by Kurath and Cohen in the $p$-shell \cite{Lawson,Kurath}.
As
the dimensions of the model spaces
in the $sd$-shell
are still
not too large, the conventional shell model approaches are well suited
and have been quite extensively and successfully applied.
In particular, Wildenthal has performed
systematic studies of all $sd$-shell nuclei and,
using a specially designed effective interaction \cite{BW},
has been able
to
reproduce energies and transition moments in these nuclei
\cite{Wild88}. However, Figure 2.1 illustrates that accurate results can be
obtained only in an untruncated basis.

\begin{figure}
$$\epsfxsize=4truein\epsffile{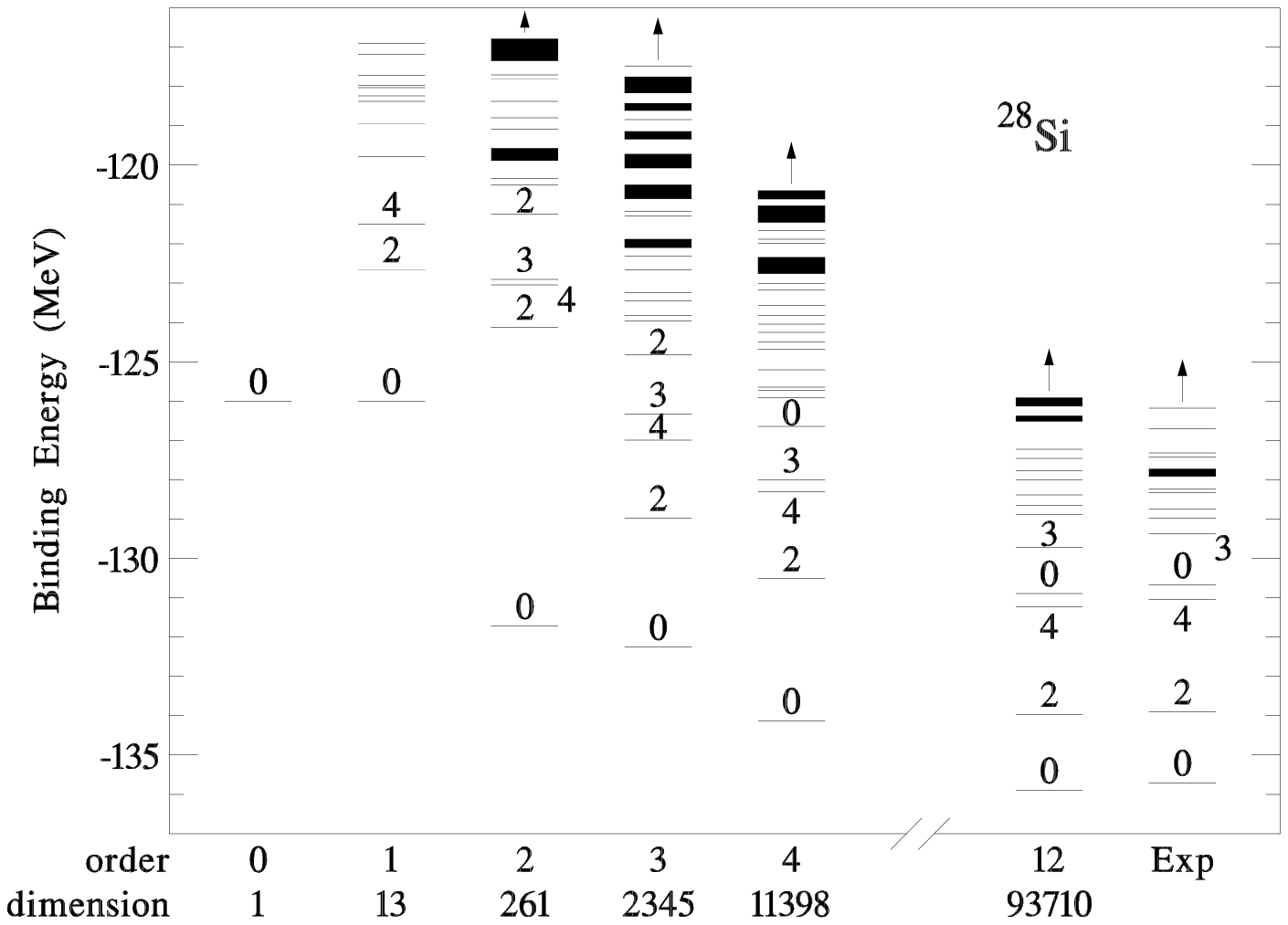}$$
{FIG~2.1
Convergence of the spectrum of $^{28}$Si with increasing dimension
of the model space; ``order 12'' corresponds to the complete
$sd$-shell space \protect\cite{Wild88}.
}
\end{figure}

Beyond the $sd$-shell, applications of the conventional shell model
become more problematic as the large dimensions of the model space
make  diagonalization impossible. To date, systematic
studies of the $A=48$ isobars, involving
model spaces with about $10^6$ Slater determinants, have been the limit
for the conventional shell model approaches \cite{Caurier94}. Nevertheless
the successful description of these nuclei underlines again that
the shell model concept is also the method of choice in heavier
nuclei -- if such applications were only possible. As noted by
Caurier {\it et al.} \cite{Caurier94}, it took
two generations of hardware and software development to extend
shell model calculations from $A=44$ to $A=48$. Thus, conventional
shell model calculations for nuclei heavier than $A=50$ appear out of
reach in the near future, even assuming an optimistic increase
of computer capabilities.

Facing the impracticality of complete shell
model calculations, Poves, Zuker, and collaborators \cite{Poves} have
used a truncation method for the shell model in the $1p0f$ space.
When applied to observables like the ground state energy
or Gamow-Teller strength,
the convergence appears fast enough to obtain
reasonable approximations to the (complete) shell model results after
a few truncation scheme iterations. As an example, Fig. 2.2 shows
the Gamow-Teller strengths in $^{54}$Fe and $^{56}$Fe
calculated at various levels of truncation; there
is clear convergence to the complete result obtained with
the SMMC methods (see Section 8).

\begin{figure}
\begin{center}
$$\epsfxsize=4truein\epsffile{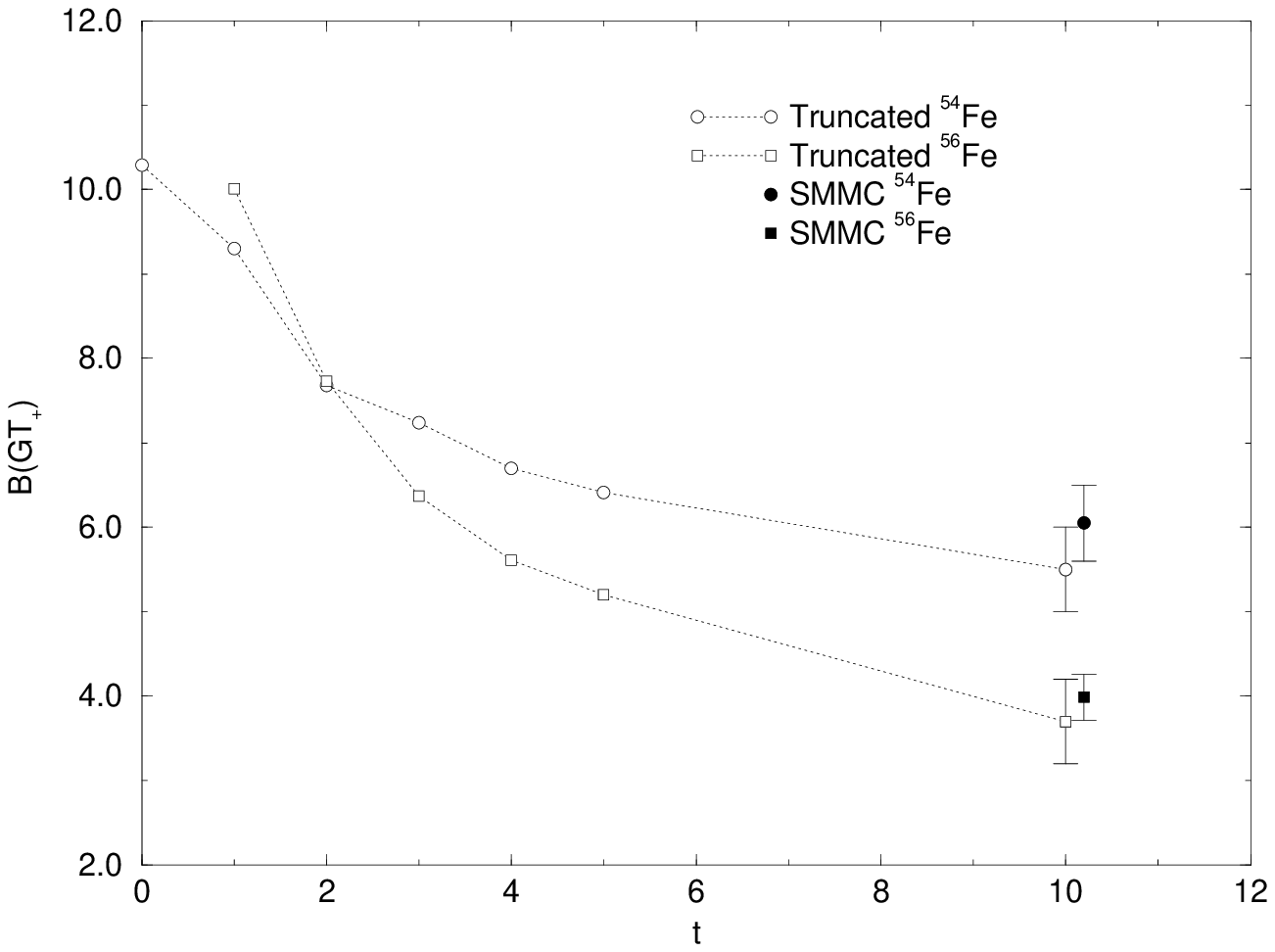}$$
\end{center}
\label{Fig22}
{FIG~2.2
Comparison of the total Gamow-Teller strengths B(GT$_+$) for $^{54,56}$Fe
in a series of direct diagonalizations with decreasing level of truncation
\protect\cite{Poves} with the complete results obtained with the
SMMC method (solid symbols at $t=10$)
\protect\cite{Langanke95}. The open symbols at $t=10$ show the extrapolated
no-truncation results of \protect\cite{Poves}.
}
\end{figure}

\section{Overview of Monte Carlo Methods}

In this section, we give an overview of SMMC methods. In particular, we
describe what the methods are (and are not) capable of calculating. We then
discuss the essence of the HS transformation, which is at
the heart of the calculations.

The SMMC methods rely on an ability to deal with the imaginary-time many-body
evolution operator, $\exp (-\beta\hat H)$, where $\beta$ is a real
$c$-number. While this does not result in a complete solution to the many-body
problem in the sense of giving all eigenvalues and eigenstates of $\hat H$,
it can result in much useful information. For example, the expectation value
of some observable $\hat \Omega$ in the grand canonical ensemble can be
obtained
by adding to $\hat H$ a term $-\mu_n \hat N-\mu_p \hat Z$, ($\mu_n$ and
$\mu_p$ are the neutron and proton chemical potentials) and then calculating
\begin{equation}
\langle \hat \Omega\rangle=
{{\rm Tr}\, e^{-\beta\hat H}\hat \Omega\over
{\rm Tr}\, e^{-\beta\hat H}}\;.
\end{equation}
Here, $\beta\equiv T^{-1}$ is interpreted as the inverse of the temperature
$T$, and the many-body trace is defined as
\begin{equation}
{\rm Tr}\,\hat X\equiv\sum_i \langle i\vert \hat X\vert i\rangle\;,
\end{equation}
where the sum is over {\it all}
many-body states of the system. Similarly, if $\hat
P_A= \delta(A-\hat N)$ is the projector onto states with $A$ nucleons
(actually the product of separate neutron and proton projectors), the
canonical ensemble is defined by
\begin{equation}
{\rm Tr}_A\,\hat X\equiv \sum_i \langle i\vert \hat P_A\hat X\vert
i\rangle\;,
\end{equation}
and the associated expectation value is
\begin{equation}
\langle \hat \Omega\rangle_A=
{{\rm Tr}_A\, e^{-\beta\hat H}\hat \Omega\over
{\rm Tr}_A\, e^{-\beta\hat H}}\;.
\end{equation}

In the limit of low temperature ($T\rightarrow0$ or
$\beta\rightarrow\infty$), both the grand-canonical and canonical traces
reduce to ground state expectation values. Alternatively, if $\vert
\Phi\rangle$ is a many-body trial state not orthogonal to the exact ground
state, $\vert\Psi\rangle$, then $e^{-\beta\hat H}$ can be used as a filter to
refine $\vert\Phi\rangle$ to $\vert\Psi\rangle$ as $\beta$ becomes large. An
observable can be calculated in this ``zero temperature'' method as
\begin{equation}
{\langle\Phi\vert e^{-{\beta\over2}\hat H}\hat \Omega e^{-{\beta\over2}\hat H}
\vert\Phi\rangle\over
\langle\Phi\vert e^{-\beta\hat H}\vert\Phi\rangle}
{}~\hbox{$\longrightarrow\hskip-20pt{}^{\beta\rightarrow\infty}$}
{\langle\Psi\vert\hat \Omega\vert\Psi\rangle\over
\langle\Psi\vert\Psi\rangle}\;.
\end{equation}

If $\hat \Omega$ is the hamiltonian, then (3.5) at $\beta=0$ is the variational
estimate of the energy, and improves as $\beta$ increases. Of course, the
efficiency of the refinement for any observable depends upon the degree to
which $\vert\Phi\rangle$ approximates $\vert\Psi\rangle$.

Beyond such static properties, $e^{-\beta\hat H}$ allows us to obtain some
information about the dynamical response of the system. For operators $\hat
\Omega^\dagger$ and $\hat \Omega$, the response
function $R_\Omega(\tau)$ in the canonical
ensemble is defined as
\begin{equation}
R_\Omega(\tau)\equiv
{{\rm Tr}_A\, e^{-(\beta-\tau)\hat H} \hat
\Omega^\dagger e^{-\tau\hat H}\hat \Omega\over
{\rm Tr}_A\, e^{-\beta\hat H}}\equiv
\langle\hat \Omega^\dagger(\tau)\hat \Omega(0)\rangle_A,
\end{equation}
where
$\hat \Omega^\dagger(\tau)\equiv e^{\tau\hat H}\hat
\Omega^\dagger e^{-\tau\hat H}$
is the
imaginary-time Heisenberg operator. As we shall see in section~8 below,
interesting choices for $\hat{\Omega}$ are the $a_j$ for particular orbitals,
the
Gamow-Teller, $M1$, or quadrupole moment, etc.
Inserting complete sets of $A$-body eigenstates of $\hat H$ ($\{\vert
i\rangle,\vert f\rangle\}$ with energies $E_{i,f}$) shows that
\begin{equation}
R_\Omega(\tau) ={1\over Z}\sum_{if} e^{-\beta E_i}
\vert\langle f\vert\hat \Omega\vert i\rangle\vert^2
e^{-\tau(E_f-E_i)},
\end{equation}
where $Z=\sum_i e^{-\beta E_i}$ is the partition function.
Thus, $R_\Omega(\tau)$
is the Laplace transform of the strength function $S_\Omega(E)$:
\begin{timeqnarray}
R_\Omega(\tau)& = & \int^\infty_{-\infty} e^{-\tau E} S_\Omega(E)dE\;; \\
S_\Omega(E)&=&  {1\over Z}\sum_{fi} e^{-\beta E_i}
\vert\langle f\vert \hat \Omega\vert i\rangle\vert^2
\delta(E-E_f+E_i)\;.
\end{timeqnarray}

\noindent Hence, if we can calculate $R_\Omega(\tau)$, $S_\Omega(E)$
can be determined. Short of
a full inversion of the Laplace transform (which is often numerically
difficult; see section~6.4 below), the behavior of $R_\Omega(\tau)$ for small
$\tau$ gives information about the energy-weighted moments of $S_\Omega$. In
particular,
\begin{equation}
R_\Omega(0)=\int^\infty_{-\infty} S_\Omega (E) dE=
{1\over Z}\sum_i e^{-\beta E_i}
\vert\langle f\vert \hat{\Omega}\vert i\rangle\vert^2=
\langle \hat{\Omega}^\dagger\hat{\Omega}\rangle_A
\end{equation}
is the total strength,
\begin{equation}
-R_\Omega^\prime (0)=\int^\infty_{-\infty} S_\Omega (E) EdE=
{1\over Z}\sum_{if} e^{-\beta E_i}
\vert\langle f\vert\hat \Omega\vert i\rangle\vert^2
(E_f-E_i)
\end{equation}
is the first moment,
\begin{equation}
R^{\prime\prime}_\Omega(0)= \int^\infty_{-\infty} S_\Omega (E) E^2 dE=
{1\over Z}\sum_{if} e^{-\beta E_i}
\vert\langle f\vert\hat \Omega\vert i\rangle\vert^2
(E_f-E_i)^2
\end{equation}
is the second moment, and so on. (In these expressions, the prime denotes
differentiation with respect to $\tau$.)

It is important to note that we cannot usually obtain detailed spectroscopic
information from SMMC calculations.
Rather, we can calculate expectation values of
operators in the thermodynamic ensembles or the ground state. Occasionally,
these can indirectly furnish properties of excited states. For example, if
there is a collective $2^+$ state absorbing most of the $E2$ strength, then
the centroid of the quadrupole response function will be a good estimate of
its energy. But, in general, we are without the numerous specific excitation
energies and wavefunctions that characterize a direct diagonalization. This
is both a blessing and a curse. The former is that for the very large model
spaces of interest, there is no way in which we can deal explicitly with all
of the wavefunctions and excitation energies. Indeed, we often don't need to,
as experiments only measure average nuclear properties at a given excitation
energy. The curse is that comparison with detailed properties of
specific levels is difficult. In this sense, the SMMC method is complementary
to direct diagonalization for modest model spaces, but is the only method for
treating very large problems.

It remains, of course, to describe the Hubbard-Stratanovich ``trick'' by
which $e^{-\beta\hat H}$ is managed. In broad terms, the difficult many-body
evolution is replaced by a superposition of an infinity of tractable one-body
evolutions, each in a different external field, $\sigma$. Integration over
the external fields thus reduces the many-body problem to quadrature. The
following schematic discussion serves to illustrate the central idea;
important details will be added in the following sections.

With some rearrangement, as discussed in Section 4,
the many-body Hamiltonian (2.1) can be written
schematically as
\begin{equation}
\hat H=\varepsilon\hat {\cal O} +{1\over2}V\hat{\cal O}\hat{\cal O}\;,
\end{equation}
where $\hat{\cal O}$ is a density operator of the form $a^\dagger a$, $V$
is the strength of the
two-body interaction, and $\varepsilon$ a single-particle energy. In the full
problem, there are many such quantities with various orbital indices that are
summed over, but we omit them here for the sake of clarity.

All of the difficulty arises from the two-body interaction, that term in
$\hat H$ quadratic in $\hat{\cal O}$. If $\hat H$ were solely linear in
$\hat{\cal O}$, we would have a one-body quantum system, which is readily
dealt with (see the Appendix). To linearize the evolution, we employ the
Gaussian identity
\begin{equation}
e^{-\beta\hat H}=
\sqrt{\beta \mid V\mid \over 2\pi} \int^\infty_{-\infty}
d\sigma e^{-{1\over2}\beta \mid V\mid \sigma^2} e^{-\beta\hat h};\qquad
\hat h= \varepsilon \hat{\cal O} +s V\sigma\hat{\cal O}\;.
\end{equation}
Here, $\hat h$ is a one-body operator associated with a $c$-number field
$\sigma$, and the many-body evolution is obtained by integrating the one-body
evolution $\hat U_\sigma\equiv e^{-\beta\hat h}$ over all $\sigma$ with a
Gaussian weight. The phase, $s$, is $1$ if $V<0$ or $i$ if
$V>0$.
Equation (3.13) is easily verified by completing the square
in the exponent of the integrand and then doing the integral; since there is
only a single operator $\hat{\cal O}$, there is no need to worry about
non-commutation.

With an expression of the form (3.13), it is now straightforward to write
observables as the ratio of two integrals. For example, the canonical
expectation value (3.4) becomes
\begin{equation}
\langle\hat \Omega\rangle_A=
{\int d\sigma e^{-{\beta\over2}\mid V\mid \sigma^2}{\rm Tr}_A\,
\hat U_\sigma\hat \Omega\over
\int d\sigma e^{-{\beta\over2}\mid V\mid\sigma^2}{\rm Tr}_A\, \hat U_\sigma}\;,
\end{equation}
which can be more conveniently written as
\begin{timeqnarray}
&\langle\hat \Omega\rangle_A &=
{\int d\sigma W_\sigma \Omega_\sigma\over
\int d\sigma W_\sigma};
\\
W_\sigma = G_\sigma {\rm Tr}_A\, \hat U_\sigma;
\qquad\qquad G_\sigma &=&e^{-{\beta\over2}\mid V\mid \sigma^2}\;,
\qquad\quad \Omega_\sigma =
{{\rm Tr}_A\,\hat U_\sigma\hat \Omega\over {\rm Tr}_A\,\hat U_\sigma}
\qquad\qquad\;.
\end{timeqnarray}

\noindent
Thus, the many-body observable is the weighted average (weight $W_\sigma$) of
the observable $\Omega_\sigma$ calculated in a canonical ensemble
involving only the one-body evolution $\hat U_\sigma$. Similar expressions
involving two $\sigma$ fields (one each for $e^{-\tau\hat H}$ and
$e^{-(\beta-\tau)\hat H}$) can be written down for the response function
(3.6) and all are readily transcribed to the grand-canonical ensemble (3.1)
or zero-temperature case (3.5).

An expression of the form (3.15) has a number of attractive features. First,
the problem has been reduced to quadrature---we need only calculate the ratio
of two integrals. Second, all of the quantum mechanics
(which appears in $\Omega_\sigma$) is
of the one-body variety, which is simply handled by the
algebra of $N_s\times N_s$ matrices. The price to pay is that we
must treat the one-body problem for all possible $\sigma$ fields.

For a realistic hamiltonian, there will be many non-commuting density
operators $\hat{\cal O}_\alpha$ present, but we will always be able to reduce
the two-body term to diagonal form. Thus for a general two-body interaction
in a general time-reversal invariant form we write
\begin{equation}
\hat{H}=\sum_\alpha
\left(\epsilon^\ast_\alpha \hat{\bar{\cal O}}_\alpha+
\epsilon_\alpha \hat {\cal O}_\alpha\right)+
{1\over2}\sum_\alpha V_\alpha
\left\{ \hat {\cal O}_\alpha, \hat {\bar{\cal O}}_\alpha\right\}\;,
\end{equation}
where $\hat {\bar{\cal O}}_\alpha$ is the time reverse of
$\hat {\cal O}_\alpha$.
Since in general, $[\hat {\cal O}_\alpha, \hat{\cal O}_\beta]\not=0$
we must split the interval $\beta$ into $N_t$ ``time slices'' of length
$\Delta\beta\equiv\beta/N_t$,
\begin{equation}
e^{-\beta\hat H}= [e^{-\Delta\beta\hat H}]^{N_t},
\end{equation}
and for each time slice $n=1, \ldots, N_t$ perform a linearization similar
to (3.16) using auxiliary fields $\sigma_{\alpha n}$:
\begin{eqnarray}
e^{-\Delta\beta\hat H}& \approx
\int^\infty_{-\infty} \prod_n
\left(d\sigma_{\alpha n}d\sigma_{\alpha n}^{*}\Delta\beta \mid V_\alpha\mid
\over2\pi\right)
e^{-\Delta\beta \sum_\alpha \mid V_\alpha\mid
\mid\sigma_{\alpha n}\mid^2}
e^{-\Delta\beta\hat h_n}\;;
& \nonumber\\
\hat h_n &=  \sum_\alpha
\left(\varepsilon_\alpha^{*} +
s_\alpha V_\alpha \sigma_{\alpha n}\right)\hat{\bar{\cal O}}_\alpha+
\left(\varepsilon_\alpha +
s_\alpha V_\alpha \sigma_{\alpha n}^{*}\right)\hat{\cal O}_\alpha .
\end{eqnarray}
Note that because the various $\hat {\cal O}_\alpha$ need not commute, (3.18)
is
accurate only through order $\Delta\beta$ and that
the representation of $e^{-\Delta\beta\hat h}$ must be accurate through order
$\Delta\beta^2$
to achieve that accuracy.

Upon composing (3.18) many times according to (3.17), we can write
expressions for observables as the ratio of two field integrals. For example,
(3.15) becomes
\begin{timeqnarray}
\langle\hat \Omega\rangle_A =
 {\int{\cal D} \sigma W_\sigma \Omega_\sigma\over
 \int{\cal D} \sigma W_\sigma},\qquad\qquad\qquad\qquad
\\
\noalign{\noindent where\hfill}
W_\sigma =  G_\sigma {\rm Tr}_A\,\hat U\;;\qquad
G_\sigma=e^{-{\Delta\beta}\sum_{\alpha n}\mid V_\alpha\mid
\mid\sigma_{\alpha n}\mid^2}\;;\nonumber \\
 \Omega_\sigma= {{\rm Tr}_A\, \hat U\hat \Omega\over
 {\rm Tr}_A\, \hat U}\;;\qquad
{\cal D} \sigma \equiv
 \prod^{N_t}_{n=1}\prod_\alpha d\sigma_{\alpha n}d\sigma_{\alpha n}^{*}
 \left(\Delta\beta\vert V_\alpha\vert\over 2\pi\right),
 \\
\noalign{\noindent and\hfill}
\hat U=
 \hat U_{N_t}\ldots \hat U_2\hat U_1\;;\qquad
 \hat U_n= e^{-\Delta\beta\hat h_n};\nonumber \\
\hat h_n =\sum_\alpha
\left(\varepsilon_\alpha^{*} +
s_\alpha V_\alpha \sigma_{\alpha n}\right)\hat{\bar{\cal O}}_\alpha+
\left(\varepsilon_\alpha +
s_\alpha V_\alpha \sigma_{\alpha n}^{*}\right)\hat{\cal O}_\alpha\;.
 \end{timeqnarray}

\noindent This is, of course, a discrete version of a path integral over
$\sigma$.
Because there is a field variable for each operator at each time slice, the
dimension of the integrals ${\cal D} \sigma$ can be very large, often
exceeding $10^5$. Note that because of the errors in (3.18), the errors in
Eqs.~(3.19) are of order $\Delta\beta$, so that high accuracy requires large
$N_t$ and perhaps extrapolation to $N_t=\infty$ ($\Delta\beta=0$).

Two steps are then necessary to implement a SMMC calculation. First, we must
learn how to write expressions like (3.16, 3.19) for a realistic hamiltonian.
Second, we must learn how to efficiently compute the resulting ratio of
high-dimensional integrals. These tasks are, respectively, the subjects of
the following two sections.

\section{Decomposition of the Hamiltonian}

To realize the HS transformation, the two-body parts of $\hat{H}$ must be
cast as a quadratic form in one-body density operators
$\hat{\cal{O}}_\alpha$. As discussed in Ref.~\cite{Lang}, there is considerable
freedom in doing so. In the simplest example, let us consider an individual
interaction term,
\begin{equation}
\hat{H}=a_1^\dagger a_2^\dagger a_4 a_3\;,
\end{equation}
where $a_i^\dagger, a_i$ are anti-commuting fermion creation and annihilation
operators.

There are then two ways to proceed: we can group $(1,3) $ and $(2,4)$ to get
\begin{timeqnarray}
\hat{H} &=& \overbrace{a_1^\dagger a_3}
\underbrace{a_2^\dagger a_4}
-a_1^\dagger a_4 \delta_{23}
\\
&= & -a_1^\dagger a_4 \delta_{23}+
{1 \over 2}[a_1^\dagger a_3,
a_2^\dagger a_4]+{1\over4}
(a_1^\dagger a_3+a_2^\dagger a_4)^2
-{1\over 4} (a_1^\dagger a_3 -a_2^\dagger a_4)^2\;,
\end{timeqnarray}
or group $(1,4)$ and $(2,3)$ to get
\begin{timeqnarray}
\hat{H}& =& -\overbrace {a_1^\dagger a_4}
\underbrace{ a_2^\dagger
a_3}+a_1^\dagger a_3 \delta_{24}
\\
&= & a_1^\dagger a_3 \delta_{24} -
{1\over 2} [a_1^\dagger a_4,
a_2^\dagger a_3]-{1\over 4}(a_1^\dagger
a_4+a_2^\dagger a_3)^{2}
+{1\over 4} (a_1^\dagger a_4 -
a_2^\dagger a_3)^2\;.
\end{timeqnarray}

\noindent The commutator terms in both (4.2b, 4.3b) are one-body operators, but
now the
quadratic forms are squares of density operators. We refer to Eq.~(4.2b) as
the `direct' decomposition and Eq.~(4.3b) as the `exchange' decomposition.  At
the formal level, it is also possible to consider a pairing decomposition
\cite{Lang}, although no practical calculations have been done with that
scheme.

For any general two-body Hamiltonian, there is the freedom to choose between
the direct and exchange formulations and it is particularly convenient to use
quadratic forms of density operators that respect rotational invariance,
isospin symmetry, parity conservation,
and the shell structure of the system. Although the exact
path integral result is independent of the scheme used, different schemes
will lead to different results under certain approximations (e.g.,~mean field
or static path approximation \cite{SPA}).
The choice of decomposition will also affect the rate of convergence
of numerical results as $N_t \rightarrow \infty$, as well as the statistical
precision of the Monte Carlo evaluation.

We note that the two-body Hamiltonian for fermion systems is completely
specified by the set of anti-symmetrized two-body matrix elements
$V_J^A(ab,cd)$ of Eq. (2.5)
that are the input to many standard shell model codes such as
OXBASH \cite{Oxbash1}.
Indeed, we can add to the $V_J^A(ab,cd)$ any set of (unphysical)
symmetric two-body matrix elements $V_J^S(ab,cd)$ satisfying
\begin{equation}V_J^S(ab,cd) = (-1)^{j_c+j_d-J}V_J^S(ab,dc)\;,
\end{equation}
without altering the action of $\hat{H}_2$ on any many-fermion wave function.
However, note that although the $V_J^S(ab,cd)$ do not alter the eigenstates
and eigenvalues of the full Hamiltonian, they can (and do) affect the
character of the decomposition of $\hat{H}_2$, as is shown below. In what
follows, we define the set of two-body matrix elements $V_J^N(ab,cd)$ that
may have no definite symmetries as
\begin{equation}V_J^N(ab,cd) = V_J^A(ab,cd)+V_J^S(ab,cd)\;,
\end{equation}
allowing us to write the two-body Hamiltonian (2.2) as
\begin{equation}\hat{H}_2 ={1 \over 4} \sum_{abcd}\sum_J
\left[ (1+\delta_{ab})(1+\delta_{cd}) \right
]^{1/2}V_{J}^{N}(ab,cd)\sum_{M}
\hat{A}_{JM}^\dagger (ab)\hat{A}_{JM}(cd)\;.
\end{equation}

To decompose $\hat{H}_2$, we perform a Pandya
transformation to recouple $(a,c)$ and $(b,d)$ into density operators with
definite multipolarity,
\begin{equation}\hat{\rho}_{KM}(ab)=\sum_{m_a,m_b}
(j_a m_aj_b m_b|K M)
a^\dagger _{j_a m_a}
\tilde{a}_{j_b m_b}\;,
\end{equation}
where $\tilde{a}_{j_a m_a} = (-1)^{j_a+m_a}
a_{j_a-m_a}$.
Then $\hat{H}_2$ can be rewritten as
\begin{equation}\hat{H}_2=\hat{H}'_2 +\hat{H}'_1 \;;
\end{equation}
\begin{equation}\hat{H}_2 '={1 \over 2}\sum_{abcd}\sum_K E_K
(ac,bd)\sum_M (-1)^{M}
\hat{\rho}_{K-M}(ac)\hat{\rho}_{KM}(bd)\;,
\end{equation}
where the particle-hole matrix elements of the interaction are
\begin{eqnarray}
E_{K}(ac,bd) = & (-1)^{j_b+j_c} \sum_J (-1)^J (2J+1)
\left\{
\matrix{ j_a & j_b & J \nonumber \\
j_d & j_c & K}\right\} & \nonumber \\
\times & {1\over 2} V^{N}_{J}(ab,cd) \sqrt{(1+\delta_{ab}) (1+\delta_{cd})}\;,
&
\end{eqnarray}
and $\hat{H}_1 '$ is a one-body operator given by
\begin{equation}\hat{H}'_1 = \sum_{ad}
\epsilon'_{ad}\hat{\rho}_{0\, 0}(a,d)\;,
\end{equation}
with
\begin{equation}\epsilon'_{ad} = -{1\over 4} \sum_b \sum_J
(-1)^{J+j_a+j_b} (2J+1){1 \over \sqrt{2 j_a +1}}
V^{N}_{J}(ab,bd)\sqrt{(1+\delta_{ab}) (1+\delta_{cd})}\;.
\end{equation}
Note that adding symmetric matrix elements is equivalent to using the
exchange decomposition for some parts of the interaction. The freedom in
choosing the combinations of direct and exchange decomposition is then
embodied in the arbitrary symmetric part of the matrix elements $V^N_J$.

Introducing the shorthand notation $i=(ac),j=(bd)$, we can write Eq.~(4.9)
as
\begin{equation}\hat{H}'_2 = {1\over 2} \sum_{ij} \sum_{K}
E_{K}(i,j) (-1)^M
\hat{\rho}_{KM}(i)\hat{\rho}_{K-M}(j)\;.
\end{equation}
Upon diagonalizing the matrix $E_K(i,j)$ to obtain eigenvalues
$\lambda_{K\alpha}$ and associated eigenvectors $v_{K\alpha}$, we can
represent $\hat{H}'_2 $ as
\begin{equation}\hat{H}'_2 ={1\over 2}\sum_{K\alpha}\lambda_{K}(\alpha)
(-1)^M \hat{\rho}_{KM}(\alpha) \hat{\rho}_{K-M}(\alpha)\;,
\end{equation}
where
\begin{equation}\hat{\rho}_{KM}(\alpha)=\sum_i
\hat{\rho}_{KM}(i)v_{K\alpha}(i)\;.
\end{equation}

Finally, if we define
\begin{timeqnarray}
\hat{Q}_{KM}(\alpha) \equiv & {1 \over \sqrt{2(1+\delta_{M0})}}
\left(\hat{\rho}_{KM}(\alpha)+(-1)^{M}\hat{\rho}_{K-M}(\alpha)
\right)\;,&\\
\hat{P}_{KM}(\alpha) \equiv & -{i\over \sqrt{2(1+\delta_{M0})}}
\left(\hat{\rho}_{KM}(\alpha)-(-1)^{M}\hat{\rho}_{K-M}(\alpha)
\right)\;,&
\end{timeqnarray}
then $\hat{H}_2 '$ becomes
\begin{equation}\hat{H}'_2={1 \over 2}\sum_{K \alpha}\lambda_K
(\alpha)\sum_{M\geq 0}
\left(\hat{Q}_{KM}^{2}(\alpha)+\hat{P}_{KM}^{2}
(\alpha)\right)\;.
\end{equation}
This completes the representation of the two-body interaction as a diagonal
quadratic form in density operators. We then couple auxiliary fields
$\sigma_{KM}(\alpha)$ to $\hat{Q}_{KM}$ and $\tau_{KM}(\alpha)$ to
$\hat{P}_{KM}$ in the HS transformation. (The latter are not to be confused
with the ``imaginary time'' $\tau$.)

In the treatment thus far, protons and neutrons were not distinguished from
each other. Although the original Hamiltonian $\hat{H}_2$ conserves proton
and neutron numbers, we ultimately might deal with one-body operators
$\rho_{KM}(a_p,b_n)$ and $\rho_{KM}(a_n,b_p)$ ({\it n,p} subscripts denoting
neutron and proton) that individually do not do so. The one-body hamiltonian
$\hat{h}_\sigma$ appearing in the HS transformation then mixes neutrons and
protons. The single-particle wavefunctions in a Slater determinant then
contain both neutron and proton components and neutron and proton numbers are
not conserved separately in each Monte Carlo sample; rather the conservation
is enforced only statistically.

It is, of course, possible to recouple so that only density operators
separately conserving neutron and proton numbers ($\hat{\rho}_{KM}(a_p,b_p)$
and $\hat{\rho}_{KM}(a_n,b_n)$) are present. To do so, we write the two-body
Hamiltonian in a manifestly isospin-invariant form,
\begin{equation}\hat{H}_2={1\over 4} \sum_{abcd}\sum_{JT}
\left[(1+\delta_{ab})(1+\delta_{cd})\right]^{1/2}
V_{JT}^{N}(ab,cd)\sum_{MTz} {\hat A^\dagger} _{JT;MTz}
(ab) {\hat A}_{JT;M Tz}(cd)\;,
\end{equation}
where, similar to the previous definition (4.6), the pair operator is
\begin{equation}{\hat A}^\dagger _{JT;MTz}(ab)=\sum_{m_a,m_b}
(j_a m_a j_b m_b|JM)({1 \over 2} t_a {1 \over 2} t_b|T
T_z)a^\dagger _{j_b m_b
t_b}a^\dagger _{j_a m_a t_a}\;.
\end{equation}
Here $({1 \over 2},t_a)$, etc. are the isospin indices with $t_a=-{1 \over
2}$ for proton states and $t_a={1 \over 2}$ for neutron states, and $(T T_z)$
are the coupled isospin quantum numbers. The two-body Hamiltonian can now be
written solely in terms of density operators that conserve the proton and
neutron numbers. Namely,
\begin{equation}\hat{H}_2 = \hat{H}'_1 + \hat{H}'_2\;,
\end{equation}
where
\begin{equation}\hat{H}'_1 = \sum_{ad} \sum_{t = p,n}
\epsilon'_{ad} \rho_{0\, 0 t}(a,d)\;,
\end{equation}
with
\begin{equation}\epsilon'_{ad} = -{1\over 4} \sum_b \sum_J
(-1)^{J+j_a+j_b} (2J+1){1 \over \sqrt{2 j_a +1}}
V^{N}_{J T=1}(ab,bd)\sqrt{(1+\delta_{ab})(1+\delta_{cd})}\;,
\end{equation}
and
\begin{equation}\hat{H}'_2 = {1\over 2} \sum_{abcd} \sum_{K, T=0,1}
E_{KT}(ac,bd)
[\hat{\rho}_{KT}(i)\times\hat{\rho}_{KT}(j)]^{J=0}\;.
\end{equation}
Here, we define $\hat{\rho}_{KMT}$ as
\begin{equation}\hat{\rho}_{K\,MT} = \hat{\rho}_{K\, Mp} +(-1)^T
\hat{\rho}_{K\, Mn}\;,
\end{equation}
and the $E_{KT}$ are given by
\begin{eqnarray}
E_{KT=0}(ac,bd) = & (-1)^{j_b+j_c} \sum_J (-1)^J (2J+1)
\left\{ \matrix{ j_a & j_b & J \cr j_d & j_c & K \cr}\right\}
\sqrt{(1+\delta_{ab})(1+\delta_{cd})} & \nonumber \\
  \times &
{1 \over 2} \left[
V^N_{JT=1}(ab,cd) +{1\over
2}(V_{JT=0}^{A}(ab,cd)-V_{JT=1}^{S}(ab,cd))
\right]\;,
&
\end{eqnarray}
\begin{eqnarray}
E_{KT=1}(ac,bd) = & -(-1)^{j_b+j_c} \sum_J (-1)^J (2J+1)
\left\{ \matrix{ j_a & j_b & J \cr j_d & j_c & K \cr}\right\}
\sqrt{(1+\delta_{ab})(1+\delta_{cd})} \nonumber \\
\times &
{1 \over 4} \left(
V^A_{JT=0}(ab,cd) -V^S_{JT=1}(ab,cd)
\right).
&
\end{eqnarray}

In this isospin formalism, since ${\hat A}_{JT;MTz}(ab) = (-1)^{j_a+j_b -J+T}
{\hat A}_{JT;MTz} (ba)$, the definitions of the
symmetric and antisymmetric parts of
$V_{JT}^{N} (ab,cd)$, $V^S_{JT} (ab,cd)$, and $V^A_{JT} (ab,cd)$ become
\begin{equation}V^{S/A}_{JT} (ab,cd) \equiv
{1 \over 2}
\left[
V^{N}_{JT} (ab,cd) \pm (-1)^{J +j_a+j_b +T-1}V^{N}_{JT}(ba,cd)
\right]\;.
\end{equation}

Note that these expressions allow less freedom in manipulating the
decomposition since we have to couple proton with proton and neutron with
neutron in forming the density operators. Also note that
$E_{KT=0}(ac,bd)-E_{KT=1}(ac,bd)$ is an invariant related only to the
physical part of the interactions, $(V_{JT=1}^{A}+V_{JT=0}^{A})$. We can
choose all $E_{KT=1}$ to be zero in the above (by setting $V_{JT=1}^{S}=
V_{JT=0}^{A}$) leaving $E_{KT=0}$ completely determined by the physical
matrix elements. In that case, we can halve the number of fields to be
integrated.

If we now diagonalize the $E_{KT}(i,j)$ as before and form the operators
\begin{timeqnarray}
\hat{Q}_{KMT}(\alpha)\equiv & {1 \over \sqrt{2(1+\delta_{M0})}}
(\hat{\rho}_{KMT}(\alpha)+(-1)^{M}\hat{\rho}_{K-MT}(\alpha))\;,&\\
\hat{P}_{KMT}(\alpha)\equiv & -{i\over \sqrt{2(1+\delta_{M0})}}
(\hat{\rho}_{KMT}(\alpha)-(-1)^{M}\hat{\rho}_{K-MT}(\alpha))\;,&
\end{timeqnarray}

\noindent the two-body part of the Hamiltonian can finally be written as
\begin{equation}\hat{H}_2' = {1\over 2}
\sum_{KT} \sum_\alpha
\lambda_{KT}(\alpha) \sum_{M \ge 0} \left(
\hat{Q}^2_{K \,MT}(\alpha) + \hat{P}^2_{K\, MT}(\alpha) \right)\;.
\end{equation}

In this decomposition, the one-body hamiltonian $\hat{h}$ of the HS
transformation does not mix protons and neutrons. We can then represent the
proton and neutron wavefunctions by separate determinants, and the number of
neutrons and protons will be conserved rigorously during each Monte Carlo
sample. For general interactions, even if we choose nonzero $E_{KT=1}$
matrix elements, the number of fields involved is half that for
a decomposition that mixes
neutrons and protons, and the matrix dimension is also halved.
These two factors combine to speed up the computation significantly. In this
sense, an isospin formalism is more favorable, although at the cost of
limiting the degrees of freedom embodied in the symmetric matrix elements
$V_J^S$.

\section{Monte Carlo quadrature}

The manipulations of the previous sections have reduced the shell model to
quadrature. That is, thermodynamic expectation values are given as the ratio
of two multidimensional integrals over the auxiliary fields. The dimension
$D$ of these integrals is of order $N_s^2N_t$, which can exceed $10^5$ for
the problems of interest. Monte Carlo methods are the only practical means of
evaluating such integrals. In this section, we review those aspects of Monte
Carlo quadrature relevant to the task at hand. Our discussion is adapted from
that of Ref.~\cite{Koonin}, where a more general exposition can be found.

Monte Carlo quadrature can be a very efficient way of evaluating integrals of
high dimension. The name ``Monte Carlo'' arises from the random or ``chance''
character of the method and the famous casino in Monaco. The essential idea
is not to evaluate the integrand at every one of a large number of quadrature
points, but rather at only a representative random sampling of fields. This
is analogous to predicting the results of an election on the basis of a poll
of a small number of voters.

To apply Monte Carlo quadrature to the shell model (i.e., expressions such as
(3.19)), the weight function $W_\sigma$ must be real and non-negative. While
the properties of $W$ clearly depend upon the interaction and decomposition
used, these conditions are not satisfied for the natural decompositions of
most realistic hamiltonians. However, a broad class of schematic interactions
and all realistic interactions satisfy these conditions closely enough (or,
occasionally, exactly) so that Monte Carlo quadrature can be used. Thus, for
the purposes of the present discussion, we assume $W_\sigma\geq0$ and
postpone a discussion of the so-called ``sign problem'' to section~7 below.

To understand the Monte Carlo method, we recast the ratio of integrals in
Eq.~(3.19) as
\begin{equation}\langle\hat \Omega\rangle = \int d^D\sigma P_\sigma
\Omega_\sigma\;,
\end{equation}
where
\begin{equation}P_\sigma = {W_\sigma\over \int d^D\sigma W_\sigma}\;.
\end{equation}
Since $\int d^D\sigma P_\sigma=1$ and $P_\sigma\geq0$, we can think of
$P_\sigma$ as a probability density and $\langle\hat \Omega\rangle$ as the
average
of $\Omega_\sigma$ weighted by $P_\sigma$. Thus, if $\{\sigma_s, s=1, \ldots,
S\}$ are a set of $S$ field configurations randomly chosen with probability
density $P_\sigma$, we can approximate $\langle\hat \Omega\rangle$ as
\begin{equation}\langle\hat \Omega\rangle\approx
{1\over S}\sum^S_{s=1} \Omega_s\;,
\end{equation}
where $\Omega_s$ is the value of $\Omega_\sigma$ at the field configuration
$\sigma_s$.
Since this estimate of $\langle\hat \Omega\rangle$ depends upon the randomly
chosen field configurations, it too will be a random variable whose average
value is the required integral. To quantify the uncertainty of this estimate,
we consider each of the $\Omega_s$ as a random variable and invoke the central
limit theorem to obtain
\begin{equation}\sigma^2_{\langle\hat \Omega\rangle} =
{1\over S}\int d^D\sigma P_\sigma
(\Omega_\sigma-\langle\hat \Omega\rangle)^2\approx
{1\over S^2}\sum^S_{s=1}
(\Omega_s-\langle\hat \Omega\rangle)^2\;.
\end{equation}
Equation (5.4) reveals two very important aspects of Monte Carlo quadrature.
First, the uncertainty in the estimate of the integral decreases as
$S^{-1/2}$. Hence, if more samples are used, we will get a more precise
answer, although the error decreases very slowly with the number of samples
(a factor of four more numerical work is required to halve the uncertainty).
The second important point to realize from (5.4) is that the precision of the
result depends upon the extent to which $\Omega_\sigma$ deviates from its
average
value over the region of integration; the best case is when $\Omega_\sigma$ is
as
smooth as possible.

The discussion above shows that Monte Carlo quadrature involves two basic
operations: generating field configurations randomly distributed according to
the specified distribution $P_\sigma$ (or, equivalently, $W_\sigma$, to which
it is proportional) and then evaluating the observable $\Omega_\sigma$ for
these
fields. The second operation is straightforward (details are discussed in the
following section and the appendix),
but it is less obvious how to generate the required field
configurations.

When the probability distribution has a simple analytical form, there are a
number of methods that can be invoked to generate independent random samples
\cite{Recipe}. However, in the present case $P_\sigma$ has a complicated
dependence
upon the auxiliary fields that is not amenable to these methods. Rather, we
employ the Metropolis, Rosenbluth, Rosenbluth, Teller, and Teller algorithm
\cite{Metropolis}, which requires only the ability to calculate the weight
function for a
given value of the integration variables.

Although the algorithm of Metropolis {\it et al.}\ can be implemented in a
variety of ways, we begin by describing one simple realization. Suppose that
we want to generate a set of samples distributed according to a (not
necessarily normalized) weight function $W_\sigma$. The Metropolis algorithm
generates an arbitrarily long sequence of samples $\{\sigma_k,
k=0,1,2,\ldots\}$ as those field configurations visited successively by a
random walker moving through $\sigma$-space; as the walk becomes longer and
longer, the points it connects approximate more closely the desired
distribution.

The rules by which the random walk proceeds through $\sigma$-space are as
follows. Suppose that the walker is at a point $\sigma_k$ in the sequence. To
generate $\sigma_{k+1}$, it makes a trial step to a new point $\sigma_t$.
This new point can be chosen in any convenient manner, for example uniformly
at random within a multi-dimensional cube of small side $\delta$ about
$\sigma_k$. This trial step is then ``accepted'' or ``rejected'' according to
the ratio
\begin{equation}r= {W_t\over W_k}\;,
\end{equation}
where $W_t$ is $W$ evaluated at $\sigma_t$ and $W_k$ is $W$ evaluated at
$\sigma_k$. If $r$ is larger than one, then the step is ``accepted'' (i.e.,
we put $\sigma_{k+1}= \sigma_t$), while if $r$ is less than one, the step is
accepted with probability $r$. This latter is conveniently accomplished by
comparing $r$ with a random number $\eta$ uniformly distributed in the
interval $[0,1]$ and accepting the step if $\eta<r$. If the trial step is not
accepted, then it is ``rejected,'' and we put $\sigma_{k+1}= \sigma_k$. These
rules generate $\sigma_{k+1}$, and we can then go on to generate
$\sigma_{k+2}$ by the same process, making a trial step from $\sigma_{k+1}$.
Any arbitrary point, $\sigma_0$, can be used as the starting point for the
random walk.

To prove that the algorithm described above does indeed generate a sequence
of points distributed according to $W$, let us consider a large number of
walkers starting from different initial points and moving independently
through $\sigma$-space. If $N_k (\sigma)$ is the density of these walkers at
step $k$, then the net number of walkers moving from field configuration
$\sigma$ to field configuration $\rho$ in the next step is
\begin{eqnarray}
\Delta N(\sigma) =&
N_k(\sigma) P(\sigma\rightarrow\rho)-
N_k(\rho) P(\rho\rightarrow\sigma)  \nonumber \\
=&
N_k(\rho) P(\sigma\rightarrow\rho)
\left[ {N_k(\sigma)\over N_k(\rho)}-
{P(\rho\rightarrow\sigma)\over P(\sigma\rightarrow\rho)}\right]\;.
\end{eqnarray}
Here, $P(\sigma\rightarrow\rho)$ is the probability that a walker will make a
transition to $\rho$ if it is at $\sigma$. This equation shows that there is
equilibrium (no net change in population) when
\begin{equation}{N_k(\sigma)\over N_k(\rho)}=
{P(\rho\rightarrow\sigma)\over P(\sigma\rightarrow\rho)}=
{N_e(\sigma)\over N_e(\rho)}
\end{equation}
and that changes in $N(\sigma)$ when the system is not in equilibrium tend to
drive it toward equilibrium (i.e., $\Delta N(\sigma)$ is positive if there
are ``too many'' walkers at $\sigma$, or if $N_k(\sigma)/N_k(\rho)$ is
greater than its equilibrium value). Hence it is plausible (and can be
proved) that, after a large number of steps, the population of the walkers
will settle down to its equilibrium distribution, $N_e$.

It remains to be shown that the transition probabilities of the Metropolis
algorithm lead to an equilibrium distribution of walkers $N_e(\sigma)\sim
W_\sigma$. For the random walk governed by the rules described above, the
probability of making a step from $\sigma$ to $\rho$ is
\begin{equation}P(\sigma\rightarrow\rho)=
T(\sigma\rightarrow\rho) A(\sigma\rightarrow\rho)\;,
\end{equation}
where $T$ is the probability of making a trial step from $\sigma$ to $\rho$
and $A$ is the probability of accepting that step. If $\rho$ can be reached
from $\sigma$ in a single step (i.e., if it is within a cube of side $\delta$
centered about $\sigma$), then $T(\sigma\rightarrow\rho)=
T(\rho\rightarrow\sigma)$, so that the equilibrium distribution of the
Metropolis random walkers satisfies
\begin{equation}{N_e(\sigma)\over N_e(\rho)}=
{A(\rho\rightarrow\sigma)\over A(\sigma\rightarrow\rho)}\;.
\end{equation}
If $W_\sigma> W_\rho$, then $A(\rho\rightarrow\sigma)=1$ and
$A(\sigma\rightarrow\rho)= W_\rho/W_\sigma$, while if $W_\sigma< W_\rho$ then
$A(\rho\rightarrow\sigma)= W_\sigma/W_\rho$ and $A(\sigma\rightarrow\rho)=1$.
Hence, in either case, the equilibrium population of Metropolis walkers
satisfies
\begin{equation}{N_e(\sigma)\over N_e(\rho)}=
{W_\sigma\over W_\rho}
\end{equation}
so that the walkers are indeed distributed with the correct distribution.

Note that although we made the discussion concrete by choosing $\sigma_t$ in
the neighborhood of $\sigma_k$, we can use {\it any} transition and
acceptance rules that satisfy
\begin{equation}{T(\rho\rightarrow\sigma) A(\rho\rightarrow\sigma)\over
T(\sigma\rightarrow\rho) A(\sigma\rightarrow\rho)}=
{W_\sigma\over W_\rho}\;.
\end{equation}
Indeed, one limiting choice is $T(\sigma\rightarrow\rho)= W_\rho$,
independent of $\sigma$, and $A=1$. This is the most efficient choice, as no
trial steps are ``wasted'' through rejection. However, this choice is
somewhat impractical, because if we knew how to sample $W$ to take the trial
step, we wouldn't need to use the algorithm to begin with.

An obvious question is ``If trial steps are to be taken within a neighborhood
of $\sigma_k$, how do we choose the step size, $\delta$?'' To answer this,
suppose that $\sigma_k$ is at a maximum of $W$, the most likely place for it
to be. If $\delta$ is large, then $W_t$ will likely be very much smaller than
$W_k$ and most trial steps will be rejected, leading to an inefficient
sampling of $W$. If $\delta$ is very small, most trial steps will be
accepted, but the random walker will never move very far, and so also lead to
a poor sampling of the distribution. A good rule of thumb is that the size of
the trial step should be chosen so that about half of the trial steps are
accepted.

One bane of applying the Metropolis algorithm is that the field
configurations that make up the random walk, $\sigma_0, \sigma_1, \ldots$,
are not independent of one another, simply from the way in which they were
generated; that is, $\sigma_{k+1}$ is likely to be in the neighborhood of
$\sigma_k$. Thus, while the samples might be distributed properly as the walk
becomes very long, they are not statistically independent of one another, and
some care must be taken in using them to evaluate observables. In particular,
if we estimate $\langle \hat \Omega\rangle$ from Eq.~(5.3) using the successive
field configurations in the random walk, the estimate of the variance given
by Eq.~(5.4) is invalid because the $\Omega_k$ are not statistically
independent.
This error can be quantified by calculating the auto-correlation function
\begin{equation}C_i=
{\langle \Omega_k\Omega_{k+i}\rangle -\langle \Omega_k\rangle^2\over
\langle \Omega_k^2\rangle -\langle \Omega_k\rangle^2}\;,
\end{equation}
where $\langle\ldots\rangle$ indicates average over the random walk. Of
course, $C_0=1$, but the non-vanishing of $C_i$ for $i>0$ means that the
$\Omega_k$'s are not independent. What can be done in practice is to compute
the
observable and its variance using samples along the random walk separated by
a fixed interval, the interval being chosen so that there is effectively no
correlation between the field configurations used. An appropriate sampling
interval can be estimated from the value of $i$ for which $C_i$ becomes small
(say $<0.1$).

Another issue in applying the Metropolis algorithm is where to start the
random walk; i.e., what to take for $\sigma_0$. In principle, any location is
suitable and the results will be independent of this choice, as the walker
will ``thermalize'' after some number of steps. In practice, an appropriate
starting point is a probable one, where $W$ is large. Some number of
thermalization steps then can be taken before actual sampling begins to
remove any dependence on the starting point.

\section{Numerical methods and computational issues}

\subsection{Number projection}

For large quantum systems in which many valence fermions participate, the
grand canonical ensemble (with a fixed chemical potential, but fluctuating
particle number) is perfectly acceptable, and simple formulas exist for
calculating the required one-body observables (e.g., Eqs.~(A.15,16)). But in
nuclei, where only a few valence fermions are important in determining the
low lying states, and the properties
of neighboring even-$A$ and odd-$A$ systems differ dramatically, the
canonical (fixed-number) ensemble must be used. In this section, we discuss
how to project the fixed-number trace from the grand-canonical ensemble. For
simplicity, we will consider only a single species of fermions with number
operator $\hat N$; the required generalization to two species (protons and
neutrons) is straightforward, as the HS transformation used does not involve
density operators that mix the species, so that $\hat U_\sigma$
factors into separate evolution operators for neutrons and protons.

The grand canonical partition function for a one-body evolution operator
$\hat U_\sigma$ is given by Eq.~(A.12) as
\begin{equation}Z={\rm Tr}\, e^{\beta\mu\hat N}\hat U_\sigma=
\det (1+e^{\beta\mu}{\bf U}_\sigma)\;,
\end{equation}
where $\mu$ is the chemical potential. Note that since the matrix
corresponding to $\hat N$ is simply the unit matrix, we may include the
chemical potential term in each of the $\hat h$ defining $\hat
U_\sigma$, or write it as a common prefactor, as we have done in (6.1). If
we write the $N_s$ eigenvalues of ${\bf U}_\sigma$ as
$e^{-\beta\varepsilon_\lambda}$, where $\varepsilon_\lambda$ may be complex,
we can express (6.1) as
\begin{equation}Z=\prod_\lambda (1+e^{\beta(\mu-\varepsilon_\lambda)})\;,
\end{equation}
which is a form reminiscent of the usual fermi-gas expressions.

The corresponding canonical partition function for $A$ particles in the
valence space is
\begin{equation}Z_A={\rm Tr}_A\, \hat U_\sigma= {\rm Tr}\,\hat P_A\hat
U_\sigma\;,
\end{equation}
where $\hat P_A=\delta(A-\hat N)$ is the number projector. To effect the
operation of $\hat P_A$, we can write it as the exponential of the one-body
operator $\hat N$ using the fact that $\hat N$ has integer eigenvalues
$0,1,\ldots,N_s$:
\begin{equation}\hat P_A=
e^{-\beta\mu A}\int^{2\pi}_0 {d\phi\over 2\pi} e^{-i\phi A}
e^{(\beta\mu+i\phi)\hat N}\;,
\end{equation}
where $\mu$ is an arbitrary $c$-number, which will be specified below.

The canonical trace thus becomes
\begin{timeqnarray}
Z_A&=&
 e^{-\beta\mu A}\int^{2\pi}_0 {d\phi\over2\pi} e^{-i\phi A}
 {\rm Tr}\, e^{(\beta\mu+i\phi)\hat N} \hat U_\sigma
 \\
&=&
 e^{-\beta\mu A} \int^{2\pi}_0 {d\phi\over 2\pi}
 e^{-i\phi A} \prod_\lambda
 (1+ e^{i\phi} e^{\beta(\mu-\varepsilon_\lambda)} )\;.
\end{timeqnarray}

{}From (6.5b), we can see that the effect of the number projection is to sum
all different products of $A$ terms of the form
$e^{-\beta\varepsilon_\lambda}$. Similar expressions can be written for
expectation values of few-body operators following Eqs.~(A.15, A.16). For
example, for a one-body operator $\hat{\Omega}$,
\begin{equation}{\rm Tr}_A\,\hat U_\sigma \hat{\Omega}=
e^{-\beta\mu A} \int_0^{2\pi} {d\phi\over2\pi} e^{-i\phi A}
\det(1+e^{\beta\mu+i\phi}{\bf U}_\sigma)
{\rm tr}\,
{1\over 1+e^{\beta\mu+i\phi}{\bf U}_\sigma}
e^{\beta\mu+i\phi}{\bf U}_\sigma{\Omega}\;.
\end{equation}

The parameter $\mu$ is arbitrary and hence can be chosen for numerical
convenience. Since the grand-canonical trace in the integrand of (6.5b)
contains contributions from canonical ensembles with all $0\leq A\leq N_s$,
and these vary strongly with $A$, $\mu$ should be chosen to emphasize that
contribution from the particular value of $A$ desired; a poor choice can lead
to numerical difficulties in the projection. This can be done in analogy with
usual thermodynamic treatment. In particular, if the eigenvalues of
${\bf U}_\sigma$ are ordered so that ${\rm Re}\,\varepsilon_1\leq {\rm
Re}\,\varepsilon_2\leq\ldots
\leq {\rm Re}\,\varepsilon_{N_s}$, then a good choice for $\mu$ is $({\rm
Re}\,\varepsilon_A+{\rm Re}\,\varepsilon_{A+1})/2$.

The numerical effort implied by Eqs.~(6.5) or (6.6) need not be
intimidating. First, because the integer eigenvalues of $\hat N$ range from 0
to $N_s$, the maximal Fourier component in the $\phi$ integration is
$e^{iN_s\phi}$, so that an $N_s$-point quadrature can be exact. That is,
for $f(\phi)$ one of the required integrands,
\begin{equation}\int^{2\pi}_0 {d\phi\over2\pi} f(\phi)=
{1\over N_s} \sum^{N_s}_{m=1}
f(\phi_m);\qquad
\phi_m\equiv {m\cdot2\pi\over N_s}\;.
\end{equation}
Second, the matrix algebra associated with expressions like (6.6) is
minimized by computing
only once both the eigenvalues $e^{-\beta\varepsilon_\lambda}$
and the diagonalizing
transformation matrix ${\bf T}_{\alpha\lambda}$ of ${\bf U}_\sigma$ and
then using these to evaluate the integrand at each quadrature point $\phi_m$.
Thus, for example, (6.5b) is simply evaluated, and the matrix trace required
in the integral of (6.6) is evaluated as
\begin{equation}
\sum_{\lambda=1}^{N_s} {1\over 1+e^{\beta(\mu-\varepsilon_\lambda)+i\phi}}
e^{\beta(\mu-\varepsilon_\lambda)+i\phi}
\sum^{N_s}_{\alpha\beta=1} {\bf T}_{\lambda\alpha}{\Omega}_{\alpha\beta}
{\bf T}_{\beta\lambda}\;.
\end{equation}

\subsection{Auxiliary field quadrature}

When the number of auxiliary field variables becomes large, naive application
of the sampling algorithm of Metropolis {\it et al.}\ as described in
Section~5 becomes inefficient. For example, in the ${}^{170}$Dy calculation
described in Section~8.8 where there are some 10$^5$
fields, when the step size
$\delta$ (taken to be the same for each field variable) is adjusted so that
the acceptance ratio is approximately 0.5, the auto correlation length for
the energy (as computed from Eq.~(5.12)) is more than 200 sweeps (i.e., more
than 200 updates of the field variables at each time slice).

The Metropolis efficiency in generating uncorrelated field configurations can
be improved significantly by approximating the continuous integral over each
$\sigma_{\alpha n}$ by a discrete sum derived from a Gaussian quadrature
\cite{Dean93}.
In particular, the relation
\begin{equation}e^{\Delta\beta V \hat {\cal O}^2/2}\approx
\int^\infty_{-\infty} d\sigma f(\sigma)
e^{\Delta\beta V\sigma \hat{\cal O}}
\end{equation}
is satisfied through terms in $(\Delta\beta)^2$ if
\begin{equation}f(\sigma)={1\over6}
\left[\delta(\sigma-\sigma_0)+\delta(\sigma+\sigma_0)+
4\delta(\sigma)\right]\;,\end{equation}
where {$\sigma_0=(3/V\Delta\beta)^{1/2}$}. (Note that the commutator terms
render the HS transformation accurate only through order $\Delta\beta$
anyway.)

In this way, each $\sigma_{\alpha n}$ becomes a 3-state variable and the
integrals in (3.19) become (very large) sums, which can be sampled using the
algorithm of Metropolis {\it et al.} In particular, to
update the fields at a given time slice, we select at random a fraction $q$
of them and assign each of these to ($-\sigma_0$, 0, or $\sigma_0$) with
probability ($1/6$, $2/3$, and $1/6$). To satisfy the condition (5.11), the
move is then accepted or rejected according to the ratio of the old and trial
values of ${\rm Tr_A}\,{\hat U}_\sigma$.
The fraction $q$ is adjusted to give an acceptance ratio of about 0.5.
When used in the ${}^{170}$Dy calculation,
this discretization reduces the energy autocorrelation length to only five
sweeps, a factor of 40 improvement relative to the continuous case, with no
loss of accuracy. Similar, albeit less substantial, savings accrue in smaller
calculations.

\subsection{Stabilization at low temperatures}

The excitation energy of the first excited state
in even-even nuclei with $A\leq 80$
is 1-2 MeV. Thus, one may calculate
at a temperature of $T=0.5$~MeV to get a reasonable description of the
ground state. However, in odd-$A$ and odd-odd
systems the excitation energy can be as low as
tens of keV, so that cooling to the ground state
requires calculations at $\beta > 2.0$~MeV$^{-1}$.
Unfortunately, numerical instabilities arise for larger $\beta$
due to the multiplication of many ${\bf U}$ matrices together;
the product becomes increasingly illconditioned, with exponentially
divergent numerical scales, as illustrated below.

Physically, the one-body evolution operator $\hat U_n$ amplifies low-energy
states for any particular field configuration, and attenuates high-energy
states. The most important states, those near some intermediate
Fermi surface, are buried exponentially by the states at the bottom
of the spectrum,
and extraction of relavent information
becomes increasingly difficult with lower temperatures.
This technical problem, resulting from the limits of finite numerical precision
achievable on any digital computer,
can be circumvented using singular value decompositon
(SVD) methods, which were first applied in Green's Function Monte Carlo
simulations of interacting electron systems \cite{Guber92}. Here we adapt
the SVD methods to the canonical calculations of the nuclear
problem.

Let us suppose that a matrix ${\bf A}$ can be decomposed into three matrices,
${\bf A}={\bf SDV}$,
where ${\bf S}^\dagger {\bf S}={\bf 1}$,
${\bf V}^\dagger {\bf V}={\bf 1}$, and ${\bf D}$ is a real,
diagonal matrix. Schematically, this decomposition looks like
\begin{eqnarray}
{\bf SDV} = &
\pmatrix{
x & x & x \cr
x & x & x \cr
x & x & x }
& \pmatrix{
{\textstyle X}\cr
&& {\scriptstyle X}\cr
&&& x }
\pmatrix{
x & x & x\cr
x & x & x\cr
x & x & x }  \nonumber \\
  = & \pmatrix{\displaystyle
{\large X} & {\large X} & {\large X}\cr
{\large X} & {\large X} & {\large X} \cr
{\large X} & {\large X} & {\large X} }\;, &
\end{eqnarray}
where the size of the symbols
$({\large X}, {\scriptstyle X}, x)$ indicates the
magnitude of the elements.
Note that ${\bf S}$ and ${\bf V}$
are well controlled matrices shown schematically as having unit
scale, and that all the
scales of the original matrix ${\bf A}$
are contained in ${\bf D}$. Note also that once the multiplication
has been performed all elements of the resulting matrix are of the
same (largest) scale, and the product is essentially an outer product
of the first column of ${\bf S}$ and the first row of ${\bf V}$. The smallest
numerical scales exist only implicitly as differences of large matrix
elements, and as such are difficult to recover on a computer with finite
precision.

In contrast, a matrix that explicitly displays its small scales,
such as a column stratified matrix ${\bf M}$, can be factored stably.
\begin{eqnarray}
{\bf S}^{-1}{\bf M}{\bf V}^{-1} = &
\pmatrix{
x & x & x \cr
x & x & x \cr
x & x & x \cr}
 \pmatrix{
 {\textstyle X} & {\scriptstyle X} & x\cr
 {\textstyle X} & {\scriptstyle X} & x\cr
 {\textstyle X} & {\scriptstyle X} & x\cr
}
\pmatrix{
x & x & x\cr
x & x & x\cr
x & x & x\cr}  \nonumber \\
=    &
\pmatrix{
  {\textstyle X}\cr
& {\scriptstyle X}\cr
&& x\cr}
  =  {\bf D} \; .&
\end{eqnarray}

Multiplication on the left of ${\bf M}$ by a transformation matrix combines
only elements in a given column which are all of the same scale. Thus there
is no loss of information. Multiplication on the right by ${\bf V}^{-1}$
combines columns of different scales, but does not overwrite any small scale
as long as large scaled columns are scaled down appropriately before the
addition.

These problems are quantified by the
condition number, defined as the ratio of the
largest element of ${\bf D}$ to the smallest after the
transformation has taken place.  When this ratio is greater than
the machine accuracy (usually about 10$^{12}$ for double precision), then
the calculation will become unstable, and techniques of stable
matrix multiplication must be employed.
Numerical problems first occur in the multiplication of
${\bf U}_{n}$ matrices to
create ${\bf U}=\Pi_{n=1}^{N_t} {\bf U}_{n}$.
Although each ${\bf U}_{n}$ has a good
condition number, the final product can have an unacceptable condition number.
The above examples can be combined to
disentangle the scales inherent in the multiplication of the ${\bf U}_n$
matrices.

SVD matrix multiplication works by isolating the scales of the problem at
each multiplication step.
Let the matrix
${\bf W}_n=\Pi_{m=1}^{n\le N_t}{\bf U}_m$, be the partial product
of the matrix multiplications, with ${\bf W}_0={\bf 1}$,
${\bf W}_1={\bf U}_1$,
${\bf W}_2={\bf U}_2{\bf U}_1$, etc.
Suppose we have already decomposed the
matrix ${\bf W}_n$ in SVD format.
We want to multiply the well-conditioned matrix
${\bf U}_n$ with ${\bf W}_{n-1}$. The multiplication proceeds as follows:
\begin{equation}{
{\bf U}_n{\bf W}_{n-1}=
\left({\bf U}_n{\bf S}_{n-1}{\bf D}_{n-1}\right){\bf V}_{n-1}=
\left[{\bf U}_n{\bf S}_{n-1}
\pmatrix{
{\textstyle X}\cr
& {\scriptstyle X}\cr
&& x\cr}\right]
{\bf V}_{n-1} \; .
}
\end{equation}

Multiplication of the matrices in brackets yields a column stratified matrix
that can be put into SVD format (i.e., ${\bf S}' {\bf D}' {\bf V}'$)
without loss of scale. Thus
\begin{equation}{ {\bf U}_n{\bf W}_{n-1}  =
\pmatrix{
{\textstyle X} & {\scriptstyle X} &  x\cr
{\textstyle X} & {\scriptstyle X} &  x\cr
{\textstyle X} & {\scriptstyle X} &  x\cr}
{\bf V}_{n-1}
=
({\bf S}'{\bf D}'{\bf V}'){\bf V}_{n-1}
={\bf S}'{\bf D}'({\bf V}'{\bf V}_{n-1})=
{\bf S}_n {\bf D}_n {\bf V}_n \; ,
} \end{equation}
where in the last step we have defined
${\bf S}_n={\bf S}'$,
${\bf D}_n={\bf D}'$,
and ${\bf V}_n={\bf V}'{\bf V}_{n-1}$.
Provided that the ${\bf V}$ matrices are sufficiently well-conditioned
that we can multiply many of them together without scale problems,
we are able to perform stable matrix multiplication of all the
one-body evolution operators.

For evaluation of the quantum mechanical trace, we need to
calculate ${\bf 1}+{\bf U}_\sigma$
consistent with SVD. We assume that SVD has previously been performed on
${\bf U}_\sigma$ such that ${\bf U}_\sigma={\bf SDV}$. The goal is to
keep all of the large scales in the problem separated from the small ones.
This is done by calculating
\begin{equation}
\det\left({\bf 1}+{\bf U}_\sigma\right)=
\det\left[{\bf S}\left({\bf S}^{-1}{\bf V}^{-1}+{\bf D}\right){\bf V}\right]=
\det {\bf S}\tilde{{\bf S}}\tilde{{\bf D}}\tilde{{\bf V}}{\bf V}=
\det {\bf S}'\det {\bf D}'\det {\bf V}'\;. \end{equation}
Here we have performed a new SVD on the matrix in parenthesis
to get $\tilde {\bf S} \tilde {\bf D} \tilde {\bf V}$ (after
the addition of ${\bf S}^{-1}{\bf V}^{-1}$ to ${\bf D}$), and we define
${\bf S}'={\bf S}\tilde{\bf S}$ and ${\bf V}'={\bf V}\tilde{\bf V}$
The other quantity we need is
\begin{equation}
\left({\bf 1}+{\bf U}_\sigma\right)^{-1}{\bf U}_\sigma=
{\bf 1}-{{\bf 1}\over{{\bf 1}+{\bf U}_\sigma}}=
{\bf 1}-{\bf V}'^{-1}{\bf D}'^{-1}{\bf S}'^{-1}\;,
\end{equation}
where we have used the inverses of the matrices calculated in Eq.~(6.15).
These steps can be carried over exactly into the number projection method
described in section (6.1).

As an example we consider the matrix multiplications
to obtain ${\bf U}_\sigma$ for $^{54}$Fe
in the course of the calculations described in section 8.1.
Shown in Fig. 6.1 is an example of the condition number with and
without SVD for various $\beta$, with $\Delta\beta=1/16$~MeV$^{-1}$.
Numerical round-off causes the condition number to virtually saturate in the
non-SVD case. Shown in Fig. 6.2 are the elements of the diagonal matrix
${\bf D}$ using SVD for all multiplies, compared to using SVD only on the
full ${\bf U}_\sigma$. Note that the symmetry of ${\bf U}_\sigma$
(that eigenvalues come in complex
conjugate pairs, see section 7.1) is evident when the matrix-multiplies are
accomplished through SVD, where the {\it real} elements
of ${\bf D}$ come in pairs.
This is not the case without SVD multiplication.
The numerical eigenvalues of ${\bf U}_\sigma$ violate the symmetry
only after $\beta\ge 2.0$ (and at larger $\beta$ in larger systems).

We have made limited use of SVD in the calculations presented in this
Report. In test calculations we find that we can increase $\beta$ almost
indefinitely (e.g. to $\beta=6$ in $^{22}$Ne) with SVD, whereas without
the stabilization, the maximum attainable $\beta$ is 2.0 MeV$^{-1}$.
In the $^{170}$Dy
and Xe calculations described in section 8,
$\beta=5$ MeV$^{-1}$ is attainable without SVD.

\subsection{Maximum entropy techniques}

The response functions that can be evaluated in SMMC are
given by
Eq.~(3.6). Calculation of the strength function $S_\Omega(E)$ for a given
operator can be difficult due to the ill-posed inversion of the
Laplace transform required
by Eq.~(3.8a). In this section, we discuss one
particular inversion technique, Maximum
Entropy (MaxEnt), which we have used to find the Gamow-Teller
strength functions presented in section~8.6.

We wish to find the function $S(E)$ from
the function $R(\tau)$ using the kernel $K(\tau,E)$:
\begin{equation}
R(\tau)=\int dE K(\tau,E)S(E) \;.\end{equation}
The function $R(\tau)$ is known at
$N$ grid points $\tau_i$ ($i=1,\cdots, N$),
where the values
of $R$ are given at $\tau_i$ by $r_i$ with errors $\sigma_i$.
Fluctuations
in different data points may be correlated, in which case the
covariance matrix $C_{ij}=\langle (r_i - \bar{r}_i)(r_j-\bar{r}_j) \rangle$
is diagonalized to find the independent modes, and $r_i$ and
the kernel are then transformed to this new basis.
The goal is to find the best $S(E)$ by analysing the
$\chi^2$ figure of merit,
\begin{equation}
\chi^2\left\{S\right\}=\sum_i\left({{r_i - R_i\left\{S\right\}}\over{\sigma_i}}
\right)^2\; , \end{equation}
of the data $r_i$ from the fit values $R_i$ produced by the
trial inverse in the points $\tau_i$,
\begin{equation}
R_i\left\{S\right\}=\int dE K(\tau_i,E)S(E)\;.\end{equation}

Direct minimization of $\chi^2$ is numerically stable in only the
simplest of circumstances. Most of the modern ways to regularize the problem
are based on combining $\chi^2$ with some other
auxiliary well-conditioned functional, $P\left\{S\right\}$. This functional
should have a minimum at a smooth function $S(E)$ and assign a high
cost to strongly oscillating functions. Thus we minimize the joint
functional
\begin{equation}
{{1}\over{2}}\chi^2\left\{S\right\}+P\left\{S\right\} \; .\end{equation}

We use the information-theoretic entropy to define
$P\left\{S\right\}$ such that
\begin{equation}
P\left\{S\right\}=\alpha \int dE \left[ m(E)-S(E)+
S(E)\ln\left({{S(E)}\over{m(E)}}\right)\right]\;,
\end{equation}
where $m(E)$, the default model, and $\alpha$ are chosen to describe
the {\it a priori} knowledge about the shape of the
original $S(E)$. The choice of $m(E)$ and $\alpha$ is more doubtful than
the MaxEnt technique itself. In our application to the Gamow-Teller
strength functions
we make the simplifying
assumption that $m(E)$ is
a Gaussian function with appropriately chosen center (see section 8.6),
and $\alpha=\left[\int dE m(E)\right]^{-1}$.

In order to minimize the functional (6.20)  we employ the technique
of Ref.~\cite{Meshkov94}, which involves an iterative
sequence of linear programming problems. To do this we first
expand Eq.~(6.21) to second order in $S(E)$ about some positive function
$f(E)$, to obtain
\begin{equation}
P\left\{f\mid S\right\}=\alpha \int dE \left\{\left(m-{{f}\over{2}}\right)
+\left[\ln\left({{f}\over{m}}\right)-1\right]S
+{{S^2}\over{2f}}\right\} \; .
\end{equation}
If the true minimum $S(E)$ of the non-quadratic functional in
Eq.~(6.21) is taken as a point of expansion $f(E)$ in (6.22), then
it also gives the minimum of the corresponding quadratic functional
\begin{equation}
S(E)=\min_a \left[{{1}\over{2}}\chi^2\left\{a\right\} +
P\left\{S\mid a\right\}\right] \; . \end{equation}
We transform this identity to a recursive equation that, upon iteration,
leads to the desired solution. To ensure positivity we keep a fraction
of the result of the previous iteration in the next one, and minimize the
functional with the restriction that $S\ge 0$. Defining $n$ as the
iteration step, we obtain,
\begin{equation}
S^{(n+1)}=\min_{S\ge 0}\left[{{1}\over{2}}\chi^2\left\{S\right\}+
P\left\{f^{(n)}\mid S\right\}\right]\;, \end{equation}
with
\begin{equation}
f^{(n)}(E)=\xi S^{(n-1)}(E)+(1-\xi)S^{(n)}(E)\;, \end{equation}
and the default model is taken as the starting approximation to $S$,
\begin{equation}
S^{(0)}(E)=S^{(-1)}(E)=m(E)\;. \end{equation}
The rate of convergence and stability are controlled by the
mixing parameter $\xi$;
a value of $\xi=0.3$ is a reasonable choice
to gaurantee stability.

Each iteration is a linear regularization task with a quadratic
regularization functional reinforced by an additional stabilizing
effect of the constraints $S(E)\ge 0$. Thus, each iteration gives
a nonnegative function $S^{(n)}(E)$ and a positive definite function
$f^{(n)}(E)$. Test cases of simulated data for which the answer is
known give convergence within twenty iterations.

We have used the
technique described above
to obtain the strength functions from the Gamow-Teller response
functions, as presented in section 8.6.
A different approach to MaxEnt was outlined in \cite{Lang},
but has proven less useful due to difficulties in obtaining a converged
value of $\alpha$. The advantage of the present method
is that knowledge of $\alpha$ is not necessary, and that the
scheme provides an iterative method of converging to the most likely
solution.

\subsection{Computational considerations}
\medskip

The present SMMC code is roughly the fourth major
revision of a program whose development began in late 1990 with a
single-species, single-$j$-shell version. It is now a modular package
of some 10,000 commented lines of {\sc{FORTRAN}}.
All floating
point computations are double precision (64-bit).

The package performs all of the functions necessary for shell model
Monte Carlo calculations: initialization, thermalization of the
Metropolis walk, generation of the Monte Carlo samples, evaluation of
static observables and response functions (canonical or
grand-canonical), MaxEnt extraction of strength functions, and  the
extrapolation in $g$ required to solve the
sign problem, as discussed in section 7.
Samples may be analyzed ``on-line'' and/or
stored for post-processing in subsequent runs. The data input and
results output use standard shell model conventions, so that it is
easy to change the two-body matrix elements of the interaction, to
incorporate additional one- or two-body observables in the analysis,
or to add or change the orbitals in the calculation.
The code has
been de-bugged and tested extensively against direct diagonalization
results in the $sd$- and lower $pf$-shells. Its operation by an
experienced user can be described as ``routine,'' although it takes
several weeks to acquire that experience.

The SMMC package is highly portable. Initial
development was largely under VAX/VMS. Subsequently, we have ported
the code to DEC ALPHA's and HP 730's operating under Unix, the Intel
i860 CPU used in the DELTA and PARAGON parallel machines,
the IBM RS-6000 processor used in both workstations and the
\hbox{SP-1} and SP-2 parallel machines, and the
Fujitsu VPP500 shared memory machine (at RIKEN). Very few problems are
encountered in porting to a new machine, and the operation generally
takes less than a day.

Shell model Monte Carlo calculations are
extraordinarily well-suited to Multiple
Instruction/Multiple Data (MIMD) architectures. Indeed, our code is
``embarrassingly parallel": separate Metropolis random walks are
started on each computational node, which then produces a specified
number of Monte Carlo samples at regular intervals during the walk.
Data from all of the nodes are sent to a central node for evaluation
of the Monte Carlo averages and their uncertainties, and perhaps for
storage of the sampled field configurations in a file for
post-analysis. The post-processing of the stored samples is also
divided among the nodes.

To date, we have implemented the parallel version of our code on the
Intel DELTA, and PARAGON machines at Caltech (each with 512
i860 processors), on the
128-processor IBM SP-1 at ANL, the 512-processor IBM SP-2 at Maui,
and on a Fujitsu VPP500 shared memory vector processor (24 CPU's).
In all cases, the
ratio of communications to computation is very low, with efficiencies
always greater than 95\%.

To circumvent the limited  memory on the DELTA (12.5 MBytes) nodes,
we have also produced a version of the code in
which the chains of $U_\sigma$ matrices effecting the time evolution
are split over two or more nodes, information being passed between
them only at the time slice where they ``join.'' However, such an
implementation is a significant sacrifice in speed, as while one
processor updates a chain, the other processors working on the same
chain must sit idle.

Table 1 shows benchmarks of our code on various
single processors.
The test calculation involved a canonical ensemble in
the full $pf$-shell ($N_s=20$ for each type of nucleon, implying
$20\times20$ matrices) using a realistic interaction. $N_t=32$ time
slices were used at $\beta=2~{\rm MeV}^{-1}$
($\Delta\beta=0.0625~{\rm MeV}^{-1}$). Thirty static observables
and seven dynamical response functions were calculated at a single
$g$-value (see section 7).
Note that the
computational speed is independent of the interaction and of the
number of nucleons occupying the shell. Beyond using library
subroutines (BLAS and LAPACK routines),
no attempt was made to optimize the assembly level code
in any of these cases. Approximately 40\% of the computational effort
is in matrix-multiplies.  A significant remainder of the effort goes into
building the one-body Hamiltonian (13\%), setting up the one-body evolution
operators (15\%), and calculating two-body observables (15\%), none
of which is easily vectorizable.

\begin{table}
\begin{center}
\begin{tabular}{|c|c|c|c|}
\hline
Processor&
Peak MF&
Average MF&
Samples/hr.\\
\hline
i860 & 35 & 9 & 44 \\
IBM-SP2 Thin66& 56 & 36 & 179 \\
ALPHA-400& 56 & 28 & 141 \\
\hline
\end{tabular}
\caption{Benchmarks of the SMMC code in various processors.}
\end{center}
\end{table}

In general, the computation time scales as $N^3_sN_t$, and is spent
roughly equally on  the dynamical response functions and the
static observable sampling. Of course, the stabilization discussed
above increases these times significantly (by at least a factor of 5).

The memory required for our calculations scales as $\bar N^2_sN_t$.
$\bar N^2_s$ is the average of the squares of the numbers of neutron
and proton single-particle states. Sample values of $N_s$ for one
isospin type are shown for various model spaces in Table 2.

\begin{table}
\begin{center}
\begin{tabular}{|c|c|}
\hline
Model Space & $N_s$ \\
\hline
$0p$       &6\\
$1s$-$0d$  &12\\
$1p$-$0f$  &20\\
$1p$-$0f$-$g_{9/2}$ &30\\
$2s$-$1d$-$0g$ &30\\
$2p$-$1f$-$0g_{9/2}$-$0i_{13/2}$ &44 \\
\hline
\end{tabular}
\end{center}
\caption{Matrix dimension for various model spaces.}
\end{table}

The code is currently structured so that a calculation with $\bar
N_s=32$, six $j$-orbitals, and
$N_t=64$ time slices will fit in 12~MB of memory (note
that all field variables are stored as 1-byte integers because of the
3-point Gauss-Hermite quadrature described in section 6.2).
A calculation in the
($1p$-$0f$)-($2s$-$1d$-$0g$) basis has $\bar N_s=50$ and would
require about 64~MB of memory for $N_t=64$ time slices and about
128~MB for $N_t=128$. For calculations with a ``good'' Hamiltonian,
as defined in the next section, these storage estimates can be reduced by a
factor of two by exploiting the time reversal symmetry of
the $U_\sigma$, albeit at some loss in speed.

\begin{figure}
$$\epsfxsize=4truein\epsffile{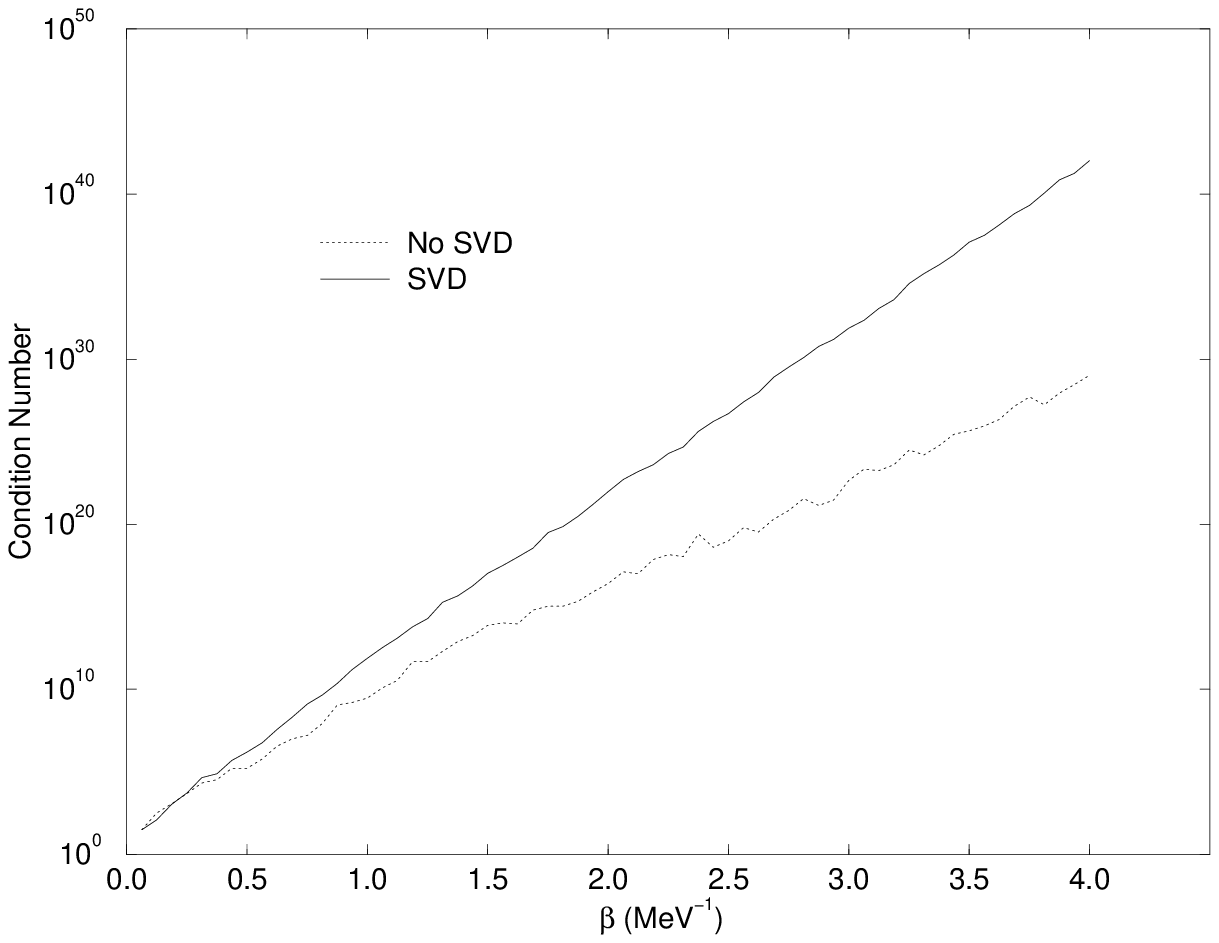}$$
{FIG~6.1
The condition number of ${\bf U}_\sigma$ with and without SVD for various
$\beta$ is plotted for $^{54}$Fe.
}
\end{figure}

\subsection{Contrast with interacting electron systems}

It is interesting to contrast the SMMC simulations of nuclear systems
with similar work on the Hubbard model in the context of high-temperature
superconductivity. The latter involve electrons hopping locally on a
two-dimensional spatial lattice interacting with each other through an
on-site repulsion. For a review of this field see \cite{Linden}.

One significant difference is that physically interesting SMMC calculations
can be performed in smaller model spaces. The $N_s$ values
quoted in the previous section are significantly smaller than the number
of sites needed to approximate the translational invariance of the CuO planes.
However, the matrices representing the
fluctuating one-body hamiltonians $\hat h$
in SMMC calculations are dense, as the nuclear hamiltonian does not have
the local spatial structure of the Hubbard model that gives rise to
sparse $\bf h$ matrices.

Another difference is that relatively higher temperatures are of interest
in the nuclear case, allowing smaller $N_t$. For a nominal temperature
of $T=300$ keV and strength of the two-body interaction $V=3$ MeV, we
have $T/V=0.1$. In the high-$\rm T_c$ case, $T \approx 10^2$ K $\approx
10^{-2}$ eV, while $V \approx 1$ eV, so $T/V \approx 10^{-2}$, an order
of magnitude smaller.

The SMMC calculations must be performed in the canonical ensemble, as the
effects of finite particle number are an important aspect of nuclear structure.
This necessitates the rather cumbersome number projection described in
section 6.1. In contrast, the Hubbard simulations can be done in the
grand-canonical ensemble.

A final and most crucial difference between the SMMC and Hubbard
hamiltonians is that the nuclear interaction is predominantly
attractive. This significantly diminishes (and in some cases
eliminates) the sign problems that plague the condensed matter work.
These problems, and their resolution in SMMC calculations,
are the subject of the following section.

\begin{figure}
$$\epsfxsize=4truein\epsffile{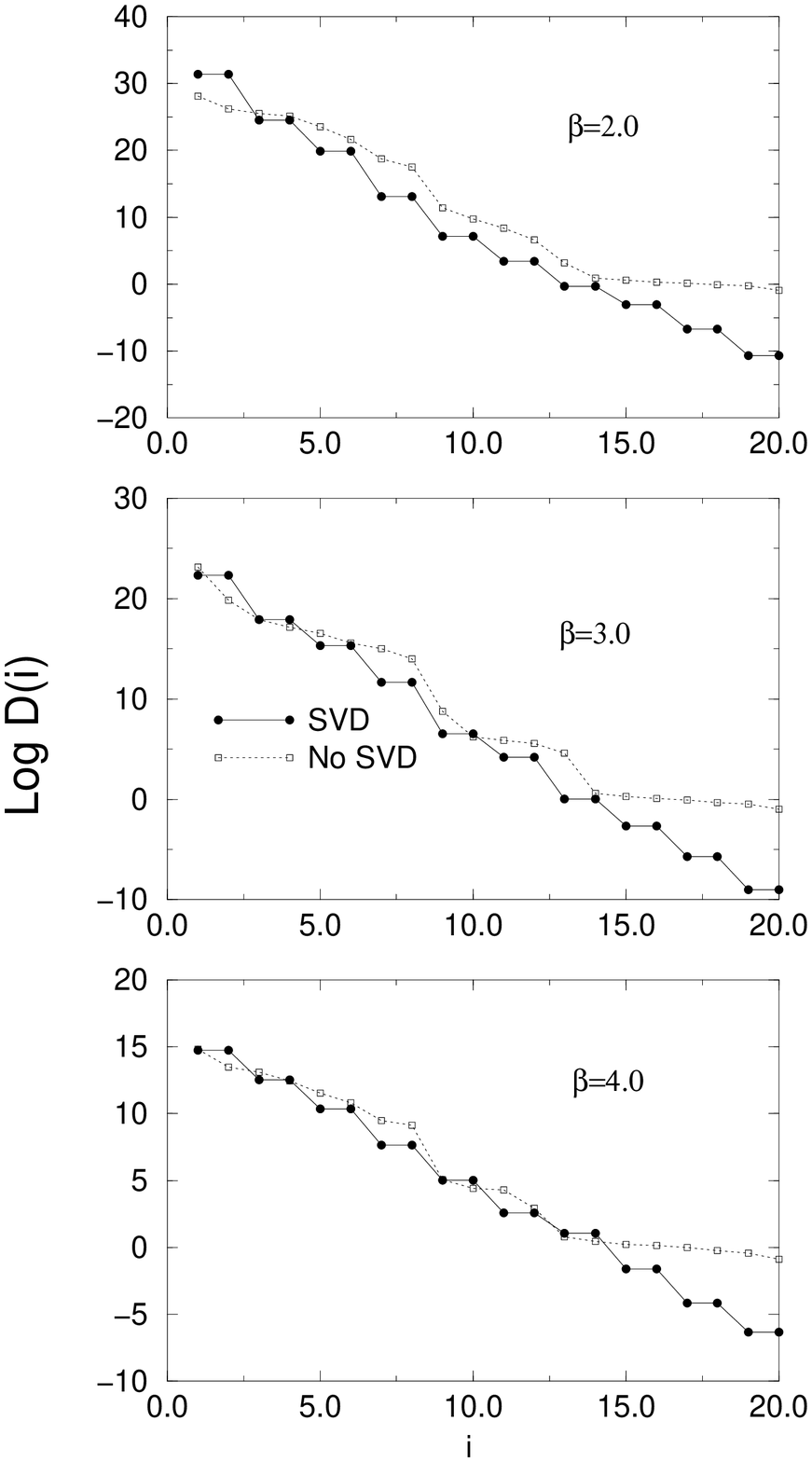}$$
{FIG~6.2.
Elements of the diagonal matrix ${\bf D}$ are
shown when
SVD is used for all multiplies of the ${\bf U}_{n}$, compared with
those obtained
when SVD is performed only on the final matrix ${\bf U}_\sigma$.
}
\end{figure}


\noindent
\section{Sign problems}

Although we have briefly alluded to  ``sign problems'' in some of the
previous chapters, virtually all of our discussion has been based on the
premise that the weight function $W_\sigma$ of Eq.~(3.19)
is non-negative for all
field configurations $\sigma$. Were this not the case, the function
$P_\sigma$
of Eq.~(5.2) could not be interpreted as a probability density, and the MC
quadrature described in section~5 could not be effected.

Unfortunately, many of the hamiltonians of physical interest suffer from a
sign problem, in that $W_\sigma$ is negative over significant fractions of the
integration volume. To understand the implications of this, let us rewrite
Eq.~(5.1) as
\begin{equation}\langle {\hat \Omega}\rangle =
\int d^D\sigma P_\sigma \Phi_\sigma \Omega_\sigma\;,
\end{equation}
where
$$
P_\sigma=
{{\mid W_\sigma\mid}\over{\int d^D\sigma\mid W_\sigma \mid \Phi_\sigma}} \;,
$$
and $\Phi_\sigma =W_\sigma/\mid W_\sigma\mid $
is the sign of the real part of $W_\sigma$. (Note that
since the partition function is real, we can neglect the imaginary part.)
Since $\mid W_\sigma \mid$ is non-negative by definition, we can interpret it,
suitably
normalized, as a probability density, so that upon rewriting (7.1) as
\begin{equation}\langle {\hat \Omega} \rangle ={{\int d\sigma \mid W_\sigma\mid
\Phi_\sigma {\Omega}_\sigma}
\over{\int d\sigma \mid W_\sigma \mid \Phi_\sigma}}=
{{\langle \Phi \hat {\Omega}\rangle}\over{\langle\Phi\rangle}}\;,
\end{equation}
we can think of the observable as a ratio in which the numerator and
denominator can be separately evaluated by MC quadrature. Leaving aside the
issue of correlations between estimates of these two quantities (they can
always be evaluated using separate Metropolis walks), the fractional variance
of $\langle {\hat \Omega}\rangle$ will be
\begin{equation}
{{\sigma_\Omega}\over{\langle {\hat \Omega}\rangle}} = \sqrt{
{{\langle {\hat \Omega}^2\rangle}\over{\langle\Phi {\hat \Omega}\rangle^2}}+
{{1}\over{\langle\Phi\rangle^2}}-2 \; ,
}
\end{equation}
which becomes unacceptably large as the average sign
$\langle\Phi\rangle$ approaches zero.
The average sign of the weight thus determines the feasibility of
naive MC quadrature.
Only a handful of interacting electron systems
are known to give rise to a positive-definite path integral \cite{Linden}:
the
one-dimensional Hubbard model, the half-filled Hubbard model, and the
attractive Hubbard model at any dimension and filling.

When $\langle\Phi\rangle$ is near unity, we can attempt MC evaluation of the
numerator and
denominator separately, as will be illustrated in the example of section~8.8
below. However, for the most general case, we must have a deeper
understanding of the sign problem.
We rewrite the canonical expectation value of an observable $\hat \Omega$ as
\begin{equation}\langle{\hat \Omega}\rangle\equiv
{{\rm Tr}\,({\hat \Omega}e^{-\beta \hat H})\over
{\rm Tr}\,(e^{-\beta \hat H})}\approx
{\int D \sigma W_\sigma \Phi_\sigma
{\Omega}_\sigma\over
\int D \sigma W_\sigma \Phi_\sigma}\;,
\end{equation}
where in the spirit of Eq.~(7.1,7.2) we have introduced a positive definite
weight
\begin{equation}
W_\sigma=G_\sigma\mid {\rm Tr}_A {\hat U}_\sigma\mid \;,
\end{equation}
and the Monte Carlo sign
\begin{equation}
\Phi_\sigma={{{\rm Tr}_A {\hat U}_\sigma}\over{\mid {\rm Tr}_A
{\hat U}_\sigma\mid}}\;.
\end{equation}
The sign problem arises because the one-body partition function
${\rm Tr}_A {\hat U}_\sigma$
is not necessarily positive, so that the Monte Carlo
uncertainty in the denominator of Eq.~(7.4) (the $W$-weighted average sign,
$\langle\Phi\rangle$) can become comparable to or larger than
$\langle\Phi\rangle$ itself. In most cases $\langle\Phi\rangle$ decreases
exponentially with $\beta$ or with the number of time slices \cite{Guber92}.

\noindent
\subsection{Hamiltonians without sign problems}

An important class of interactions (pairing+quadrupole)
free from the sign problem (i.e.,
$\Phi_\sigma\equiv1)$ was found in Ref.~\cite{Lang}.
These are realized when $V_\alpha \leq 0$
for all $\alpha$ in Eq.~(3.16). In that case, $s_\alpha=1$ for all $\alpha$,
and, in the linearized hamiltonian given by Eq.~(3.19c),
$\hat{\cal O}$ and $\hat{\bar{\cal O}}$ couple to complex conjugate fields.
In order to understand the implications of this situation, we represent
the single-particle wave function by an $N_S$-component vector of the form
\begin{equation}
\pmatrix{
jm\cr
j\bar{m}}\;,\end{equation}
with $N_S/2$ states $m>0$ in the first half of the vector, and their time
reversed
orbitals in the second half. Due to Eq.~(3.16) and the fact that
time-reversed operators are coupled to complex-conjugate
fields, the matrix $\bf {h}_n$ has the structure
\begin{equation}
\pmatrix{
{\bf A}_n & {\bf B}_n \cr
-{\bf B}^{*}_n & {\bf A}^{*}_n}\;, \end{equation}
and one can easily verify that the total evolution matrix
\begin{equation}
{\bf U}_\sigma =\prod_n \exp\left(-{\bf h}_n\Delta\beta\right)
=\pmatrix{
{\bf P} & {\bf Q} \cr
-{\bf Q}^{*} & {\bf P}^{*}}\;, \end{equation}
is of the same form. Here ${\bf A}$, ${\bf B}$, ${\bf P}$, and ${\bf Q}$ are
matrices of dimension $N_s/2$. One can show that this matrix
has pairs of complex-conjugate eigenvalues ($\varepsilon,\varepsilon^{*}$),
with respective eigenvectors $\pmatrix{u\cr v}$ and
$\pmatrix{-v^{*}\cr u^{*}}$. When $\varepsilon$ is real, it is
two-fold degenerate, since the two eigenvectors are distinct.

For the grand-canonical ensemble, the grand-canonical trace is given by
\begin{equation}
{\rm Tr}{\hat U}={\rm det}\left[\bf{1}+
\pmatrix{
{\bf P} & {\bf Q} \cr
-{\bf Q}^{*} & {\bf P}^{*}}\right] = \prod_\lambda^{N_s/2}
(1+\varepsilon_\lambda)(1+\varepsilon^{*}_\lambda) >0 \;. \end{equation}
If only particle-type conserving (neutron-proton) density operators are
present in Eq.~(3.16), each type of nucleon is represented by
a separate Slater determinant having the structure (7.9) and therefore
${\rm Tr}{\hat U}_p \cdot {\rm Tr}{\hat U}_n > 0$ since
${\rm Tr}{\hat U}_p > 0$ and ${\rm Tr}{\hat U}_n>0$.

In the zero-temperature formalism, if the trial wave function for an even
number of particles is chosen to consist of time-reversed pairs of single
particle states,
\begin{equation}
{\bf \Phi} =\pmatrix{
{\bf a} &{\bf b} \cr
-{\bf b}^{*} & {\bf a}^{*}\cr} \;, \end{equation}
where ${\bf a}$ and ${\bf b}$ are matrices with dimension
$\left( {N_s\over 2}\times {N_v\over 2}\right)$, then
${\bf\Phi}^{\dagger}{\bf U}{\bf\Phi}$ is a $N_v\times N_v$ matrix with the
structure (7.9), and the trace satisfies
\begin{equation}
{\rm Tr}{\hat U} ={\rm det}\left[{\bf\Phi}^\dagger{\bf U}{\bf\Phi}\right]\;.
\end{equation}
If only particle-type conserving operators are present, then time-reversed
pairs of trial wave functions can be
chosen for both protons and neutrons in an even-even nucleus, giving rise to
${\rm Tr}{\hat U} ={\rm Tr}{\hat U}_p\cdot{\rm Tr}{\hat U}_n >0$.
For $N=Z$ odd-odd nuclei we can choose the final neutron to be in the time
reversed orbit of the final proton, and the trace is once again positive.

Eq. (6.5b) embodies a proof that for even-even nuclei
(and $N=Z$ odd-odd systems) the partition
function is real and positive definite under the condition that all
$V_\alpha < 0$. This is most easily seen for the case where $N_s=2$, and
$A=1$ or $A=2$. From Eq.~(6.5b), and using the $N_s$-point quadrature
of Eq.~(6.7),
\begin{eqnarray}
Z_A &=& {1\over 2} \sum_{m=1}^{2} e^{\left(-i\phi_m A\right)}
\prod_{\lambda=1}^{2}\left[1+e^{(i\phi_m)} e^{(-\beta\varepsilon_\lambda)}
\right] \nonumber \\
   &=& {1\over 2}\sum_{m=1}^2\left\{
e^{(-i\phi_m A)}+e^{[-i\phi_m(A-2)]} e^{(-\beta 2{\rm Re}\varepsilon_1)}
 +   e^{[i\phi_m(A-1)]}{\rm Re}\  e^{(-\beta\varepsilon_1)} \right\}\;,
\end{eqnarray}
where we have ignored the $e^{(-\beta\mu)}$ term for simplicity, and
$\phi_m=m\pi$.
It is easy to show that for $A=2$ the first and second terms survive the
sum and that these terms are positive definite.
It is also easy to show that for $A=1$ only the third term,
${\rm Re}\  e^{(-\beta\varepsilon_1)}$ remains after the sum is taken.
While this
term is real, it is not positive definite. Hence there appear two distinct
sign problems. For even-even nuclei and
all $V_\alpha \leq 0$, the sign
problem can be overcome; however, for odd-$A$ nuclei, even with all
$V_\alpha \leq 0$,
the sign problem remains since the
contributions to $\langle\Phi\rangle$  are either positive (+1) or negative
($-1$), but not complex. This demonstration can be extended to all
$A$ in a given space. As in the zero-temperature formalism, $N=Z$ odd-odd
nuclei pose no sign problem, since the evolution is the same for both
the protons and the neutrons, and the product of the traces is thus
positive definite.

\noindent
\subsection{Practical solution to the sign problem}

For an arbitrary hamiltonian, we are not guaranteed that
all $V_\alpha \leq 0$.
However, we may expect that a {\it realistic} hamiltonian will be dominated
by terms like those of the schematic force (this is, after all, why the
schematic forces were developed) so that it is, in some sense close to a
hamiltonian for which the MC is directly applicable. Thus, the ``practical
solution'' to the sign problem presented in Ref.~\cite{Alhassid} is based on an
extrapolation of observables calculated for a ``nearby'' family of
Hamiltonians whose integrands have a positive sign. Success depends crucially
upon the degree of extrapolation required. Empirically, one finds that, for
all of the many realistic interactions tested in the $sd$- and $pf$-shells,
the extrapolation required is modest, amounting to a factor-of-two variation
in the isovector monopole pairing strength.

Based on the above observation, it is possible to decompose $\hat H$
in (3.16) into its
``good'' and ``bad'' parts, ${\hat H}= {\hat H}_G+ {\hat H}_B$, with
\begin{eqnarray}
{\hat H}_G&=&\sum_\alpha(\epsilon_\alpha^\ast\hat
{\bar {\cal O}}_\alpha+\epsilon_\alpha
{\hat {\cal O}}_\alpha)+ {1\over2}\sum_{V_\alpha<0}V_\alpha
\left\{{\hat {\cal O}}_\alpha,
\hat{\bar {\cal O}}_\alpha\right\}
\nonumber \\
{\hat H}_B&=&{1\over2}\sum_{V_\alpha>0}V_\alpha
\left\{
{\hat {\cal O}}_\alpha, \hat{\bar{\cal O}}_\alpha \right\}.
\end{eqnarray}
\noindent
The ``good'' Hamiltonian ${\hat H}_G$ includes, in addition to the one-body
terms,
all the two-body interactions with $V_\alpha \leq0$, while the ``bad''
Hamiltonian ${\hat H}_B$ contains all interactions with $V_\alpha>0$. By
construction, calculations with ${\hat H}_G$
alone have $\Phi_\sigma\equiv1$ and
are thus free of the sign problem.

We define a family of Hamiltonians ${\hat H}_g$
that depend on a continuous real
parameter~$g$ as
${\hat H}_g=f(g){\hat H}_G+g {\hat H}_B$, so that ${\hat H}_{g=1}={\hat H}$,
and
$f(g)$ is a function with $f(1)=1$ and $f(g<0)>0$
that can be chosen to make the extrapolations less
severe. (In practical applications $f(g)=1-(1-g)/\chi$ with $\chi\approx4$
has been found to be a good choice.)
If the $V_\alpha$ that
are large in magnitude are ``good,'' we expect
that ${\hat H}_{g=0}={\hat H}_G$ is a
reasonable starting point for the calculation of an
observable $\langle{{\hat \Omega}
}\rangle$. One might then hope to calculate
$\langle{{\hat \Omega}}\rangle_g={\rm
Tr}\,({\hat \Omega}e^{-\beta \hat H_g})/{\rm Tr}\,(e^{-\beta \hat H_g})$
for small $g>0$ and
then to extrapolate to $g=1$, but typically $\langle\Phi\rangle$ collapses
even for small positive $g$. However, it is evident from our construction
that ${\hat H}_g$ is characterized by $\Phi_\sigma\equiv1$ for any $g\leq 0$,
since
all the ``bad'' $V_\alpha(>0)$ are replaced by ``good'' $g V_\alpha<0$. We
can therefore calculate
$\langle{{\hat \Omega}}\rangle_g$ for any $g\leq0$ by a
Monte Carlo sampling that is free of the sign problem.
If $\langle{{\hat \Omega}
}\rangle_g$ is a smooth function of $g$, it should then be possible to
extrapolate to $g=1$ (i.e., to the original Hamiltonian) from $g\leq0$. We
emphasize that $g=0$ is not expected to be a singular point of
$\langle{{\hat \Omega}
}\rangle_g$; it is special only in the Monte Carlo evaluation.

As described in Section~4,
the matrices $E^\pi_{KT}$
are constructed from the two-body matrix elements $V^\pi_{JT}(ab,cd)$ of good
angular momentum $J$, isospin $T$, and parity $\pi=(-1)^{l_a+l_b}$
through a Pandya
transformation. For interactions that are time-reversal invariant and
conserve parity, the $E^\pi_{KT}(i,j)$ (here $\pi=(-1)^{l_a+l_c}$)
are real symmetric matrices that can
be diagonalized by a real orthogonal transformation. The eigenvectors
${\hat \rho}_{KM}(\alpha)$ play the role of
${\hat {\cal O}}_\alpha$ in Eq.~(3.16), and the
eigenvalues $\lambda_{K\pi}(\alpha)$ are proportional to $V_\alpha$. In the
Condon-Shortley \cite{Condon35}
convention $\hat{{\bar \rho}}_{KM}=\pi(-)^{K+M}{\hat \rho}_{K-M}$ so
that the ``good'' eigenvalues satisfy ${\rm sign}~[\lambda_{K\pi} (\alpha)]=
\pi(-)^{K+1}$ \cite{7D}.

As an example, we consider
the mid-$pf$ shell nucleus ${}^{54}$Fe using the realistic Kuo-Brown KB3
interaction \cite{Zuker}.
Figure~7.1 (upper) shows the eigenvalues $V_{K\pi\alpha}= \pi(-)^{K}
\lambda_{K\pi} (\alpha)$ of the KB3 interaction; only about half of
the eigenvalues are negative. However, those with the largest magnitude are
all ``good.'' It is possible to use an inverse Pandya transformation to
calculate the two-body matrix elements $V^\pi_{JT}(ab,cd)$ for the
``good'' and ``bad'' interactions, allowing the matrix elements of
${\hat H}_G$ to
be compared in Fig.~7.1 (lower) with those of the full interaction. The
greatest deviation is for $J=0,T=1$ (the monopole pairing interaction), where
${\hat H}_G$ is about twice as attractive as the physical ${\hat H}$.
In all other
channels, ${\hat H}_G$ and ${\hat H}$ are quite similar.

In Fig. 7.2 we exemplify the $g$-extrapolation procedure for several
observables, calculated again for $^{54}$Fe.
In all cases we use polynomial extrapolations from negative $g$-values
to the physical case, $g=1$. The degree of the polynomial is
usually chosen to be
the smallest that yields a $\chi^2$ per degree of freedom less than 1.
However, in several studies, like the one of the $pf$-shell nuclei
reported in section 8.1, we have conservatively chosen second-order
polynomials for all extrapolations, although in many cases a first-order
polynomial already resulted in $\chi^2$-values less than 1.
At $T=0$ the variational principle requires that
the expectation value of $\hat H$ has a minimum at $g=1$. We have
incorporated this fact in our extrapolations of ground state energies
by using a second-order polynomial with zero-derivative at $g=1$.

\begin{figure}
$$\epsfxsize=4truein\epsffile{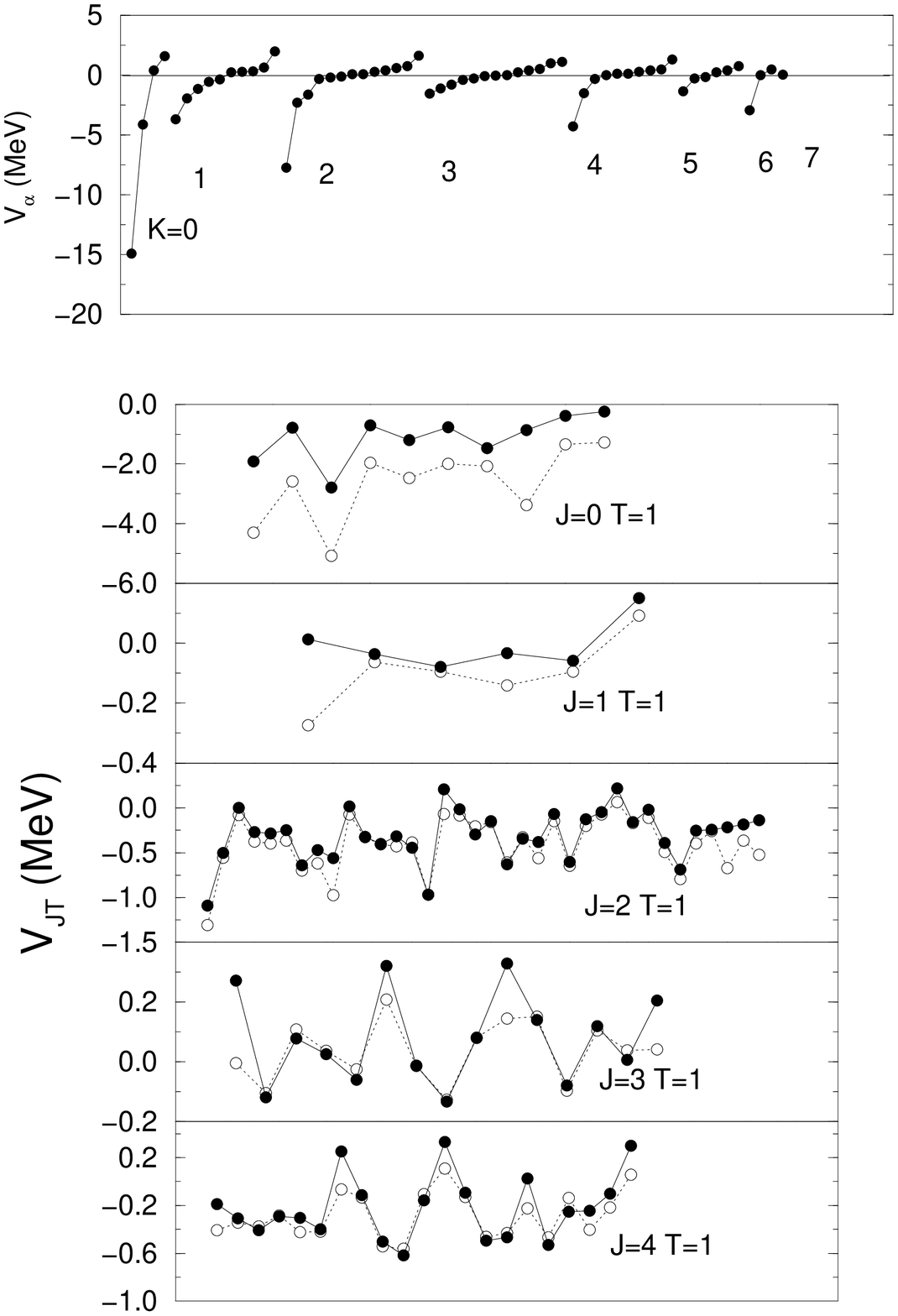}$$
{FIG~7.1.
Decomposition of the Kuo-Brown KB3 interaction. The upper panel
shows the eigenvalues $V_\alpha$ for the various particle-hole angular
momenta $K$. The lower panels show the particle-particle two-body
matrix elements in selected $JT$ channels. Solid symbols are the
full hamiltonian, while open symbols correspond to ${\hat H}_G$.
}
\end{figure}

\begin{figure}
$$\epsfxsize=4truein\epsffile{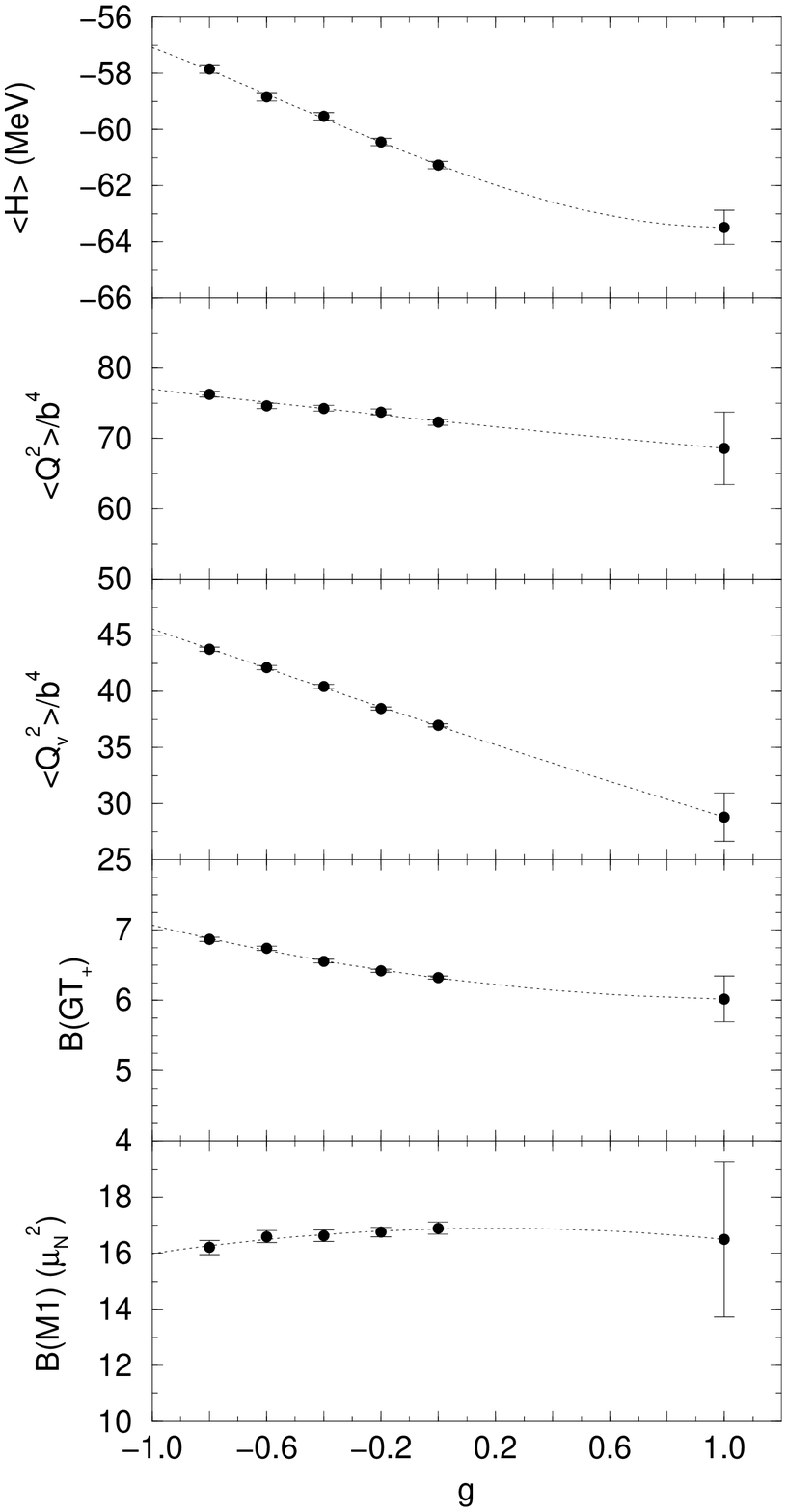}$$
{FIG~7.2.
$g$-extrapolation of several observables for $^{54}$Fe calculated
with the Kuo-Brown interaction KB3.
}
\end{figure}

\noindent
\subsection{Validation}

Confidence in SMMC results requires numerical validation of the methods used.
Separate issues are the
validation of the SMMC formalism and algorithms, as
developed in sections 3--6, and of the $g$-extrapolation required
in calculations with realistic residual interactions (see subsection 7.2).

Apart from demonstrating convergence in the various calculational parameters
($\Delta \beta$, number of samples, etc.), SMMC algorithms have been
validated by comparison with direct diagonalization.
Several examples of both the zero temperature and thermal SMMC formalisms
are given in Refs. \cite{Johnson,Lang}. These calculations
have been performed for Hamiltonians possessing good sign properties
(i.e. all $V_{\alpha} \leq 0$). For ground state and thermal properties,
as well as for strength distributions, excellent agreement between
direct diagonalization and the SMMC methods has been demonstrated \cite{Lang}.
Some examples are summarized in Fig. 7.3.

\begin{figure}
$$\epsfxsize=4truein\epsffile{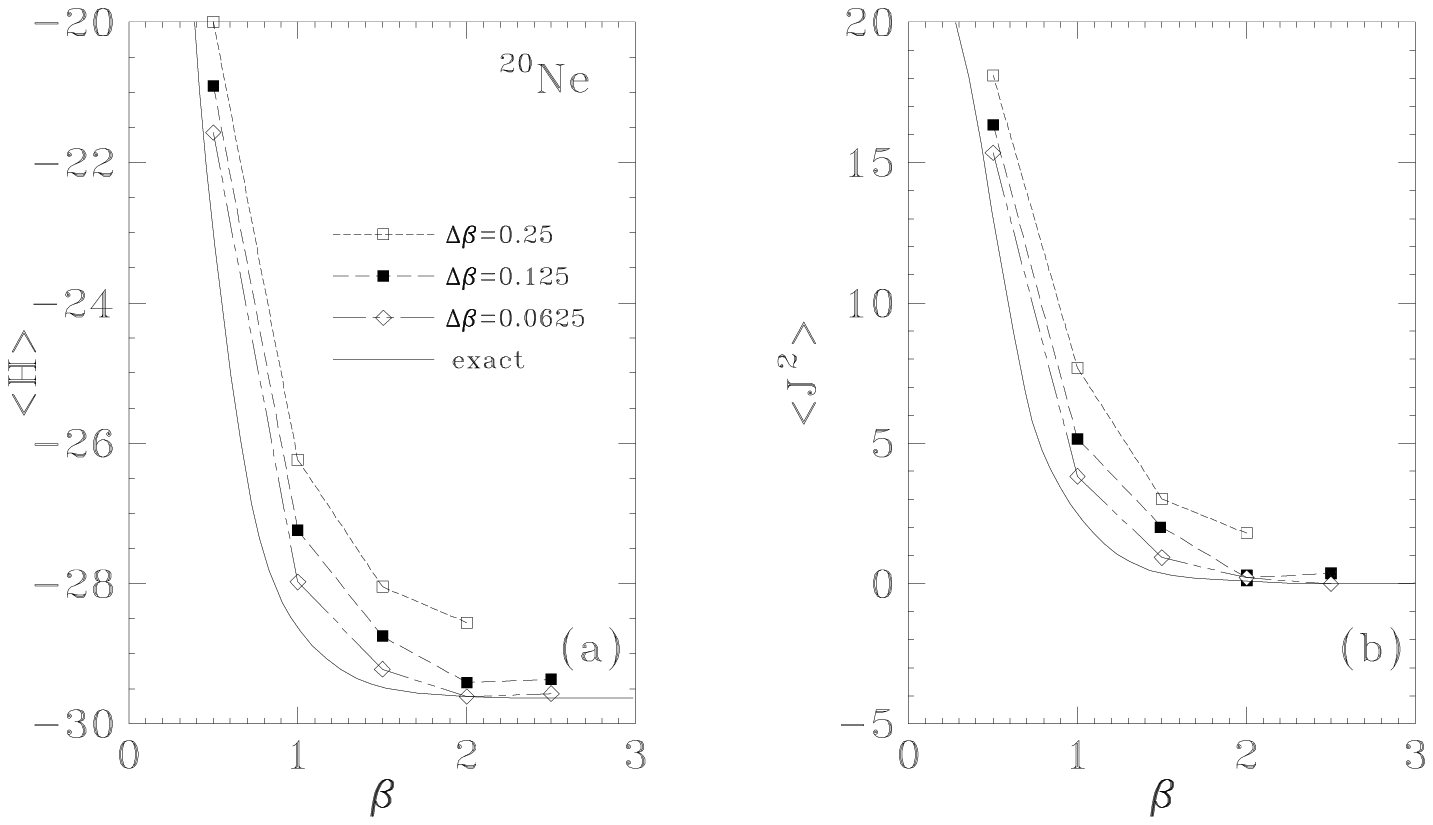}$$
{FIG~7.3.
Comparison of SMMC results for selected observables with those
obtained by conventional diagonalization \protect\cite{Lang}.
The calculations have
 been performed for
$^{20}$Ne within the $sd$ model space. A pairing+quadrupole interaction,
which is free from the sign problem discussed in section 7.2,
has been used.
}
\end{figure}

The $g$-extrapolation has been tested for several nuclei in the $sd$-shell
\cite{Alhassid} and in the low $pf$-shell \cite{Langanke95}.
For example, using the parameter
$\chi=4$ SMMC can reproduce \cite{Langanke95}
all observables (ground state energy,
total B(E2), B(M1), GT strength etc.) for those nuclei in the mass range
$A=48-56$ for which direct diagonalization studies \cite{Caurier94}
or estimates on the basis
of a series of truncation calculations \cite{Poves} exist. An example for
the nuclei $^{54,56}$Fe is given in Fig. 2.2.

The extrapolation procedure in thermal SMMC studies
has been validated for
$^{44}$Ti by comparison to a direct diagonalization
calculation. The code CRUNCHER and the residual FPVH interaction
were used to calculate
the lowest
750 eigenstates up to an excitation energy of 43 MeV, which
is sufficient to calculate the energy expectation value as a function
of temperature up to $T=2$ MeV. We note that exact reproduction
of the CRUNCHER results was obtained only after a $\Delta \beta \rightarrow 0$
extrapolation which lowers the energies slightly compared to the calculation
with finite $\Delta \beta = 1/32$ MeV$^{-1}$. Fig. 7.4 compares the
($\Delta \beta$- and $g$-extrapolated) SMMC results with those obtained
by direct diagonalization. Note that we cannot
validate thermal properties
requiring the calculation of transition matrix elements,
as the latter are not numerically practical in direct diagonalization
for thermal ensembles.

\begin{figure}
$$\epsfxsize=4truein\epsffile{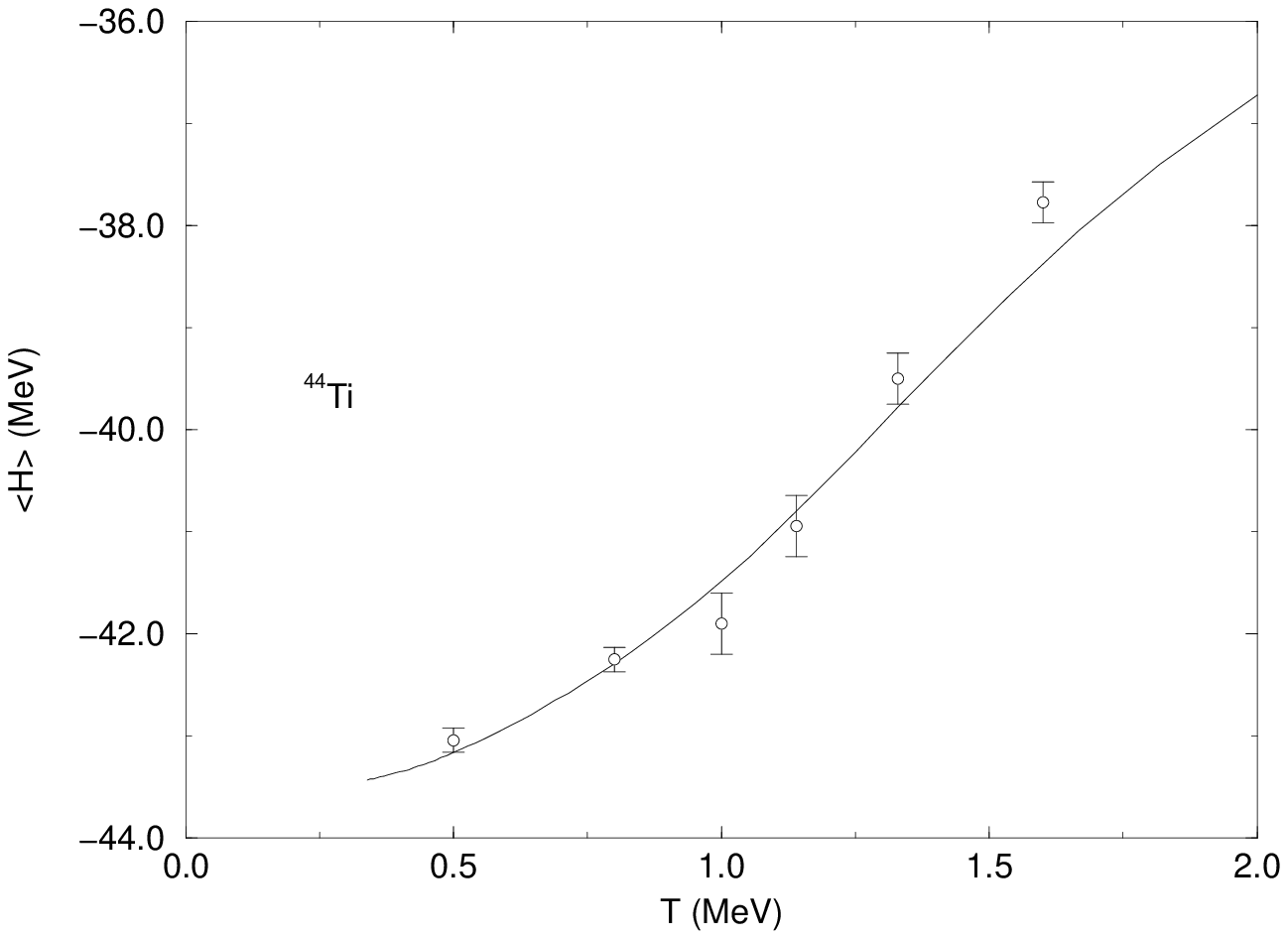}$$
{FIG~7.4.
Comparison of SMMC results for the temperature dependence
of the internal energy in $^{44}$Ti with those
obtained by conventional diagonalization code CRUNCHER.
}

\end{figure}

\noindent
\section{Selected results}

The previous sections have outlined the motivation, formalism,
implementation, and validation of Monte Carlo methods for treating the
nuclear shell model. Of course, the ultimate goal of this work is to gain
insight into the properties of real nuclei under a variety of conditions. In
this section, we present a sampling of such results and their interpretation.
The calculations presented should be viewed only as an indication of the
power and potential of the method; undoubtedly, more work of this sort,
abetted by ever-increasing computer power, will follow in the future.

\noindent
\subsection{Ground state properties of $pf$-shell nuclei}

While complete $0\hbar\omega$ calculations can be carried out by direct
diagonalization in the $p$- and $sd$-shells, the exponentially increasing
number of configurations limits such studies in the next ($pf$) shell to only
the very lightest nuclei \cite{Richter,Caurier94}.
SMMC techniques allow calculation of groundstate
observables in the full $0\hbar\omega$ model space for nuclei throughout the
$pf$-shell. Here, we discuss a set of such calculations that use the modified
KB3
interaction \cite{Zuker};
the single-particle basis is such that $N_s=20$ for both
protons and neutrons. A more detailed description of the calculations and
their results are found in Ref. \cite{Langanke95}.

These studies were performed for 28 even-even Ti, Cr, Fe, Ni, and Zn
isotopes; extension to heavier nuclei is dubious because of the neglect of
the $g_{9/2}$ orbital. We have used $\beta=2$ MeV$^{-1}$ with fixed
$\Delta\beta=1/32$ MeV$^{-1}$;
test calculations at varying $\beta$ show that the results
accurately reflect the ground state properties. Some 4000-5000
independent Monte
Carlo samples were taken for six equally-spaced values of $g$ between $-1$
and 0, and observables were extrapolated to $g=1$ using linear or quadratic
functions, as required. We chose the parameter $\chi=4$, as described in
Section~7.2.

Figure 8.1 shows systematic results for the mass-defects, obtained directly
from $\langle \hat H\rangle$. The SMMC results have been corrected
for the Coulomb energy,
which is not included in the KB3 interaction,
using \cite{Caurier94}
\begin{equation}
H_{\rm Coul} = {\pi(\pi-1)\over2} \cdot 0.35 - \pi \nu 0.05 + \pi \cdot 7.289.
\end{equation}
where
$\pi$ and $\nu$ are the numbers of valence protons
and neutrons, respectively, and the energy is in MeV.
As in Ref. \cite{Caurier94}, we have increased the calculated energy
expectation values by $0.014 \cdot n(n-1)$ MeV
(where $n=\pi+\nu$ is the number of valence nucleons)
to correct for
a ``tiny'' residual monopole defect in the KB3 interaction.
In general, there is
excellent agreement; the average error for the nuclei shown is $+0.45$ MeV,
which agrees roughly with the internal excitation energy
of a few hundred keV expected in our finite-temperature calculation.

\begin{figure}
$$\epsfxsize=4truein\epsffile{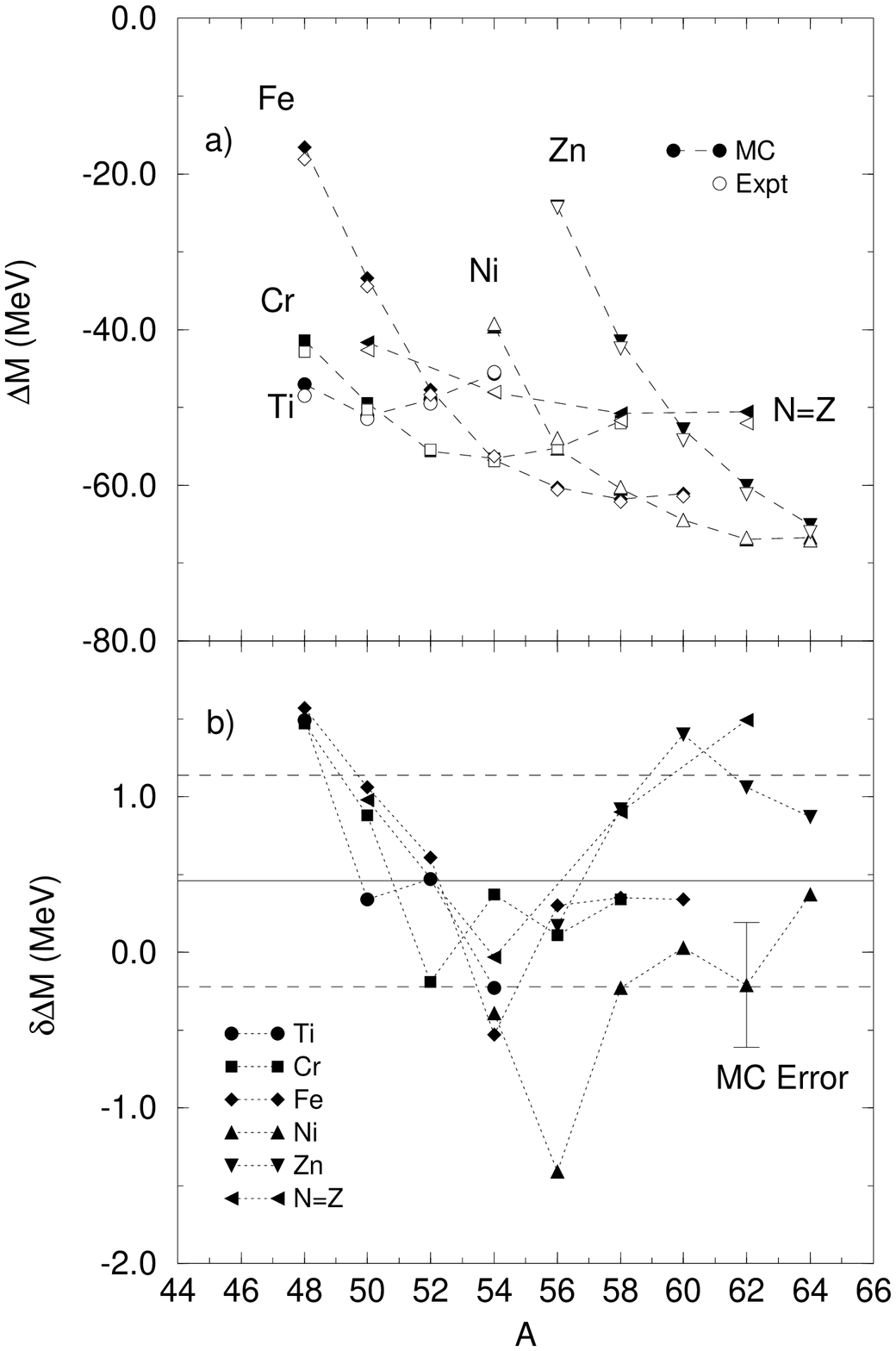}$$
{FIG~8.1.
Upper panel (a): Comparison of the mass excesses $\Delta M$ as calculated
within the SMMC approach with data. Lower panel (b): Discrepancy between
the SMMC results for the mass excesses and the data, $\delta \Delta M$.
The solid line shows the average discrepancy, 450 keV, while the dashed
lines show the rms variation about this value (from \protect\cite{Langanke95}).
}
\end{figure}

Figure 8.2 shows the calculated total $E2$ strengths for selected $pf$-shell
nuclei. This quantity is defined as
\begin{equation}
B(E2) = \langle (e_p {\hat Q}_p + e_n {\hat Q}_n )^2 \rangle \; ,
\end{equation}
with
\begin{equation}
{\hat Q}_{p(n)} = \sum_i r_i^2 Y_2 (\theta_i, \phi_i) \; ,
\end{equation}
where the sum runs over all valence protons (neutrons).
The effective charges were chosen to be $e_p=1.35$ and $e_n=0.35$,
while we used $b=1.01 A^{1/6}$ fm for the oscillator length.
Shown for comparison are the $B(E2)$ values for the $0^+_1\rightarrow 2^+_1$
transition in each nucleus; in even-even nuclei
some 20-30\% of the strength comes from higher transitions.
The overall trend is well reproduced,
For the nickel isotopes $^{58,60,64}$Ni,
the total $B(E2)$ strength is known
from $(e,e')$ data and agrees very nicely with our SMMC results.

\begin{figure}
$$\epsfxsize=4truein\epsffile{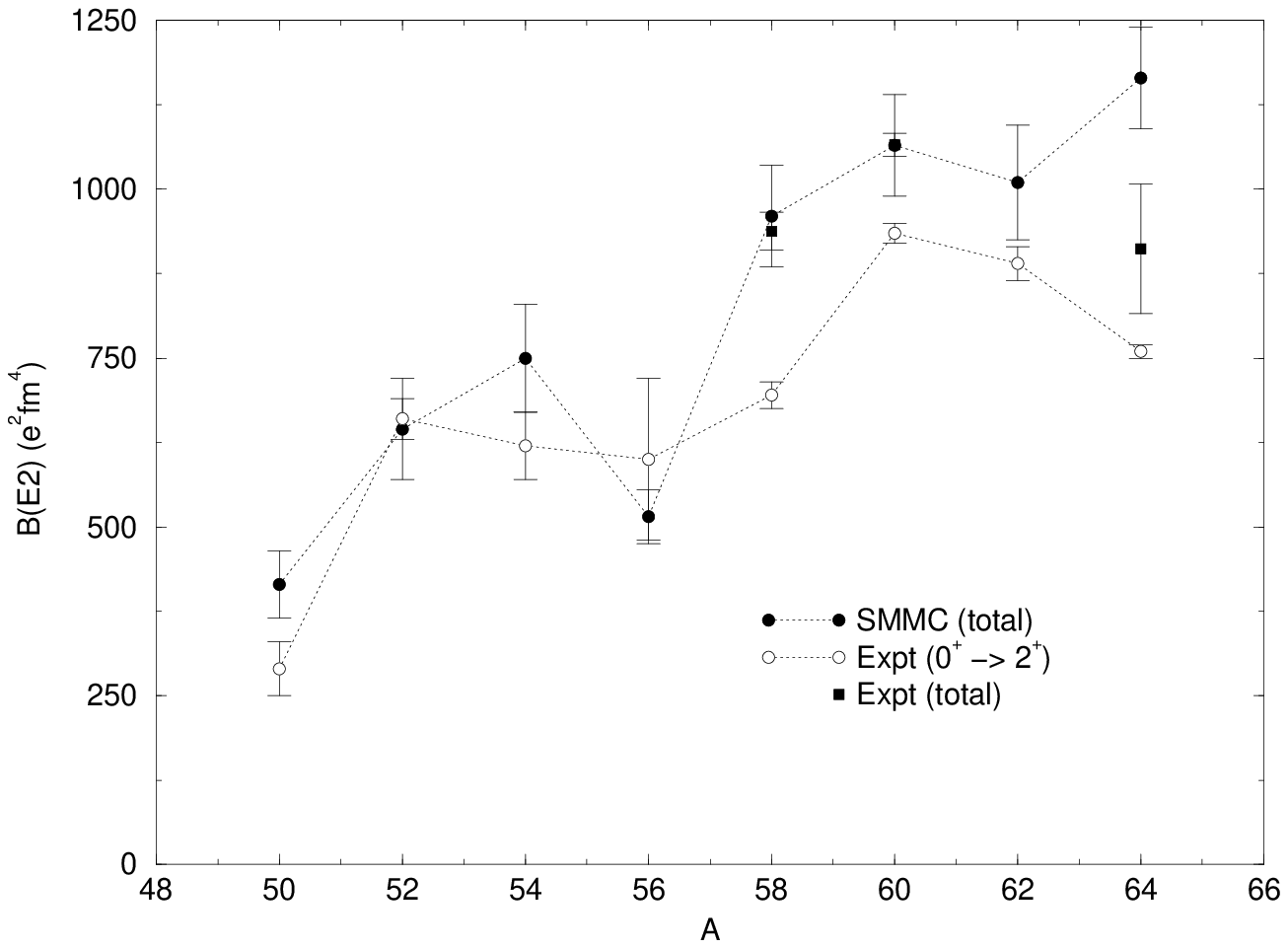}$$
{FIG~8.2.
Comparison of the experimental $B(E2, 0_1^+ \rightarrow 2_1^+)$
strengths with the total $B(E2)$ strength calculated in the SMMC approach
for various $pf$-shell nuclei having either proton or neutron
number of $28$. For the nickel isotopes $^{58,60,64}$Ni the total $B(E2)$
strength (full squares) is known from inelastic electron scattering data
(from \protect\cite{Langanke95}).
}
\end{figure}

The Gamow-Teller (GT) properties of nuclei in this region of the periodic table
are crucial for supernova physics \cite{Bethe90}.
These strengths are defined as
\begin{equation}
B(GT_{\pm}) = \langle ({\vec \sigma} \tau_{\pm} )^2 \rangle .
\end{equation}
Note that the Ikeda sum rule, $B({\rm GT}_-)-B({\rm GT}_+)=3 (N-Z)$ is
satisfied exactly by our calculations.
In Figure 8.3, we show the calculated $B({\rm GT}_+)$ for all $pf$-shell
nuclei for which
this quantity has been measured by $(n,p)$ reactions.
(For the odd nuclei $^{51}$V, $^{55}$Mn and $^{59}$Co the SMMC calculations
have been performed at $\beta =1$ MeV$^{-1}$ to avoid the odd-$A$ sign
problem.
For even-even nuclei
the GT strength calculated at this temperature is still a good
approximation for the ground state value, as is shown below).
Since the experimental
values are normalized to measured $\beta$-decay $ft$ values, and it is believed
that the axial coupling constant is renormalized from its free value
$(g_A=1.26)$
to $g_A=1$ in nuclei \cite{Wild88,Caurier94},
we have multiplied all calculated results by
$(1/1.26)^2$. Note that this results in generally excellent agreement between
the calculations and the data. Thus, we support both the statement that $g_A$
is quenched to 1 in nuclei and the statement
that complete shell-model calculations can
account for the GT strength observed experimentally.
Note that the quenching of $g_A$ in the nuclear medium is not quite
understood yet.  It is believed to be related either to a second-order
core polarization caused by the tensor force \cite{Hamomoto}
or to the screening of the
Gamow-Teller operator by $\Delta$-hole pairs \cite{Delta}

\begin{figure}
$$\epsfxsize=4truein\epsffile{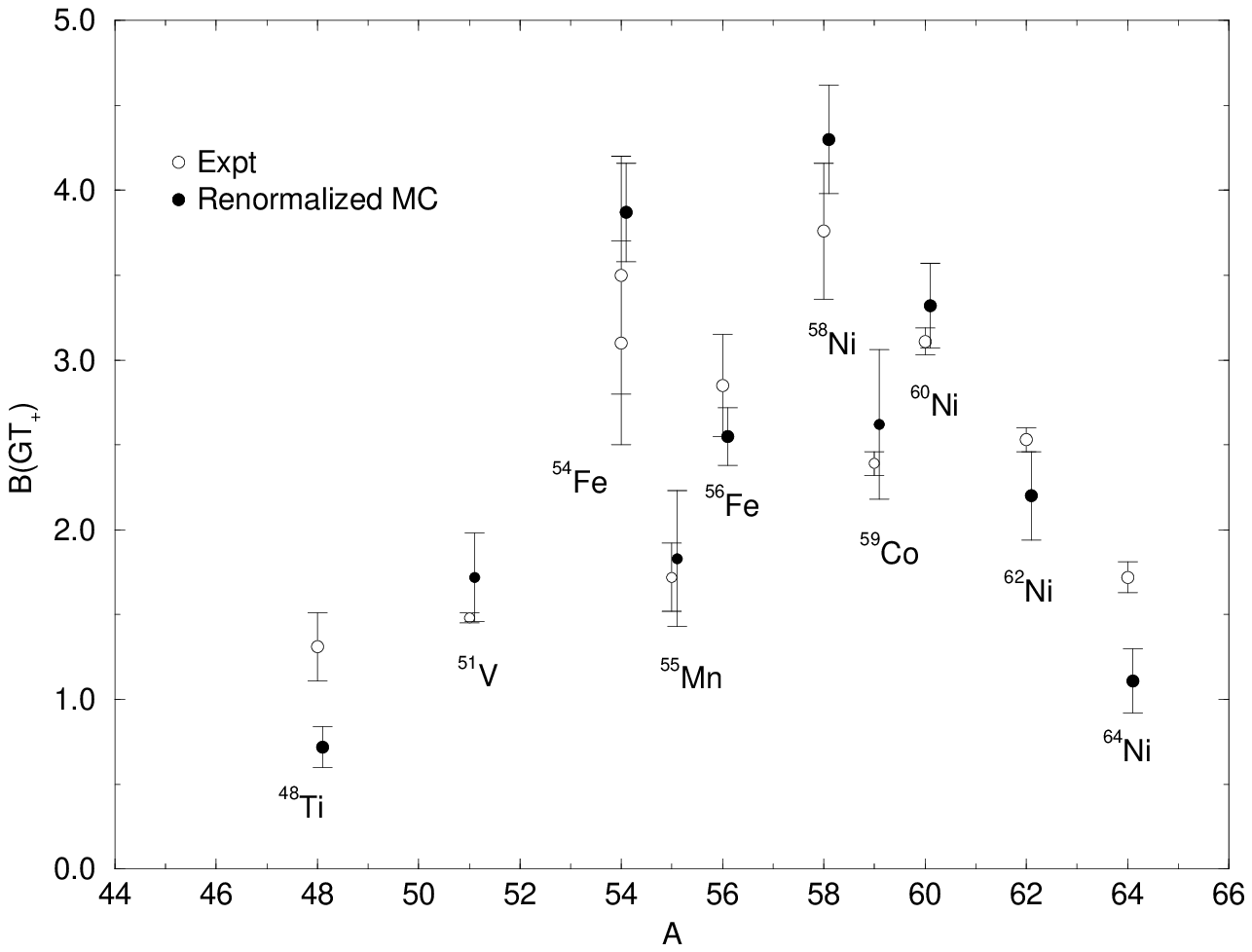}$$
{FIG~8.3.
Comparison of the renormalized total Gamow-Teller strength, as calculated
within the present SMMC approach, and the experimental $B(GT_+)$ values
deduced from $(n,p)$ data \protect\cite{gtdata1}-\protect\cite{gtdata5}.
}
\end{figure}

The calculated systematics of the $B(GT_+)$ strength
for isotope chains throughout
this region are shown in Figure~8.4, where it can be seen that ${\rm GT}_+$
increases as the number of valence neutron holes increases (unblocking), and
as the number of valence protons increases. These systematics were summarized
phenomenologically from the experimental data in Ref.~\cite{Koonin94} as
\begin{equation}
B(GT_+) = 0.045 \cdot Z_{\rm val} \cdot (20- N_{\rm val}) \; ,
\end{equation}
which is derived by assuming that the residual interaction is strong enough
to destroy all sub-shell structure, so that the entire $pf$-shell behaves as
one large orbital. Figure~8.5 shows the linearity of $B({\rm GT}_+)/Z$ with
$(N_{\rm val}-20)$ for the experimental data.

\begin{figure}
$$\epsfxsize=4truein\epsffile{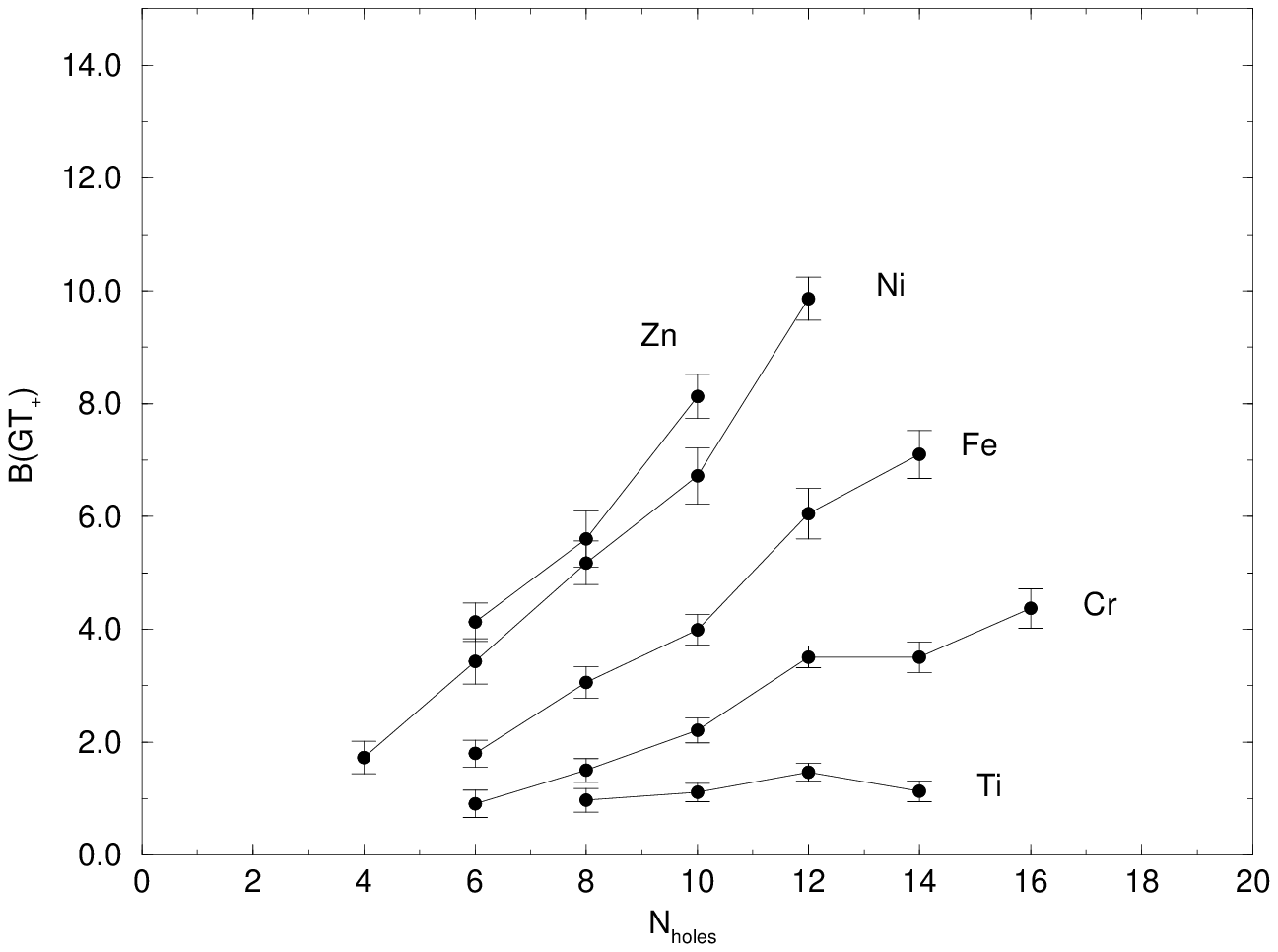}$$
{FIG~8.4.
Unrenormalized total Gamow-Teller strengths for various isotope chains
as a function of neutron holes $(N_{\rm val}-20)$ in the $pf$-shell
(from \protect\cite{Langanke95}).
}
\end{figure}

\begin{figure}
$$\epsfxsize=4truein\epsffile{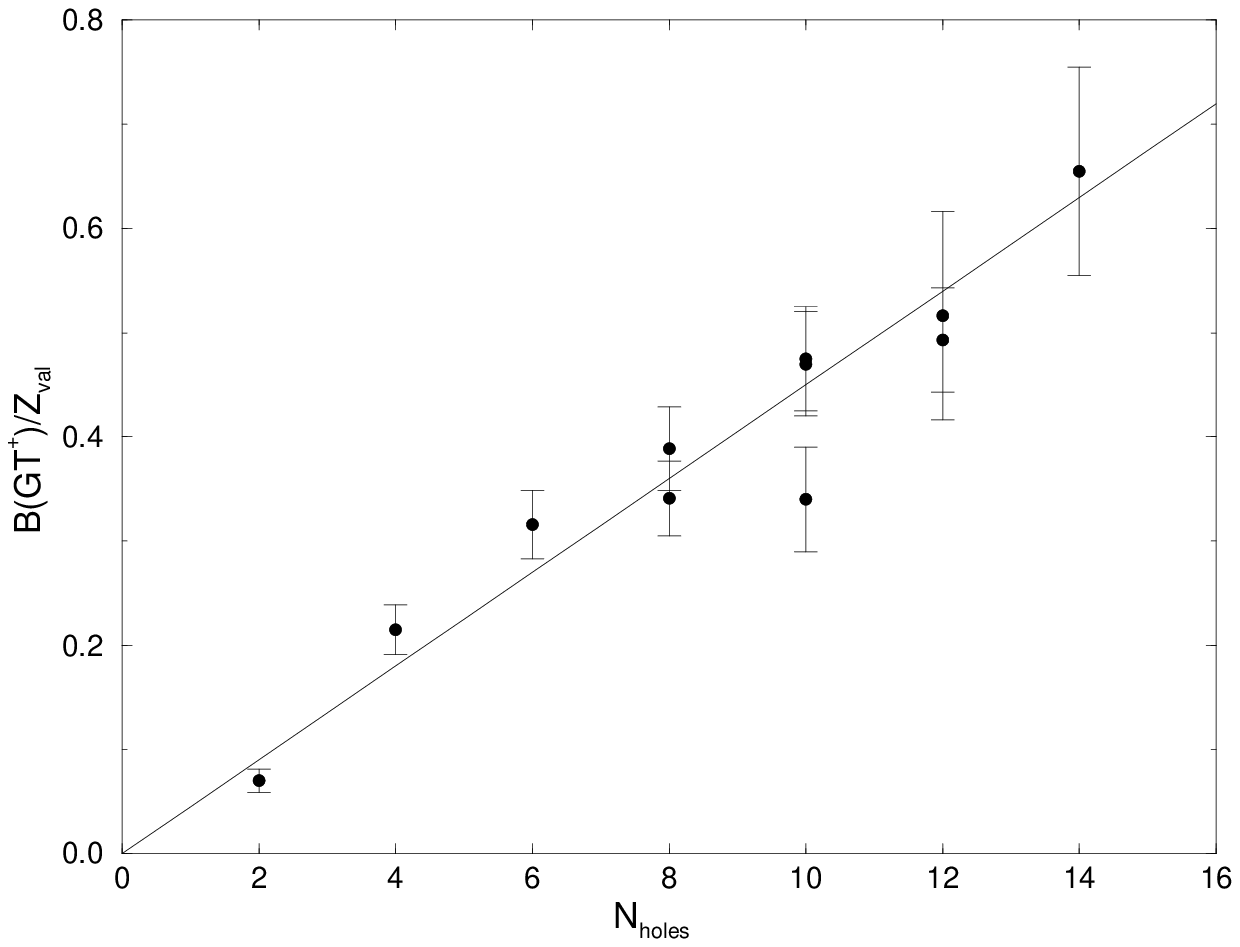}$$
{FIG~8.5.
Experimental total Gamow-Teller strengths
\protect\cite{gtdata1}-\protect\cite{gtdata5},
divided by the number of valence
protons, as a function of neutron holes in the $pf$-shell.
}
\end{figure}

The magnetic dipole strengths are defined as
\begin{equation}
B(M1) = \langle
( \sum_i \mu_N \left\{ g_l {\vec l} + g_s {\vec s} \right\} )^2
\rangle  .
\end{equation}
where $\mu_N$ is the nuclear magneton and $g_l, g_s$ are
the free gyromagnetic ratios for
angular momentum and spin, respectively
($g_l=1, g_s=5.586$ for protons, and $g_l=0, g_s=-3.826$ for neutrons).
A complete list of the total $B(M1)$ strengths as calculated in SMMC
studies of $pf$-shell nuclei
is given in Ref. \cite{Langanke95}.

Brown and Wildenthal \cite{Wild88} suggested that, as a general rule,
the strength of spin operators is renormalized in nuclei.
To account for this renormalization, it is customary to use
effective gyromagnetic ratios for the spin, ${\tilde g}_s = g_s/1.26$
in shell model calculations \cite{Wild88,Caurier94}. If we follow
this presciption, we calculate total $B(M1)$ strengths
for the $N=28$ isotones (in units of $\mu_N^2$) of $7.9\pm0.7$ ($^{50}$Ti),
$11.7\pm1.5$ ($^{52}$Cr), and $10.2\pm 2$ ($^{54}$Fe).
For these nuclei, Richter and collaborators \cite{Richter85} have used
high-resolution electron scattering to study the $B(M1)$ strength in an
energy window large enough to contain most of the strength. They find
$B(M1)=4.5\pm0.5$ ($^{50}$Ti), $8.1\pm0.8$ ($^{52}$Cr) and $6.6 \pm 0.4$
($^{54}$Fe) for excitation energies between 7 and 12 MeV \cite{Richter85}.
If one considers that this energy window should contain about $75\%$ of the
total strength \cite{Richter85}, our SMMC results appear to be consistent
with the data.

\noindent
\subsection{Pair structure of the nuclear ground state}

Numerous phenomenological descriptions of nuclear collective motion describe
the nuclear ground state and its low-lying excitations in terms of bosons.
In one such model, the Interacting Boson Model
(IBM), $L=0$ (S) and $L=2$ (D) bosons are identified with
nucleon pairs having the same quantum numbers \cite{Arima},
and the ground state can be
viewed as a condensate of such pairs.  Shell model studies of the pair
structure of the ground state and its variation with the number of valence
nucleons can therefore shed light on the validity and microscopic foundations
of these boson approaches.

For many purposes, it appears  sufficient to study the BCS pair structure
in the ground state.
The BCS pair operator for protons can be defined as
\begin{equation}
\hat{\Delta}^\dagger_p= \sum_{jm > 0} p^\dagger_{jm} p^\dagger_{j\bar m} \;,
\end{equation}
where the sum is over all orbitals with $m>0$ and
$p^{\dagger}_{j\bar{m}}=(-)^{j-m}p^{\dagger}_{jm}$ is the time-reversed
operator.
Thus, the observable
$\hat{\Delta}^\dagger \hat{\Delta}$
and its analog for neutrons are measures of the numbers of $J=0,T=1$ pairs in
the groundstate. For an uncorrelated Fermi gas, we have simply
\begin{equation}
\langle \hat{\Delta}^\dagger \hat{\Delta} \rangle
= \sum_j {n_j^2 \over {2 (2j+1)}},
\end{equation}
where the $n_j= \langle p^\dagger_{jm} p_{jm} \rangle$
are the occupation numbers, so that any
excess of $\langle \hat \Delta^\dagger \hat \Delta \rangle$
in our SMMC calculations
over the Fermi-gas value indicates pairing correlations in
the groundstate.

Fig. 8.6 shows the SMMC expectation values of the proton and neutron
BCS-like pairs,
obtained after subtraction of the Fermi gas value (8.8), for three
chains of isotopes. As expected, these excess pair correlations are quite
strong and reflect the well-known coherence in the ground states
of even-even nuclei. Note that the proton BCS-like pairing fields are not
constant within an isotope chain, showing that there are important
proton-neutron correlations present in the ground state. The shell closure
at $N=28$ is manifest in the neutron BCS-like pairing. As is demonstrated
in Fig. 8.7, the proton and neutron occupation numbers show a
much smoother behavior
with increasing $A$.

\begin{figure}
$$\epsfxsize=4truein\epsffile{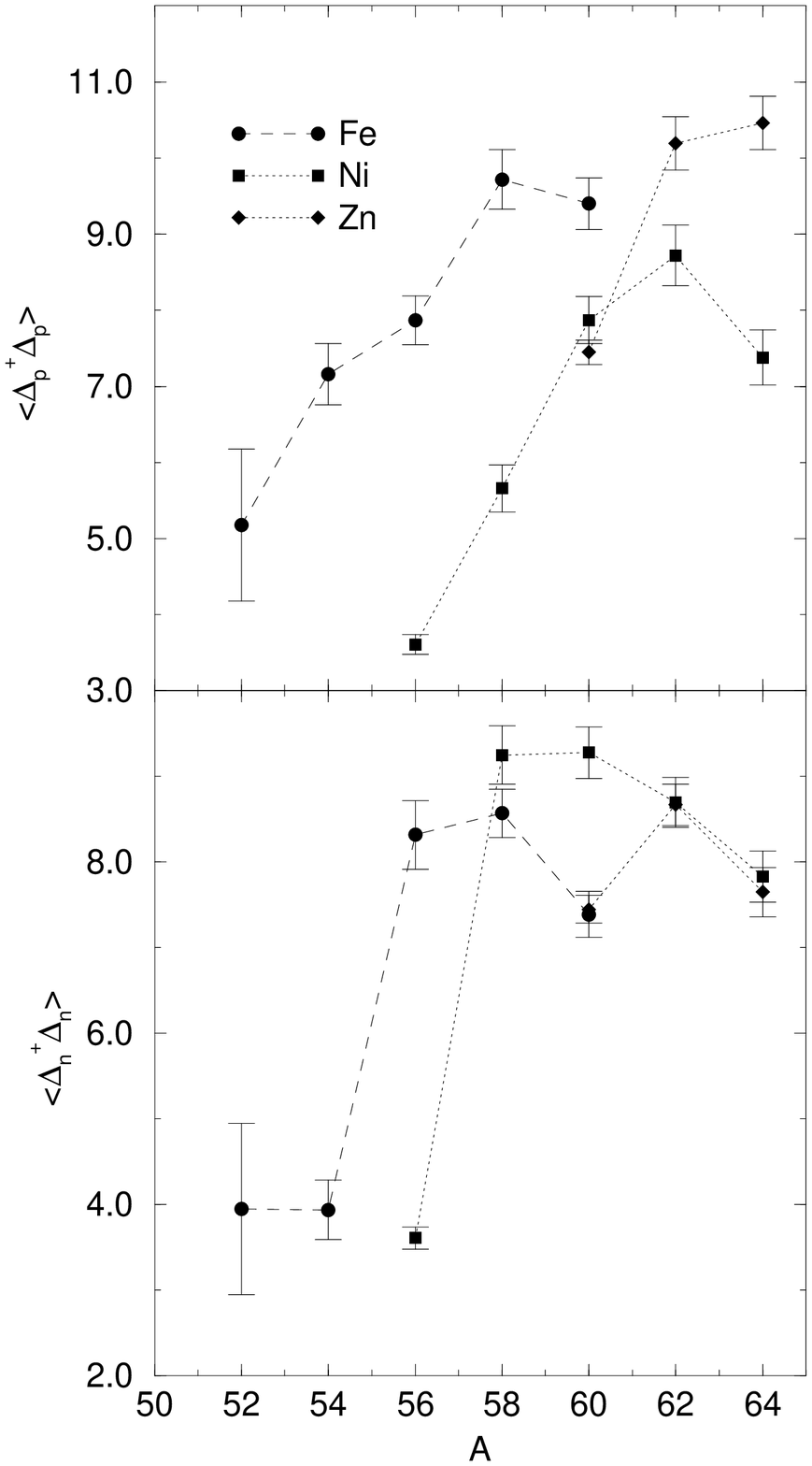}$$
{FIG~8.6.
SMMC expectation values of proton and neutron BCS-like pairs after
subtraction of the Fermi gas value
(from \protect\cite{Langanke95}).
}
\end{figure}

\begin{figure}
$$\epsfxsize=4truein\epsffile{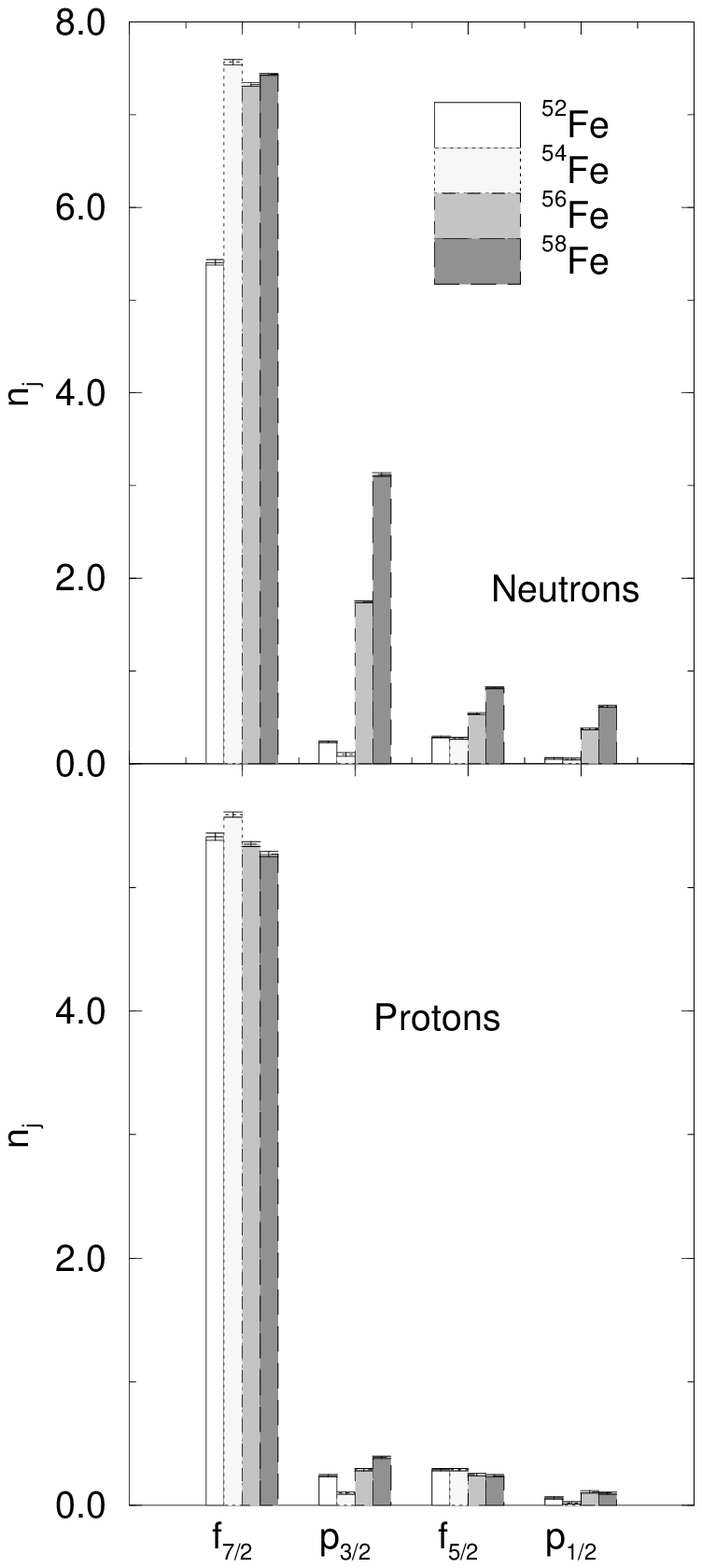}$$
{FIG~8.7.
Proton and neutron occupation numbers for various chains of isotopes
(from \protect\cite{Langanke95}).
}
\end{figure}

It should be noted that the BCS form (8.7) in which all
time-reversed pairs have equal amplitude is not necessarily the optimal one
and allows only the study of S-pair structure.
To explore the pair content of the ground state in a more general way,
we define proton pair
creation operators
\begin{equation}\hat{A}^\dagger_{J\mu}(j_aj_b)=
{1\over\sqrt{1+\delta_{ab}}}
[a^\dagger_{j_a}\times a^\dagger_{j_b}]_{J\mu}\;.
\end{equation}
These operators are boson-like in the sense that
\begin{equation}[\hat{A}^\dagger_{J\mu}
(j_aj_b),
\hat{A}_{J\mu} (j_aj_b)]=1
+{\cal O}(\hat n/2j+1)\;;
\end{equation}
i.e., they satisfy the expected commutation relations in the limit of an
empty shell.

We construct bosons $\hat{B}^\dagger_{\alpha J\mu}$ as
\begin{equation}\hat{B}^\dagger_{\alpha J\mu}=
\sum_{j_aj_b}\psi_{\alpha\lambda}(j_aj_b)
\hat{A}^\dagger_{\lambda\mu}(j_aj_b)\;,
\end{equation}
where $\alpha=1,2,\ldots$ labels the particular boson and the ``wave
function'' $\psi$ satisfies
\begin{equation}\sum_{j_aj_b} \psi^\ast_{\alpha J}(j_aj_b)
\psi_{\beta J}(j_aj_b)=\delta_{\alpha\beta}\;.
\end{equation}
(Note that $\psi$ is independent of $\mu$ by rotational invariance.)

To find $\psi$ and \hbox{$n_{\alpha J}\equiv\sum_\mu\langle
\hat{B}^\dagger_{\alpha J\mu}
\hat{B}_{\alpha J\mu}\rangle$}, the number of bosons of type
$\alpha$ and multipolarity $J$, we compute the quantity
$\sum_\mu\langle \hat{A}^\dagger_{J\mu}(j_aj_b)\hat{A}_{J\mu} (j_cj_d)\rangle$,
which can be thought of as an hermitian matrix $M^J_{\alpha\alpha'}$
in the space of orbital pairs
$(j_aj_b)$; its non-negative eigenvalues define the $n_{\alpha J}$ (we
order them so that $n_{1 J}> n_{2 J}> \ldots$), while the
normalized eigenvectors are the $\psi_{\alpha J}(j_aj_b)$.
The index $\alpha$ distinguishes the various possible bosons.
For example, in the complete $pf$-shell the square matrix $M$ has
dimension $N_J=4$ for $J=0$, $N_J=10$ for $J=1$, $N_J=13$ for $J=2,3$.

The presence of a pair condensate in a correlated ground state will be
signaled by the largest eigenvalue for a given $J$, $n_{1 J}$,
being much greater than any of the others; $\psi_{1J}$ will then be the
condensate wavefunction.
In Fig. 8.8 we show the pair matrix eigenvalues $n_{\alpha J}$
for the three isovector $J=0^+$ pairing channels as calculated for the iron
isotopes $^{54-58}$Fe. We compare the SMMC results with those of an
uncorrelated Fermi gas, where
we can compute $\langle \hat{A}^\dagger \hat{A}\rangle$ using the
factorization
\begin{equation}\langle a^\dagger_\alpha a^\dagger_\beta a_\gamma
a_\delta\rangle=
n_\beta n_\alpha (\delta_{\beta\gamma} \delta_{\alpha\delta}-
\delta_{\beta\delta}\delta_{\alpha\gamma})\;,
\end{equation}
where the $n_\beta=\langle a^\dagger_\beta a_\beta\rangle$ are the occupation
numbers. Additionally, Fig. 8.8 shows the diagonal matrix elements of the
pair matrix $M_{\alpha\alpha}$. As expected, the protons occupy mainly
$f_{7/2}$ orbitals in these nuclei. Correspondingly,
the $\langle \hat{A}^\dagger \hat{A} \rangle$
expectation value is large for this orbital and small otherwise.
For neutrons, the pair matrix is also largest for the $f_{7/2}$ orbital.
The excess neutrons in $^{56,58}$Fe occupy the $p_{3/2}$ orbital,
signalled by a strong increase of the corresponding pair matrix element
$M_{22}$ in comparison to its value for $^{54}$Fe. Upon closer inspection
we find that the proton pair matrix elements are not constant within the
isotope chain. This behavior is mainly caused by the isoscalar proton-neutron
pairing. The dominant role is played by the isoscalar $1^+$ channel, which
couples protons and neutrons in the same orbitals and in spin-orbit partners.
As a consequence we find that, for $^{54,56}$Fe, the proton pair matrix in the
$f_{5/2}$ orbital, $M_{33}$, is larger than in the $p_{3/2}$ orbital,
although the latter is favored in energy. For $^{58}$Fe, this ordering
is inverted, as
the increasing number of neutrons in the $p_{3/2}$ orbital
increases the proton pairing in that orbital.

\begin{figure}
$$\epsfxsize=4truein\epsffile{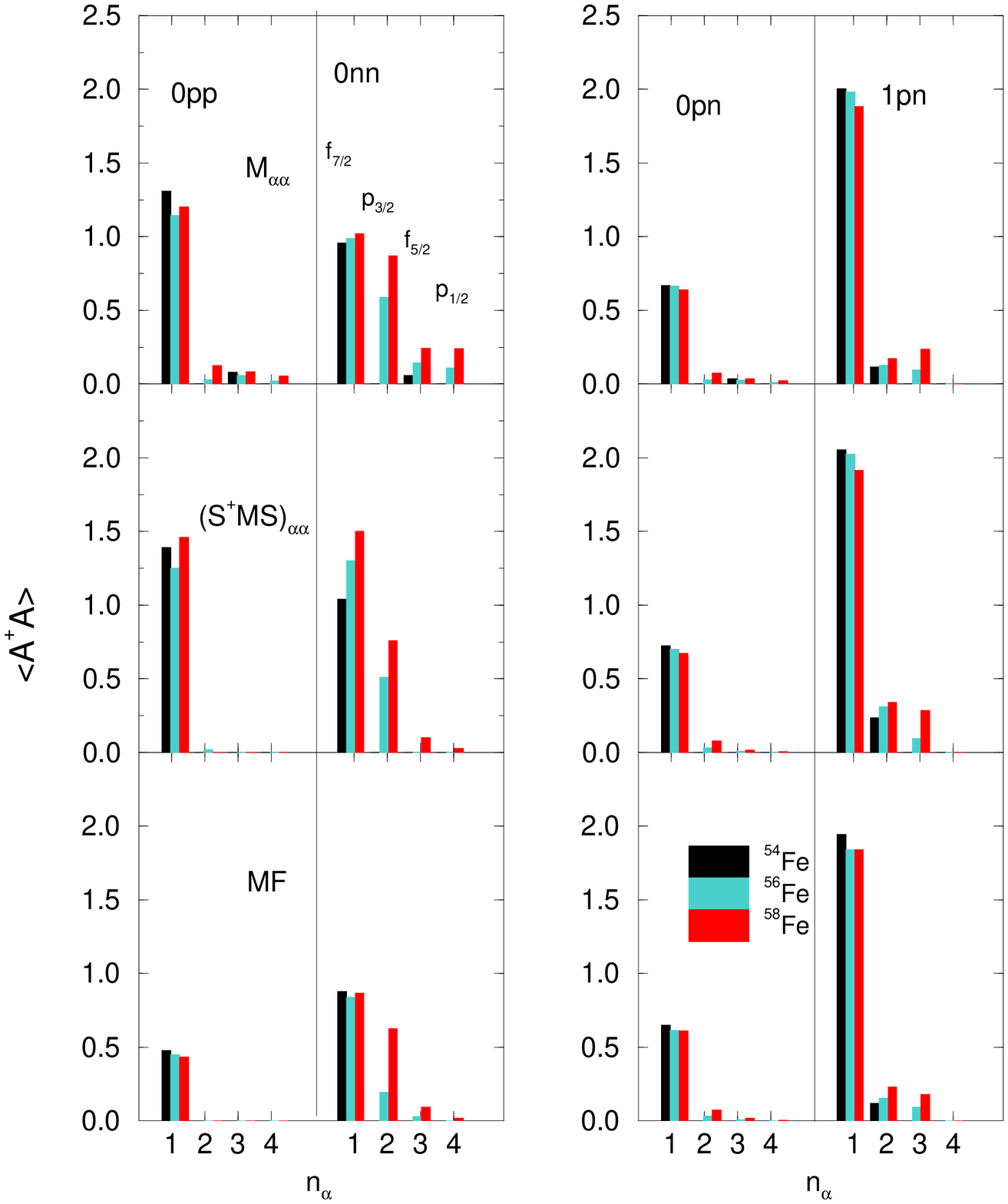}$$
{FIG~8.8. Content of isovector $0^+$ pairs and isoscalar $1^+$ pairs
in the ground states of the isotopes
$^{54-58}$Fe.
The
upper panel shows the
diagonal matrix elements of the pair matrix $M_{\alpha\alpha}$.
The index $\alpha=1,..,4$ refers to $0^+$ pairs in the
$f_{7/2}$, $p_{3/2}$, $f_{5/2}$, and $p_{1/2}$ orbitals, respectively.
For the isoscalar pairs $\alpha=1,2,3$ refers to $(f_{7/2})^2$,
$(f_{7/2}f_{5/2})$ and $(f_{5/2})^2$ pairs, respectively.
The middle panel gives the eigenvalues of the pair matrix; for the
isoscalar pairs only the 3 largest are shown. The lower panel
gives the eigenvalues of the pair matrix for the uncorrelated Fermi gas case
using Eq. (8.13) (from \protect\cite{Langanke95b}).
}
\end{figure}

After diagonalization of $M$, the $J=0$
proton pairing strength is essentially found
in one large eigenvalue. Furthermore we observe that this eigenvalue is
significantly larger than the largest eigenvalue on the mean-field level
(Fermi gas), supporting the existence of a proton pair condensate in the
ground state of these nuclei.
The situation is somewhat different for neutrons.
For $^{54}$Fe only little additional coherence is found beyond the
mean-field value, reflecting the closed-subshell neutron structure.
For the two other isotopes, the neutron pairing exhibits
two large eigenvalues. Although the larger one exceeds the mean-field
value and signals noticeable additional coherence across the subshells,
the existence of a second coherent eigenvalue shows the shortcomings of the
BCS-like pairing picture.

It has long being anticipated that $J=0^+$ proton-neutron correlations
play an important role in the ground states of $N=Z$ nuclei. To explore
these correlations we have performed SMMC calculations of the $N=Z$ nuclei
in the mass region $A=48-56$. Note that for these nuclei the pair matrix
in all three isovector $0^+$ channels
essentially exhibits only one large eigenvalue, related to the
$f_{7/2}$ orbital. We will use this eigenvalue as a convenient measure of the
pairing strength. As the even-even $N=Z$ nuclei have isospin $T=0$,
the expectation values of ${\hat A}^\dagger {\hat A}$
are identical in all three isovector
$0^+$ pairing channels. This symmetry does not hold for the odd-odd $N=Z$
nuclei in this mass range, which have $T=1$ ground states, and
$\langle \hat{A}^\dagger \hat{A} \rangle$
can be different for proton-neutron pairs
than for like-nucleons pair (the expectation values for proton pairs
and neutron
pairs are identical). We find the proton-neutron pairing strength
significantly larger for odd-odd $N=Z$ nuclei than in even-even nuclei,
while the $0^+$ proton and neutron pairing shows the opposite behavior,
in both cases leading to an odd-even staggering, as displayed in
Fig. 8.9. This staggering is caused by a constructive interference
of the isotensor and isoscalar parts of $\hat{A}^\dagger \hat{A}$
in the odd-odd
$N=Z$ nuclei, while they interfere destructively in the even-even nuclei.
The isoscalar part is related to
the pairing energy, and is found to be about constant for the nuclei
studied here.

\begin{figure}
$$\epsfxsize=4truein\epsffile{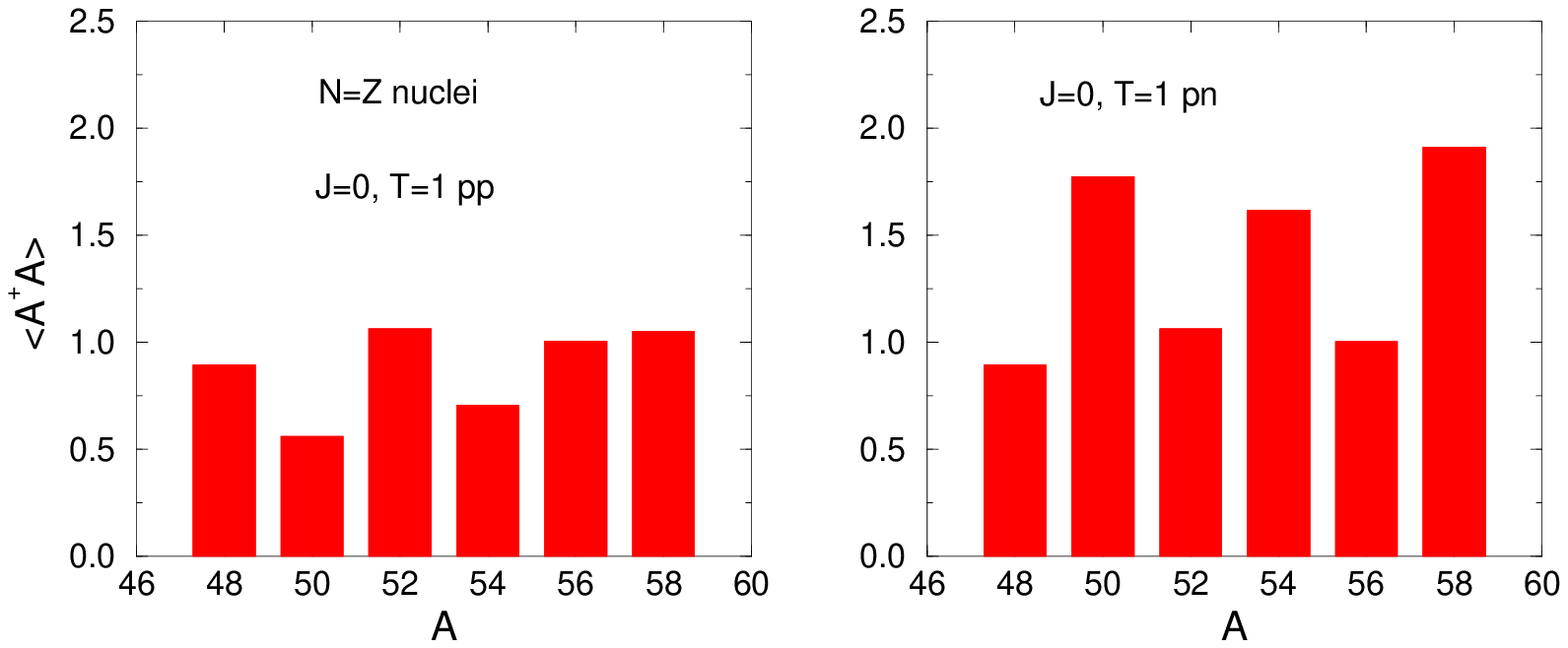}$$
{FIG~8.9. Largest eigenvalue for the various isovector $0^+$ pairs
in the $N=Z$ nuclei in the mass region $A=48-56$.
}
\end{figure}

\noindent
\subsection{Thermal properties of $pf$-shell nuclei}

The properties of nuclei at finite temperatures are of considerable
experimental \cite{Suraud,Snover}
and theoretical interest \cite{Alhassid1,Egido}.
They are clearly quite important in various astrophysical
scenarios. For example, electron capture on nuclei plays an essential
role in the early presupernova collapse \cite{Bethe90}. In that context, it is
important to know the temperature dependence of the Gamow-Teller strength.

Theoretical studies of nuclei at finite temperature have been based mainly
on mean-field approaches, and thus only consider the temperature dependence
of the most probable configuration in a given system. These approaches
have been criticized due to their neglect of quantum and statistical
fluctuations \cite{Dukelsky}. The SMMC method does not suffer this defect and
allows the consideration of model spaces large enough to account
for the relevant nucleon-nucleon correlations at low and moderate
temperatures.

We have performed SMMC calculations of the thermal properties of several
even-even nuclei in the mass region $A=50-60$ \cite{Dean95,Langanke95b}.
As a typical example, we discuss
in the following our SMMC results for the nucleus $^{54}$Fe, which
is very abundant in the presupernova core of a massive star.

Our calculations include the complete set of $1p_{3/2,1/2}0f_{7/2,5/2}$
states interacting through the realistic Brown-Richter Hamiltonian
\cite{Richter}.
(SMMC calculations using the modified KB3 interaction give essentially
the same results.) Some
$5\times10^9$ configurations of the 8 valence neutrons and 6 valence protons
moving in these 20 orbitals are involved in the canonical ensemble. The
results presented below have been obtained with a
time step of $\Delta\beta=1/32~{\rm MeV}^{-1}$ using 5000--9000 independent
Monte Carlo samples at seven values of the coupling constant $g$ spaced
between $-1$ and 0 and the value $\chi=4$.

The calculated temperature dependence of various observables is shown in
Fig.~8.10. In accord with general thermodynamic principles, the internal energy
$U$ steadily increases with increasing temperature \cite{Dean95}. It shows an
inflection point around $T \approx 1.1$~MeV, leading to a peak in the heat
capacity, $C\equiv dU/dT$, whose physical origin we will discuss below. The
decrease in $C$ for $T \gtrsim 1.4$~MeV is due to our finite model space (the
Schottky effect \cite{Schottky});
we estimate that limitation of the model space to only
the $pf$-shell renders our calculations of ${}^{54}$Fe quantitatively
unreliable for temperatures above this value (internal energies $U\gtrsim
15$~MeV). The same behavior is apparent in the level density parameter,
$a\equiv C/2T$. The empirical value for $a$ is $A/8~{\rm MeV} =6.8~{\rm
MeV}^{-1}$ which is in good agreement with our results for $T \approx
1.1$--1.5~MeV.

We also show in Fig.~8.10 the expectation values of the BCS-like
proton-proton and neutron-neutron pairing fields,
$\langle \hat \Delta^\dagger
\hat \Delta\rangle$.
At low temperatures, the pairing fields are significantly
larger than those calculated for a non-interacting Fermi gas, indicating a
strong coherence in the ground state. With increasing temperature, the
pairing fields decrease, and both approach the Fermi gas values for $T\approx
1.5$~MeV and follow it closely for even higher temperatures. Associated with
the breaking of pairs is a dramatic increase in the moment of inertia,
$I\equiv \langle J^2\rangle/3T$,
for $T=1.0$--1.5~MeV; this is analogous to the rapid increase in magnetic
susceptibility in a superconductor. At temperatures above 1.5~MeV, $I$ is in
agreement with the rigid rotor value, $10.7\hbar^2$/MeV; at even higher
temperatures it decreases linearly due to our finite model space.

In Fig.~8.11, we show various static observables. The $M1$ strength unquenches
rapidly with heating near the transition temperature. However, for
$T=1.3$--2~MeV $B(M1)$ remains significantly lower than the single-particle
estimate ($41~\mu_N^2$), suggesting a persistent quenching at temperatures
above the like-nucleon depairing. This finding is supported by the
near-constancy of the Gamow-Teller $\beta^+$ strength, $B({\rm GT}_+)$, for
temperatures up to 2~MeV. As the results of Ref.~\cite{Langanke95b}
demonstrate that
neutron-proton correlations are responsible for much of the GT quenching in
iron nuclei at zero temperature, we interpret the present results as evidence
that isovector proton-neutron correlations persist to higher temperatures
(see below). We
have verified that, in our restricted model space, both the ${\rm GT}_+$ and
$M1$ strengths unquench at temperatures above 2~MeV and that, in the
high-temperature limit, they both approach the appropriate Fermi gas values.
We note that it is often assumed in astrophysical calculations that the GT
strength is independent of temperature \cite{Aufderheide};
our calculations demonstrate
that this is true for the relevant temperature regime ($T<2$~MeV). We also
note that a detailed examination of the occupation numbers of the various
orbitals show no unusual variation as the pairing vanishes.

Although the results discussed above are typical for even-even nuclei
in this mass region
(including the $N=Z$ nucleus $^{52}$Fe), they are not for odd-odd $N=Z$
nuclei. This is illustrated in Fig. 8.12 which shows the thermal
behavior of several observables for $^{50}$Mn ($N=Z=25$), calculated
in a SMMC study within the complete $pf$-shell using the KB3 interaction.
In contrast to even-even nuclei, the total Gamow-Teller strength
is not constant at low temperatures, but increases by about $50\%$
between $T=0.4$ MeV and 1 MeV. The $B(M1)$ strength decreases
significantly in the same temperature interval, while for even-even
nuclei, it increases steadily. A closer inspection of the isovector
$J=0$ and isoscalar $J=1$ pairing correlations holds the key to the
understanding of these differences. The $J=0$ isovector correlations
are studied using the BCS pair operators (8.6), with a similar definition
for proton-neutron pairing. For the isoscalar $J=1$ correlations we have
interpreted the trace of the pair matrix $M^{J=1}$ as an overall
measure for the pairing strength,
\begin{equation}
P_{sm}^J = \sum_{\beta} \lambda_{\beta}^J = \sum_\alpha M^J_{\alpha\alpha}.
\end{equation}

Note that at the level of the non-interacting Fermi gas, proton-proton,
neutron-neutron and proton-neutron $J=0$ correlations are identical
for $N=Z$ nuclei. However, the residual interaction breaks the symmetry
between like-pair correlations and proton-neutron correlations
in odd-odd $N=Z$ nuclei. As is obvious from Fig. 8.12, at low temperatures
proton-neutron pairing dominates in $^{50}$Mn, while pairing among
like nucleons shows only a small excess over the Fermi gas values, in
strong contrast to even-even nuclei.

A striking feature of Fig. 8.12 is that the isovector
proton-neutron correlations decrease strongly with temperature
and have essentially vanished
at $T=1$ MeV, while the isoscalar pairing strength
remains about constant in this temperature region
(as it does in even-even nuclei) and greatly exceeds the Fermi gas values.
We also note that the pairing between like
nucleons is roughly constant at $T<1$ MeV.
The change of importance between isovector and isoscalar proton-neutron
correlations with temperature is nicely reflected in the isospin expectation
value, which decreases from $<\hat T^2>=2$ at temperatures around 0.5 MeV,
corresponding to the dominance
of isovector correlations, to $<\hat T^2>=0$ at temperature $T=1$ MeV, when
isoscalar proton-neutron correlations are most important.
The low-temperature behavior of the Gamow-Teller and $B(M1)$ strength
is related to the fading of the isoscalar proton-neutron correlations.
For example, the Gamow-Teller strength increases as transitions between
the same orbital (mainly $f_{7/2}$) are less quenched by weaker isovector
proton-neutron correlations.

\begin{figure}
$$\epsfxsize=4truein\epsffile{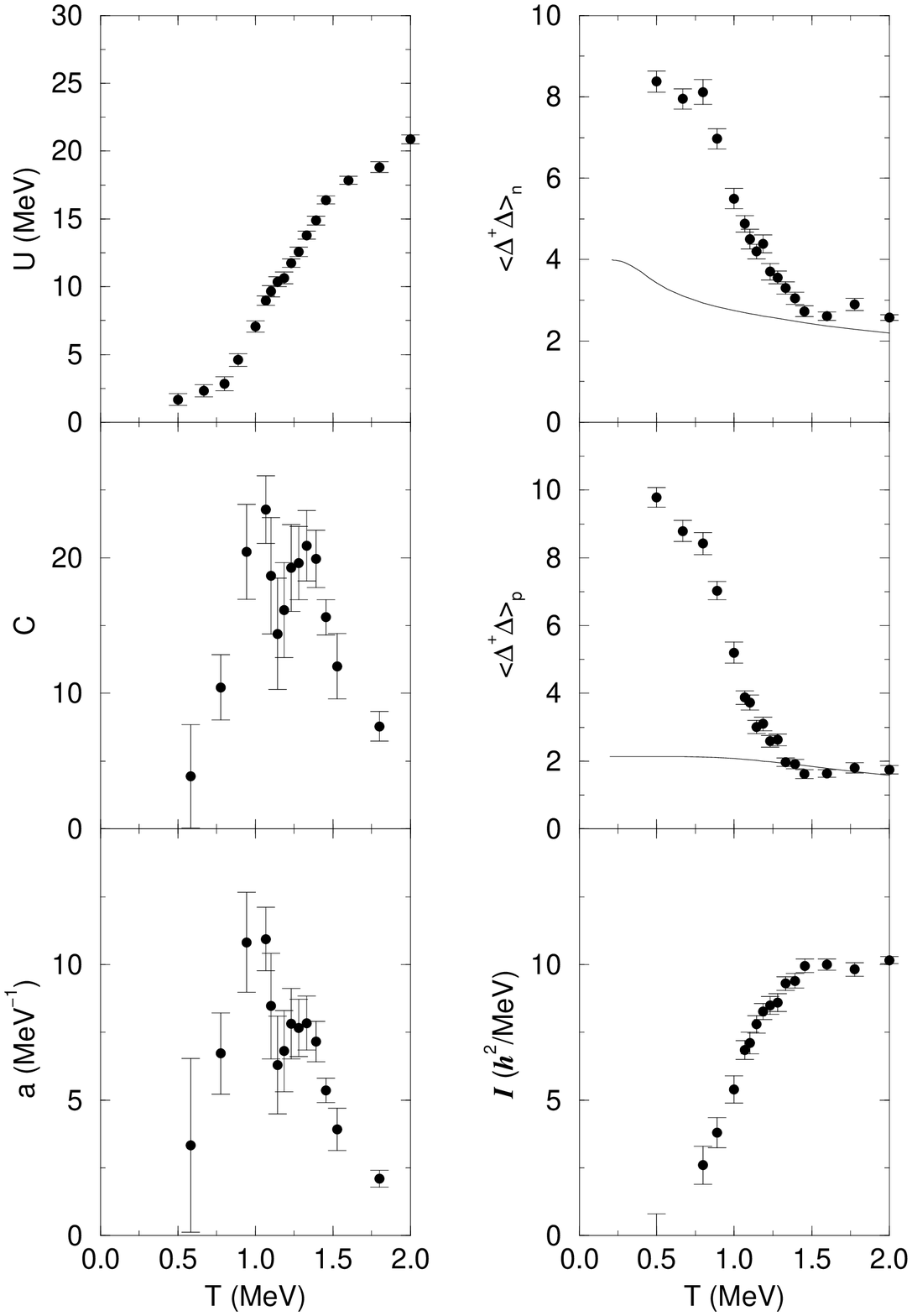}$$
{FIG~8.10.
Temperature dependence of various observables in ${}^{54}$Fe. Monte
Carlo points with statistical errors are shown at each temperature $T$. In
the left-hand column, the internal energy, $U$,
is calculated as $\langle {\hat H}
\rangle -E_0$, where ${\hat H}$ is the many-body Hamiltonian and $E_0$ the
ground
state energy. The heat capacity $C$ is calculated by a finite-difference
approximation to $dU/dT$, after $U(T)$ has been subjected to a three-point
smoothing, and the level density parameter is $a\equiv C/2T$.
In the right-hand column, we show the expectation values of the
squares of the proton and neutron BCS pairing fields.
For comparison, the pairing fields
calculated in an uncorrelated Fermi gas are shown by the solid curve. The
moment of inertia is obtained from the expectation values of the square of
the total angular momentum by $I=\beta \langle {\hat J}^2 \rangle/3$
(from \protect\cite{Dean95}).}
\end{figure}

\begin{figure}
$$\epsfxsize=4truein\epsffile{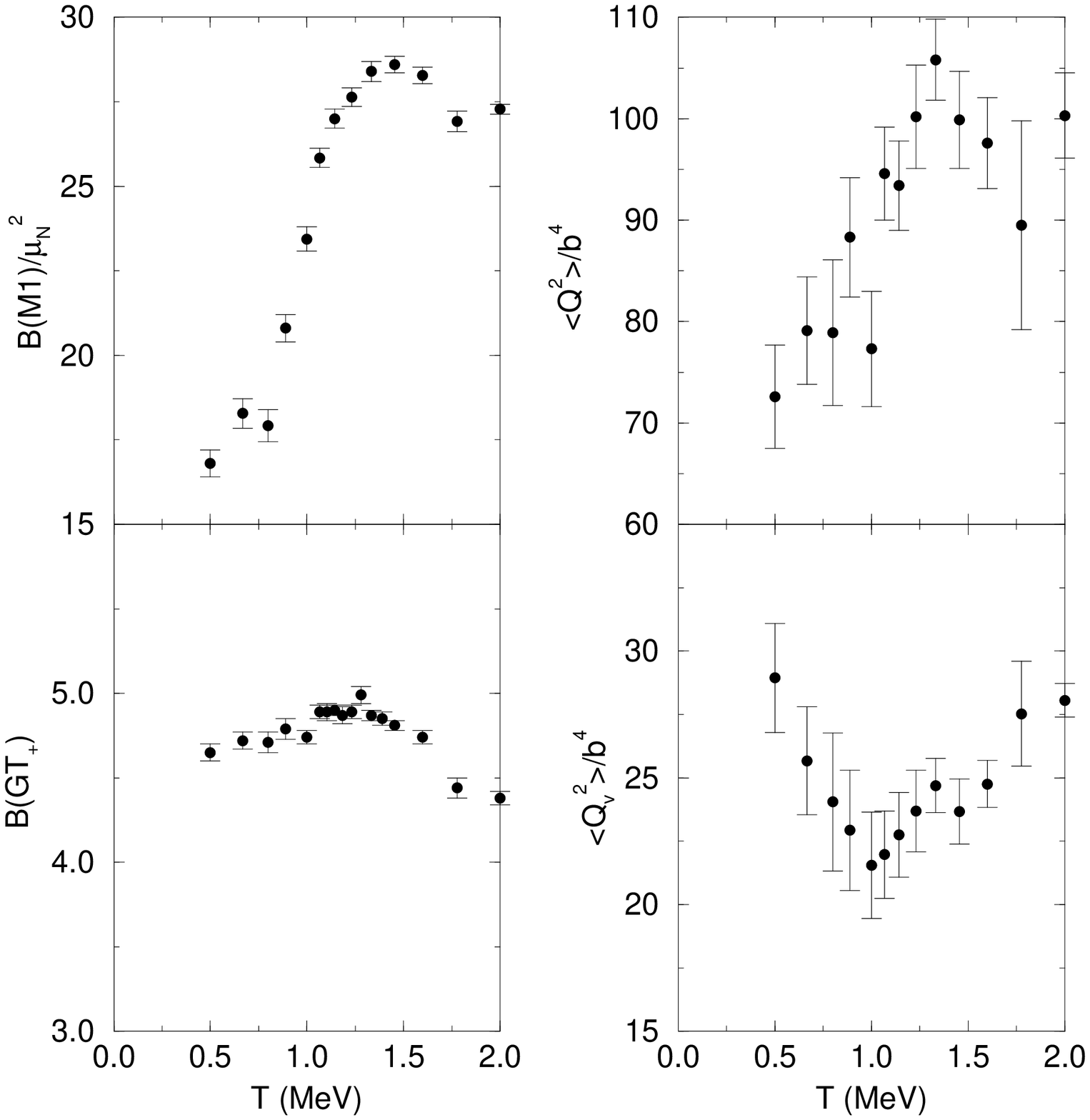}$$
{FIG~8.11.
The upper left panel shows, for ${}^{54}$Fe,
the total magnetic dipole strength,
$B(M1)$ in units of nuclear magnetons; it is calculated using free-nucleon
$g$-factors. The lower left panel shows the GT $\beta_+$ strength, while the
upper and lower right panels show the isoscalar (${\hat Q}= {\hat Q}_p
+ {\hat Q}_n$) and
isovector (${\hat Q}_{\rm v} ={\hat Q}_p - {\hat Q}_n$)
quadrupole strengths. In these latter
observables, the quadrupole operators are $r^2Y_2$, and results are given in
terms of the oscillator length, $b=1.96$~fm
(from \protect\cite{Dean95}).}
\end{figure}

\begin{figure}
$$\epsfxsize=4truein\epsffile{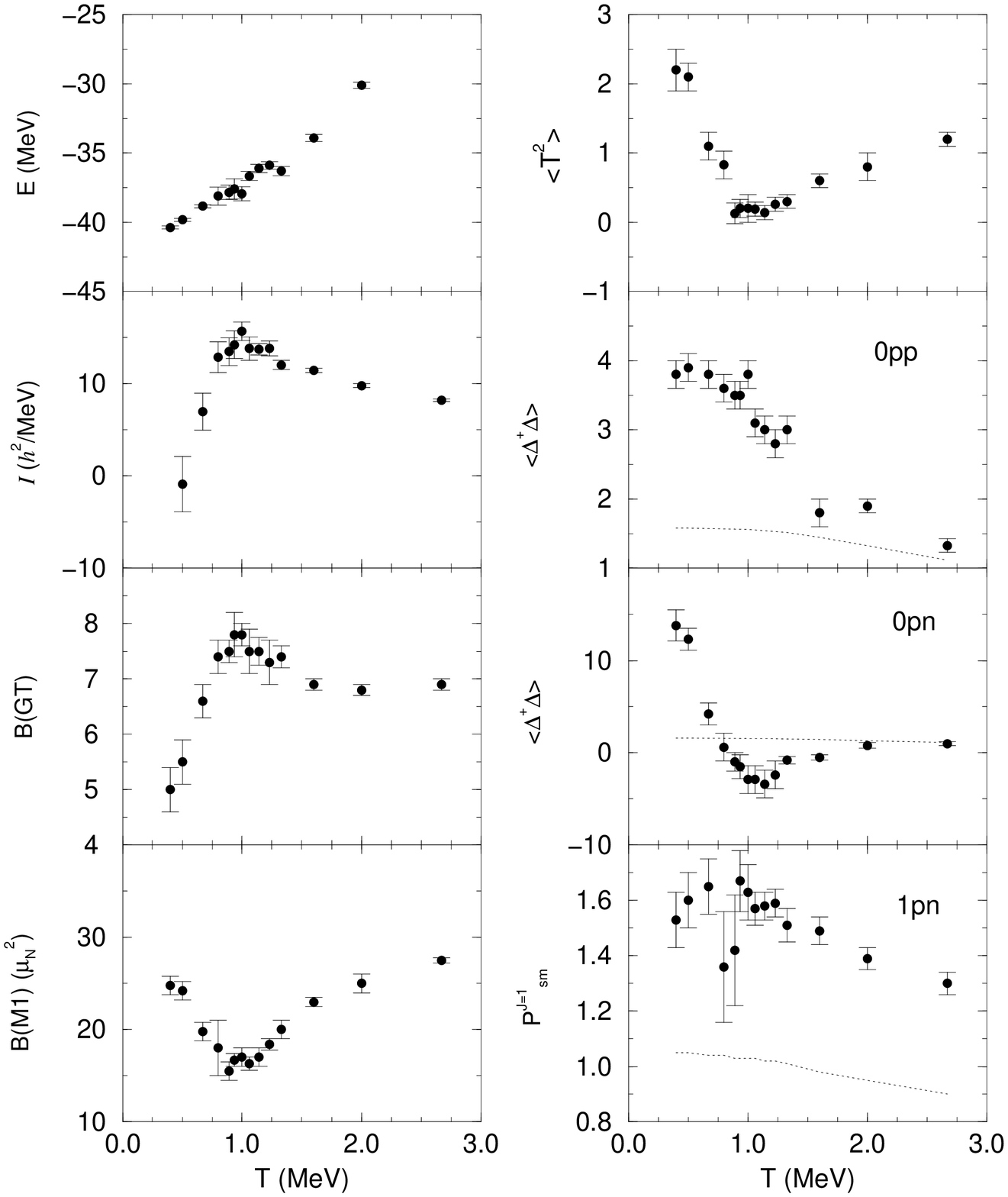}$$
{FIG~8.12
Temperature dependence of various observables in $^{50}$Mn.
The left panels show (from top to bottom) the total energy,
the moment of inertia, the $B(M1)$ strength, and the Gamow-Teller strength,
while the right panels exhibit
the expectation values of the isospin operator
$\langle \hat T^2 \rangle$, the isovector $J=0$ proton-proton and
proton-neutron BCS pairing fields, and the isoscalar $J=1$ pairing
strength, as defined in the text. For comparison,
the solid lines indicate the
Fermi gas values of the BCS pairing fields and $J=1$ pairing strength.
}
\end{figure}

\noindent
\subsection{Pair correlations and thermal response}

All SMMC calculations of even-even nuclei
in the mass region $A=50-60$ show
that the BCS-like pairs
break at temperatures around 1 MeV.
Three observables exhibit a particularly interesting behavior
at this phase transition: a) the moment of inertia rises sharply; b) the
M1 strength shows a sharp rise, but unquenches only partially; and c)
the Gamow-Teller strength remains roughly constant (and strongly quenched).
Note that
the $B(M1)$ and $B(GT_+)$ strengths unquench at temperatures larger than
$\approx 2.6$ MeV and
in the high-temperature limit
approach the appropriate values for our adopted model space.

Ref. \cite{Langanke95b}
has studied the pair correlations in the four nuclei
$^{54-58}$Fe and $^{56}$Cr for the various
isovector and isoscalar pairs up to $J=4$,
tentatively interpreting
the sum
of the eigenvalues of the matrix $M^J$ (8.14)
as an overall measure for the pairing strength.
Note that
the pairing strength, as defined in (8.12),
is non-zero at the
mean-field level. The physically relevant pair correlations
$P_{\rm corr}^J$ are then defined as the difference of the SMMC
and mean-field pairing strengths.

Detailed calculations of the pair correlations have been
performed for selected temperatures in the region between $T=0.5$ MeV and 8
MeV.
Fig. 8.13 shows the
temperature dependence of those pair correlations
that
play an important role in
understanding the thermal behavior of the moment of inertia and the
total $M1$ and Gamow-Teller strengths.

\begin{figure}
$$\epsfxsize=4truein\epsffile{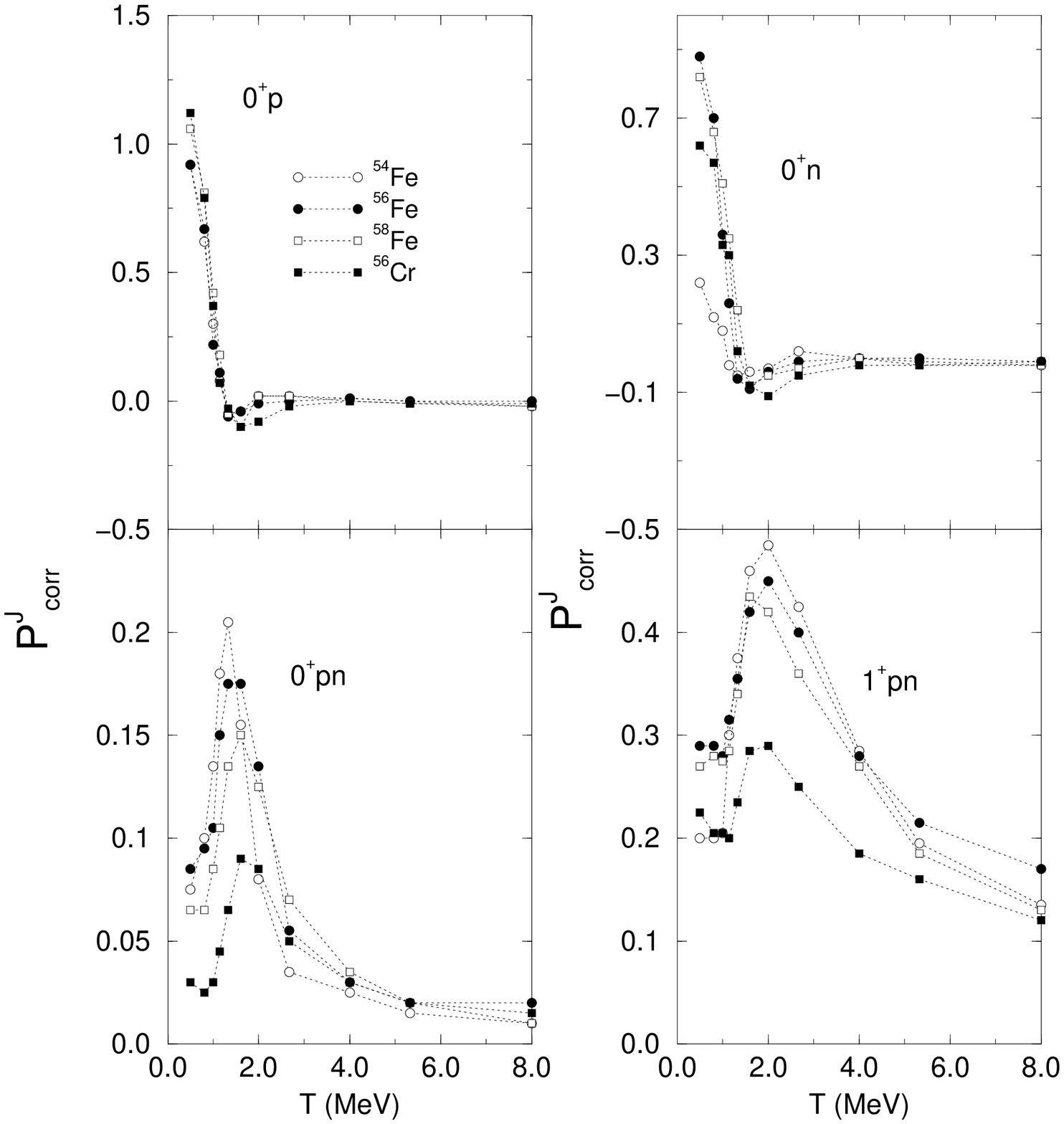}$$
{FIG~8.13
Pair correlations, as defined in Eq. (8.13),
for isovector $0^+$ and isoscalar $1^+$ pairs for $^{54-58}$Fe
and $^{56}$Cr, as functions of temperature (from \protect\cite{Langanke95b}).
}
\end{figure}

\begin{figure}
$$\epsfxsize=4truein\epsffile{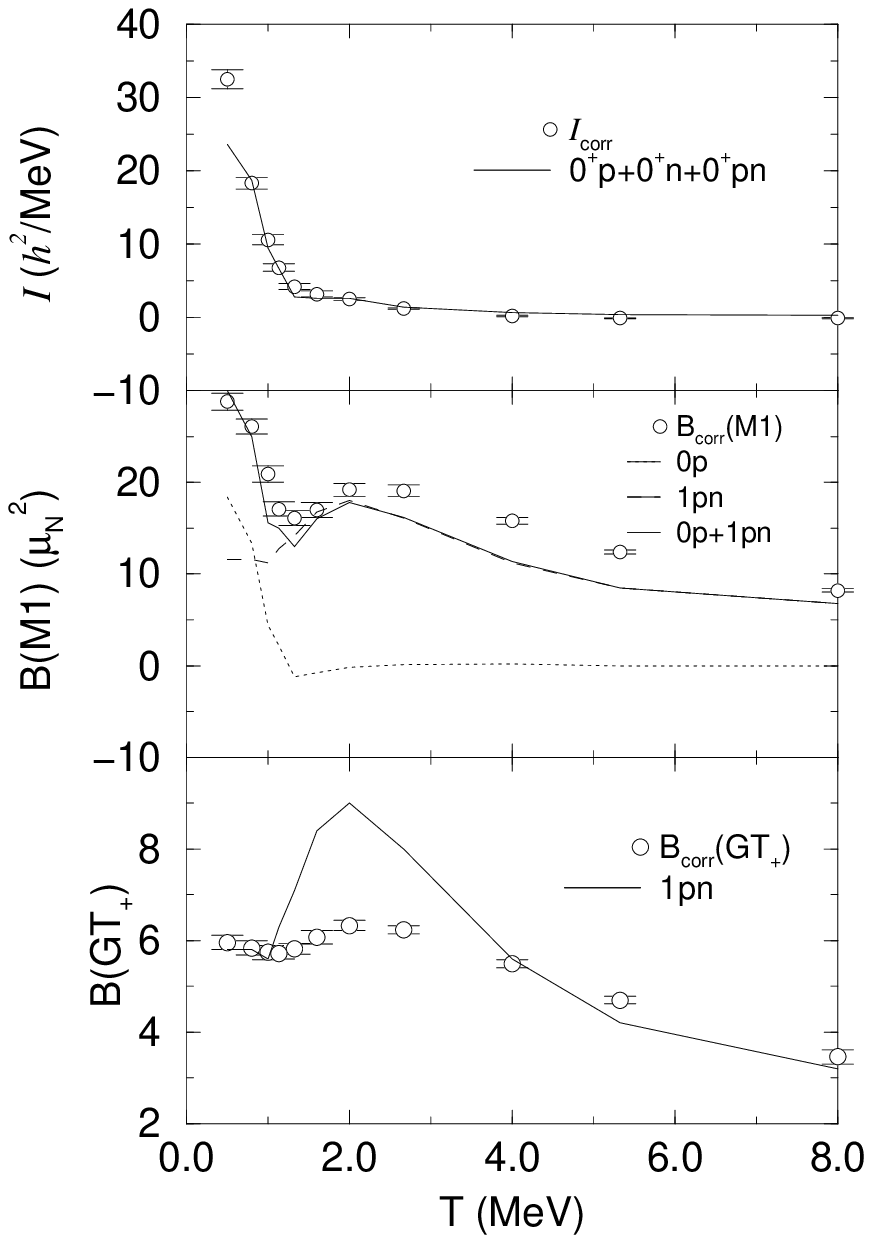}$$
{FIG~8.14.
The temperature dependence of the quenching of the moment of inertia and of
the total $B(M1)$ and $B(GT_+)$ strengths (see text)
is compared to the one
of selected pairing correlations (in arbitrary units). The results
are shown for $^{56}$Fe (from \protect\cite{Langanke95b}).
}
\end{figure}

The most interesting behavior is found in the $J=0$ proton and neutron pairs.
There is a large excess of this pairing at low
temperatures, reflecting the ground state coherence of even-even nuclei.
With increasing temperature, this excess diminishes and vanishes at
around $T=1.2$ MeV.
We observe further from Fig. 8.13 that the temperature dependence
of the $J=0$ proton-pair correlations are remarkably independent
of the nucleus, while the neutron pair correlations show interesting
differences. At first, the correlation excess is smaller in the semimagic
nucleus $^{54}$Fe than in the others. When comparing the iron isotopes, the
vanishing of the neutron $J=0$ correlations occurs at higher temperatures
with increasing neutron number.

The vanishing of the $J=0$ proton and neutron pair correlations is
accompanied by an  increase in the correlations of the other pairs.
For example, the isovector $J=0$ proton-neutron correlations
increase by about a factor 3 after the $J=0$
proton and neutron pairs have vanished. The correlation peak is reached
at higher temperatures with increasing neutron number,
while the
peak height decreases with neutron excess.

The isoscalar proton-neutron $J=1$ pairs show an interesting temperature
dependence. At low temperatures, when the nucleus is still dominated
by the $J=0$ proton and neutron pairs, the isoscalar proton-neutron
correlations show a
noticeable excess but, more interestingly, they are roughly constant and
do not directly reflect  the vanishing of the $J=0$ proton and neutron pairs.
However, at $T>1$ MeV, where the $J=0$ proton- and neutron-pairs have broken,
the isoscalar
$J=1$ pair correlations significantly increase and have their maximum
at around 2 MeV, with peak values of about twice the correlation excess
in the ground state.
In contrast to the isovector $J=0$ proton-neutron pairs, the correlation peaks
occur at lower temperatures with increasing neutron excess. We also
observe that these correlations fade rather slowly with
increasing temperature.

We note that the SMMC results for
the moment of inertia and the $B(M1)$ and $B(GT_+)$ strengths
are significantly smaller than
the mean-field values. The quenching of these observables,
defined as the difference of the mean-field and SMMC results,
can be traced back to
nucleon correlations beyond the mean-field. This is demonstrated in
Fig. 8.14,
exemplified for $^{56}$Fe.
{}From its definition
($I \sim \langle {\hat J}^2 \rangle = \langle
({\hat J}_n+{\hat J}_p)^2 \rangle$, where
${\hat J}_p, {\hat J}_n$ are the proton and neutron angular momenta,
respectively),
one expects that the moment of inertia is sensitive to the proton,
neutron and proton-neutron correlations. In fact,
the sum of the isovector $J=0$ pairing correlations shows
nearly the same temperature behavior as the moment of inertia quenching.
The drop in the quenching at $T\approx1$ MeV
clearly signals the
pairing phase transition observed in the $J=0$ proton and neutron
correlations (see Fig. 8.13).
The temperature
dependence of the $M1$ quenching is well described by that
of the sum of the $J=0$ proton and isoscalar $J=1$ proton-neutron
correlations.
These two correlations
reflect the two components in the M1 strength. The orbital part is sensitive
to the $J=0$ proton pairing correlations (the gyromagnetic moment $g_l$
is zero for neutrons), while the quenching of the spin component
is dominated by
isoscalar proton-neutron correlations.

The quenching of the $B(GT_+)$ strength is roughly constant for
$T<1$ MeV. It then increases slightly, before it slowly
dies out following a maximum at $T \approx 2.5$ MeV.
While the temperature dependence of the
Gamow-Teller strength is driven by the isoscalar proton-neutron
correlations,
the overall measure of the pairing
correlations, as applied here, is too simple for a quantitative
reproduction.

\noindent
\subsection{Gamow-Teller quenching in the $^{100}$Sn region}

The double-magic nucleus
$^{100}$Sn has recently been observed for the first time at GSI and
Ganil \cite{GSI,Ganil}. Detailed studies of this nucleus promise further
insight into
the shell structure in proton-rich nuclei and in particular into
nature of Gamow-Teller (GT) quenching.
The reasoning is as
follows: the shell gap between the
$g_{9/2}$-orbital and the
other orbitals in the $0g$-$1d$-$2s$ shell
is expected to support a closed-shell structure for both
neutrons and protons and
suppress correlations involving the higher orbitals. The valence
protons, occupying
$g_{9/2}$-orbitals,
will then be transformed with relatively less hindrance into
$g_{7/2}$-neutrons by the $\sigma \tau_+$ operator. Thus the quenching
of the GT$_+$-strength due to configuration mixing is expected to be
small.

Despite this reasoning, the observed quenching of the GT strength
in the $N=50$ isotones close to $^{100}$Sn like $^{94}$Ru, $^{96}$Pd and
$^{98}$Cd
is surprisingly large \cite{Cadmium}
Strongly truncated shell model calculations
are unable to reproduce the strong quenching,
as the configuration mixing in these restricted model spaces
was found to be insufficient \cite{Manakos,Brown94}.
The agreement between theory and experiment has been improved by
renormalization of the GT operator to consider (first-order) corrections to the
wave functions \cite{Towner,Johnstone,Manakos,Brown94}.
As shell model studies of lighter nuclei clearly stressed the need
of a complete $0 \hbar \omega$ shell model basis, it appears unlikely
that first-order corrections are sufficient to recover the full quenching
of the GT strength due to configuration mixing.

Shell model Monte Carlo
calculations have been performed for
$^{94}$Ru, $^{96}$Pd, $^{96,98}$Cd, and $^{100}$Sn in the complete
$0g$-$1d$-$2s$
major oscillator shell using a G-matrix two-body interaction built
from the Paris nucleon-nucleon interaction \cite{Dean100}.
The Coulomb-corrected SMMC mass excesses agree well with  the latest
mass compilation of Myers and Swiatecki \cite{Myers}.
Table 3 gives the SMMC
results for the Gamow-Teller strength, renormalizing the GT operator
by $1/1.26$.
The SMMC calculation
predicts a noticeable quenching of the GT strength in the $N=50$
isotones, in accord with experiment. A detailed comparison with
experiment is hampered by the fact that the measurement of the $\beta^+$-decay
can only determine the GT strength below a certain energy threshold,
so that the experiment actually provides only upper bounds for the
quenching.
In the case of $^{96}$Pd and $^{98}$Cd it is expected that the measurements
miss less than about $10\%$ of the total GT strength \cite{Manakos,Brown94},
while for $^{94}$Ru a significant amount of GT strength
is predicted at energies
that are not accessible in $\beta^+$ experiments \cite{Manakos,Brown94}.
It thus seems
that SMMC calculations are in excellent agreement with the observed GT
quenching for the nuclei $^{96}$Pd and $^{98}$Cd. For $^{94}$Ru the
SMMC result appears to be in qualitative agreement with observation.
However, a quantitative comparison requires a detailed knowledge
of the GT strength distribution in the daughter nucleus $^{94}$Tc,
which is outside the scope of the calculation of Ref. \cite{Dean100}.

The SMMC calculations \cite{Dean100} also predict a noticeable quenching of
the GT strength in the double-magic nucleus $^{100}$Sn.
Here, the total B(GT$_+$)=6.5$\pm$0.1 should be
compared to the single particle estimate
of 17.78, corresponding to a quenching factor $q=2.74\pm 0.1$.
This quenching factor is larger than predicted by recent truncated
shell model calculations \cite{Brown94}. However,
the ratio of quenching factors for $^{98}$Cd and $^{100}$Sn is found to
be less interaction dependent than the actual values \cite{Brown94}.
The SMMC calculation yields q($^{98}$Cd)/q($^{100}$Sn)=$1.18\pm0.07$,
within the range 1.23 -- 1.3 given
in \cite{Brown94}.

\begin{table}
\begin{center}
\caption{Comparison of the $B(GT)$ strengths and quenching factors $q$, as
calculated in our SMMC approach, with the data.
The single particle estimates (SPE) for the $B(GT)$ strength are also given
(from \protect\cite{Dean100}).}
\protect\vspace{1ex}
\begin{tabular}{|c|c|c|c|c|}
\hline
Nucleus & B(GT$_+$) (SPE) & B(GT$_+$) (SMMC) &
 $q$ (SMMC) &  $q$ (expt) \cite{Cadmium}
\\ \hline
$^{94}$Ru  & 7.11 & 1.48$\pm$0.1 & 4.96$\pm$0.3 & 7.2$\pm$0.6 \\
$^{96}$Pd  & 10.67 & 2.68$\pm$0.1&3.98$\pm$0.2 &4.6$^{+1.7}_{-0.8}$\\
$^{96}$Cd  & 16.18 & 5.40$\pm$0.1&2.98$\pm$0.1 & ?  \\
$^{98}$Cd  & 14.22 & 4.42$\pm$0.1&3.22$\pm$0.1 & 4.1$^{+1}_{-0.8}$\\
$^{100}$Sn & 17.78 & 6.50$\pm$0.1&2.74$\pm$0.1 & ?  \\ \hline
\end{tabular}
\end{center}
\end{table}

\noindent
\subsection{Electron capture and presupernova collapse}

The core of a massive star at the end of hydrostatic burning
is stabilized by electron degeneracy pressure as long as its mass does not
exceed the appropriate Chandrasekhar mass $M_{CH}$. If the core
mass exceeds $M_{CH}$, electrons are captured by nuclei \cite{Bethe90} to avoid
a violation of the Pauli principle:
\begin{equation}
e^- \; + \: (Z,A) \rightarrow (Z-1,A) \: + \; \nu_e
\end{equation}
The neutrinos can still escape from the core, carrying away energy.
This is accompanied by a loss of pressure
and the collapse is accelerated.

For many of the nuclei that determine the electron capture rate in
the early stage of the presupernova \cite{Aufderheide}, Gamow-Teller (GT)
transitions contribute significantly to the electron capture rate.
Due to insufficient experimental information, the GT$_+$ transition rates
have so far been treated only qualitatively in presupernova collapse
simulations, assuming the GT$_+$ strength to reside in a single resonance,
whose energy relative to the daughter ground state has been parametrized
phenomenologically \cite{FFN}; the total GT$_+$ strength has been taken
from the single particle model. Recent $(n,p)$ experiments
\cite{gtdata1}-\cite{gtdata5}, however,
show that the GT$_+$ strength is fragmented over many states,
while the total strength is significantly quenched compared to the
single particle model (see section 8.1). (A recent update of the GT$_+$ rates
for use in supernova simulations assumed a constant quenching factor of 2
\cite{Aufderheide}).

In a series of truncated shell model calculations, Aufderheide and
collaborators have demonstrated that a strong phase space
dependence makes
the Gamow-Teller contributions to the presupernova electron capture
rates more sensitive to the strength {\it distribution} in the daughter
nucleus than to the total strength \cite{Aufder1}.
In this work it also became apparent that complete $0 \hbar \omega$
studies of the GT$_+$ strength distribution are desirable.
Such studies are now
possible using the SMMC approach.

To determine the GT$_+$ strength distribution
we have calculated the
response function of the ${\bf \sigma} \tau_+$ operator, $R_{GT} (\tau)$,
 as defined
in Eq. (3.6). As the strength function $S_{GT}(E)$
is the inverse Laplace transform of $R_{GT} (\tau)$, we have
used the Maximum Entropy technique, described in section 6.4,
to extract $S_{GT}(E)$. For the default model $m(E)$ in (6.21)
we adopted a Gaussian whose centroid is
given by the slope of ${\rm ln}
R_{GT} (\tau)$ at small $\tau$ and whose width is 2 MeV.

As first examples we have studied several nuclei
($^{51}$V, $^{54,56}$Fe, $^{58,60,62}$Ni, and $^{59}$Co),
for which the Gamow-Teller strength distribution
in the daughter nucleus is known from $(n,p)$ experiments
\cite{gtdata1}-\cite{gtdata5}.
Note that the electron capture by these
nuclei, however, plays only a minor role in the presupernova collapse.
As SMMC calculates the strength function within
the parent nucleus, the results have been shifted using the experimental
$Q$-value. The Coulomb correction has been performed using Eq. (8.1).
For all nuclei, the SMMC approach calculates
the centroid and width of the strength distribution
in good agreement with data. (Following \cite{Poves}
the calculated strength distributions
have been folded with Gaussians of width 1.77 MeV to account for
the experimental resolution.)
Fig. 8.15 shows the response function $R_{GT}({\tau})$
for $^{54}$Fe and $^{59}$Co and compares
the extracted GT$_+$ strength distribution with data. (As discussed in
sections 8.1 and 8.5, the Gamow-Teller operator has been renormalized
by the factor 0.77).
The centroid of the GT$_+$ strength distributions
is found to be nearly independent of temperature (calculations for odd nuclei
could be performed only in the temperature range $T=0.8-1.2$ MeV),
while its width increases with temperature.

Following the formalism
described in Refs. \cite{FFN,Aufderheide},
the Gamow-Teller contributions to the electron capture
rates under typical presupernova conditions
have been calculated
assuming
that the electrons obey a Fermi-Dirac distribution with
a  chemical potential adopted
from the stellar trajectory
at the electron-to-nucleon ratio corresponding to the respective
nucleus \cite{Aufderheide}.
The calculation have been
performed for both the SMMC
and experimental GT$_+$ strength distributions \cite{gtdata1}-\cite{gtdata5}.
The electron capture rates so obtained agree within a factor of two
for temperatures $T=(3-5) \times 10^9$ Kelvin, which is the
relevant temperature regime in the presupernova collapse
\cite{Aufderheide} (Fig. 8.16).
Note that this
level of agreement is as good as that obtained by Aufderheide {\it et al.}
\cite{Aufder1}.
These authors calculate
the  GT$_+$ strength distribution within
strongly truncated shell model studies
by adjusting the single particle energies
to fit the observed strength distribution and then normalize the total
calculated strength to the measured B(GT$_+$) value \cite{Aufder1}.
As this approach obviously requires a knowledge of the experimental
strength distribution, the SMMC calculations in the complete $pf$ shell
are capable of {\it predicting} the desired strength distribution with
a similar accuracy. Thus, it is for the first time possible
to calculate with a reasonable accuracy the electron capture rate for
those nuclei like $^{55}$Co or $^{56}$Ni which dominate the electron capture
process in the early presupernova collapse \cite{Aufderheide}.
The SMMC calculation places the GT$_+$ strength
for $^{55}$Co at a higher energy in the daughter nucleus
than the phenemenological parametrization of Fuller {\it et al.} \cite{FFN}.
Correspondingly, the electron capture rate on this nucleus,
which is estimated to be dominated by the GT$_+$ contribution, is
smaller than previously assumed. Studies of the effect
of this smaller rate
on the dynamics of the early presupernova collapse must await the
completion of SMMC calculations
for other important $pf$-shell nuclei now
in progress.

\begin{figure}
$$\epsfxsize=5truein\epsffile{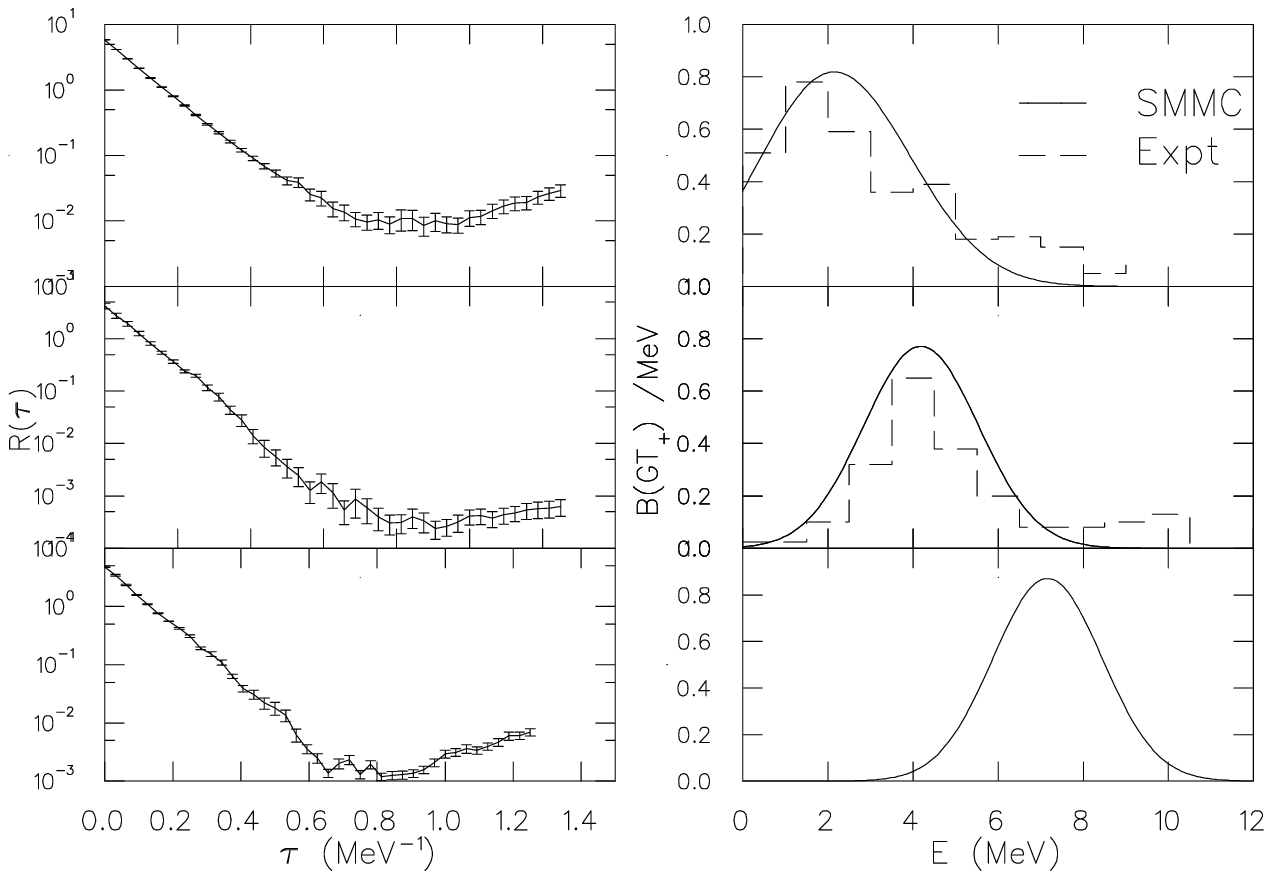}$$
{FIG~8.15.
SMMC GT$_+$ response functions (left side) and GT$_+$ strength
distributions (right side) for $^{54}$Fe (upper panel),
$^{59}$Co (middle
panel), and $^{55}$Co (lower panel).
The energies refer to the daughter nuclei.
The dashed histograms show
the experimental strength distribution as extracted from $(n,p)$ data
\protect\cite{gtdata2}.
}
\end{figure}
\begin{figure}
$$\epsfxsize=4truein\epsffile{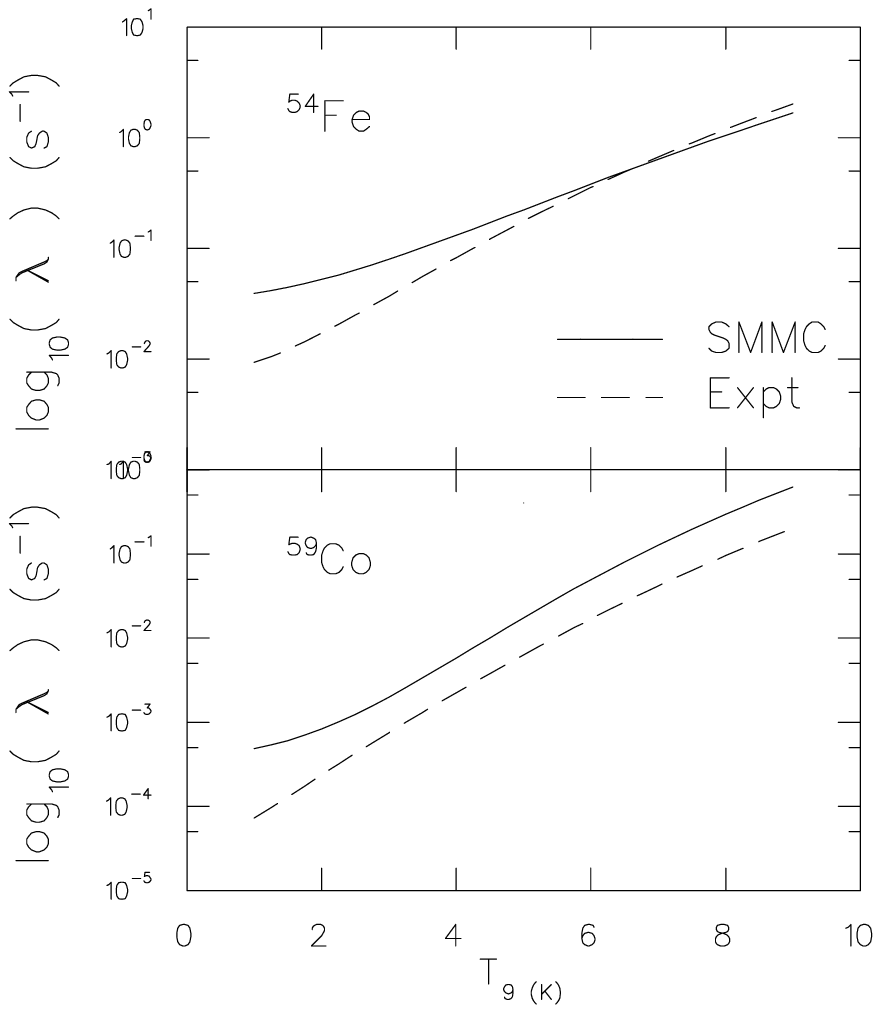}$$
{FIG~8.16.
Gamow-Teller contributions to the electron capture rates on $^{54}$Fe and
$^{59}$Co
as a function of temperature ($T_9$ measures the temperature in $10^9$ Kelvin.)
The calculations have been performed for the experimental strength
distribution \protect\cite{gtdata2} and the SMMC result.
}
\end{figure}

\noindent
\subsection{Temperature dependence of the nuclear symmetry energy}

The size of the core mantle (the difference between the iron
and homologous cores)
is a major determinant of the supernova
mechanism \cite{Bethe90}.
It has been argued recently that the mantle size might be
significantly smaller than generally calculated in supernova models
because of an overlooked temperature dependence of the electron
capture process \cite{Brown93}. The detailed reasoning is as follows
\cite{Donati}: When the nucleus is described by a static mean field
model, dynamical effects associated with collective surface
vibrations are conventionally embodied in an effective nucleon mass,
$m^\ast$ (as are spatial non-localities in the mean field). Donati
{\it et al.} \cite{Donati} studied the coupling of the mean field
single-particle levels to the collective surface vibrations within
the quasiparticle random phase approximation and found that $m^\ast$
decreases with increasing temperature for $T\le2$~MeV, the
temperatures relevant for the presupernova collapse. They attributed
this behavior to a reduction of collectivity at low excitation
energies. In the Fermi gas model, this decrease of $m^\ast$ induces
an increase in the symmetry energy contribution to the nuclear
binding energy
\begin{equation}
E_{\rm sym} (T) = b_{\rm sym} (T) {{(N-Z)^2 }\over {A}}\;,
\end{equation}
where $N$, $Z$, and $A$ are the neutron, proton, and mass numbers of
the nucleus, respectively. Quantitatively, Donati {\it et al.}
estimate that $b_{\rm sym} (T)$ increases by about 2.5~MeV as $T$
increases from 0 to 1~MeV ($b_{\rm sym} (0)\approx28$~MeV
\cite{Bohr}).
Importantly, a larger value of $E_{\rm sym} (T)$ would hinder
electron capture in the presupernova environment, and thus
reduce the size of the core
mantle, so that the shock wave would need significantly less energy to
stop the collapse and explode the star.

To explore the temperature dependence of the symmetry energy,
SMMC calculations were used \cite{symmetry} to study the thermal properties of
various pairs of isobars in the mass region $A=54-64$ which is important
for the presupernova
collapse; this includes two of the three nuclei studied in
Ref.~\cite{Donati} (${}^{64}$Zn and ${}^{64}$Ni).

The discussion focuses
on the temperature
dependence of the internal energy of a nucleus (with proton and neutron
numbers $Z$ and $N$), which is defined as
\begin{equation}
U(N,Z,T) = \langle \hat H \rangle (T) - \langle \hat H \rangle (T=0) \; ,
\end{equation}
where $\langle \hat H \rangle (T)$ is the expectation value of the Hamiltonian
in the canonical ensemble at temperature $T$. Guided by the
semi-empirical parametrization of the binding energies
(e.g., the Bethe-Weizs\"acker formula), the difference of $U(T)$
for two isobars
\begin{equation}
\Delta U(T) = U(N_1,Z_1,T) - U(N_2,Z_2,T) \;\;\;; N_1+Z_1=N_2+Z_2=A
\end{equation}
should reflect the proposed temperature dependence of the symmetry
energy contribution to the binding energies.
This proportionality is valid only if the other contributions
to the binding energy, particularly nuclear structure effects,
are negligible in $\Delta U(T)$. To investigate this point,
Dean {\it et al.} \cite{symmetry}
performed calculations
for several pairs of isobars for mass numbers
$A=54$ ($^{54}$Fe, $^{54}$Cr), $A=56$ ($^{56}$Fe, $^{56}$Cr),
$A=58$ ($^{58}$Ni, $^{58}$Fe), $A=60$ ($^{60}$Ni, $^{60}$Fe), $A=62$
($^{62}$Zn, $^{62}$Ni) and $A=64$ ($^{64}$Zn, $^{64}$Ni).

The experimental knowledge
of the excitation spectra for the nuclei studied appeared to be
sufficient to derive $U(T)$ for $T \leq 0.5$ MeV solely from data.
For higher temperatures these values were supplemented with
the appropriate SMMC results for $U(T)-U(T=0.5\;{\rm MeV})$.
Fig. 8.17 compares $\Delta U(T)$ for $T=1$ MeV as calculated
by the combination of data and SMMC results \cite{symmetry}
with the predictions of Ref.
\cite{Donati}. For all isobaric pairs with the exception of ($^{64}$Ni,
$^{64}$Zn), the values for $\Delta U(T)$ are compatible with zero
(no increase of $b_{\rm symm}$ with temperature).
However, we note that, in agreement with Ref. \cite{Donati}, the SMMC
calculation
shows an increase in $b_{\rm symm}$ for the $A=64$ pair ($^{64}$Ni, $^{64}$Zn).
However, it does not confirm
that this increase is generic.

\begin{figure}
$$\epsfxsize=4truein\epsffile{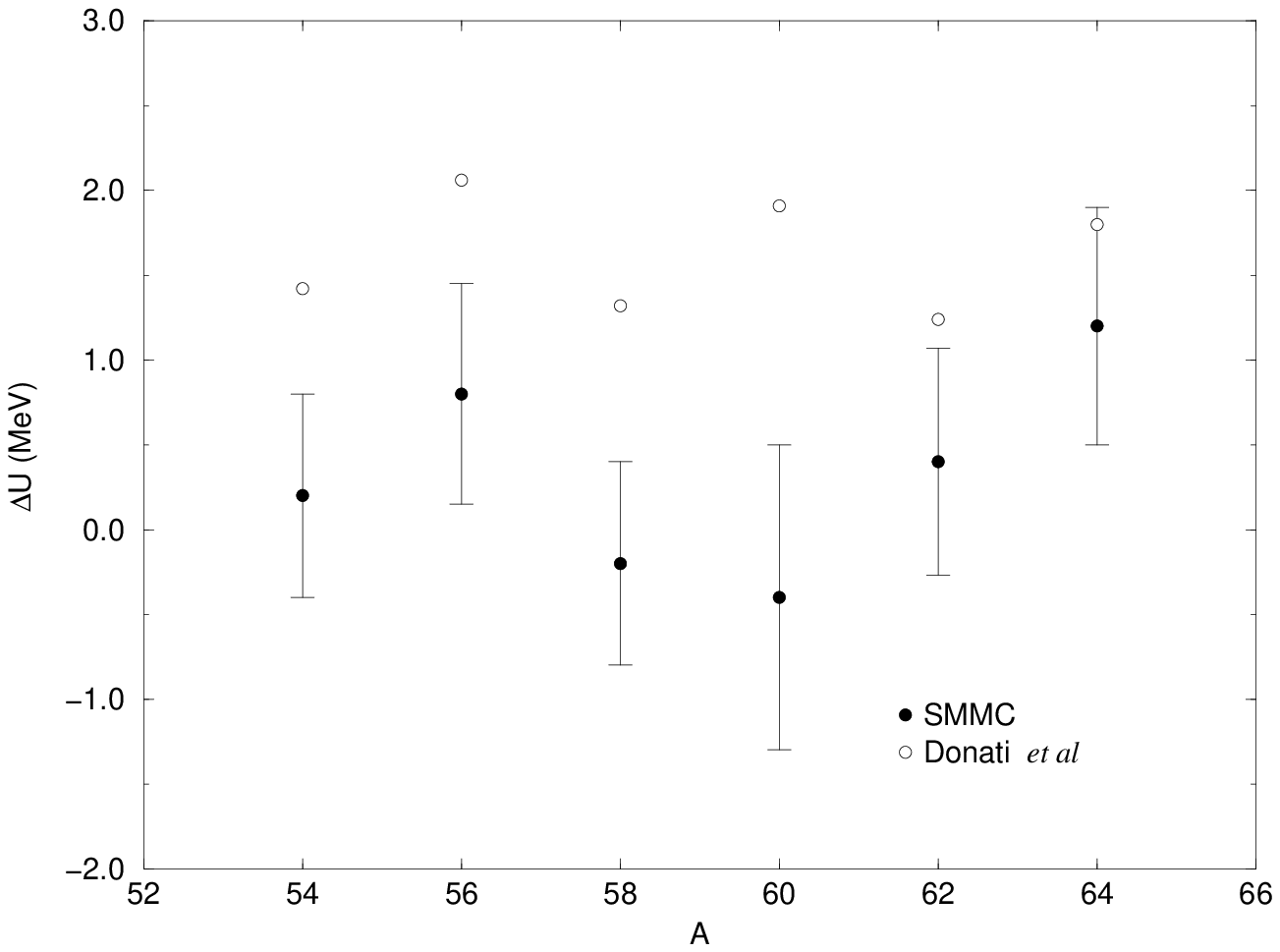}$$
{FIG~8.17.
Comparison of the values for $\Delta U(T)$ predicted by Donati {\it et al.}
(open circles, \protect\cite{Donati})
with the results calculated from
experimental data and
SMMC calculations for selected pairs of isobars, as described in the text
(full circles).
}
\end{figure}

\noindent
\subsection{${}^{170}$Dy at finite rotation and temperature}

A first SMMC calculation \cite{Dean93}
to describe rare-earth nuclei used the $Z=50$--82 shell for protons
($2s1d0g_{7/2}0h_{11/2}$) and the $N=82$--126 shell for neutrons
($2p1f0g_{9/2}0i_{13/2}$). The Hamiltonian chosen was of the pairing
plus quadrupole form given by
\begin{equation}
\hat H=\sum_\alpha\varepsilon_\alpha
a^\dagger_\alpha a_\alpha -g_p \hat P^\dagger_p \hat P_p-
g_n \hat P^\dagger_n \hat P_n- {\chi\over2} \hat Q\cdot \hat Q\;,
\end{equation}
where $\varepsilon_\alpha$ are the single particle energies,
$\hat P^\dagger_{p(n)}, \hat P_{p(n)}$ are the monopole pair creation and
annihilation
operators for protons and neutrons, and
$\hat Q=\hat Q_p+\hat Q_n$ is the quadrupole
operator.
The pairing strengths
$g_{p(n)}$, the quadrupole interaction strength $\chi$, and the single
particle energies used were taken from \cite{Baranger}.

The nucleus considered was
$^{170}$Dy.
It involved 16 valence protons and 22 valence neutrons moving among
the $N_s=32$ and 44 single-particle states, respectively, giving
a total of 10$^{21}$ $m$-scheme determinants. The calculations used
$\Delta\beta=0.0625$ MeV$^{-1}$ and $N_t=8$ to 64 time slices.
Due to
limited computer memory and cycles at that time, it was necessary to
perform canonical analyses
of the fields generated through grand-canonical sampling; all other work
in this Report uses canonical sampling.
In particular, for the $^{170}$Dy calculation,
canonical observables (subscript ``$c$'') are given in terms of the
grand-canonical sampling (subscript ``$G$'') as
\begin{equation}
\langle{\hat \Omega}\rangle_c=
{{\int D\sigma W_{G\sigma} \Phi_{c\sigma}
\left[\zeta_{c\sigma}/\zeta_{G\sigma}\right]
\langle{\Omega}\rangle_{c\sigma}}\over
{\int D\sigma W_{G\sigma} \Phi_{c\sigma}
\left[\zeta_{c\sigma}/\zeta_{G\sigma}\right]}}\;,
\end{equation}
where $\zeta_\sigma ={\rm Tr} {\hat U}_\sigma$. Although this is an exact
expression,
the fluctuations in $\left[\zeta_{c\sigma}/\zeta_{G\sigma}\right]$ determine
how precise the Monte Carlo
evaluation can be. We found  that the fluctuations are less
than 10\% (as the number dispersion in the grand-canonical ensemble is
small), so that canonical observables can be calculated with good precision.

In Fig.~8.18 we show the static observables (for the uncranked system) in both
the grand-canonical and canonical formalisms. We calculated observables
canonically, using grand canonical fields, up to $\beta=2.0$ in order to
demonstrate that, for these nuclear systems, either method may be used in this
kind of calculation. As the temperature decreases, the nucleus
becomes more deformed. Relaxation of the expectation values of
${\hat H}$ and ${\hat J}^2$
is also clearly seen. The sign in these cases is identically one for all
temperatures.

Cranking calculations in which $\hat H\rightarrow \hat H-\omega \hat J_z$
were also
performed. The systematics are shown in Fig.~8.19, where we display
the Monte Carlo sign
$\langle\Phi\rangle$, $\langle \hat H\rangle$,
$\langle \hat Q^2\rangle$, $\langle
\hat J^2\rangle$, the pairing energy
$-g\langle \hat P^{\dagger} \hat P\rangle =
-g_p\langle \hat P^\dagger \hat P\rangle_p-
g_n\langle \hat P^\dagger \hat P\rangle_n$,
and the moment of inertia,
$\langle{\cal I}_2\rangle$. Note that the sign degrades quite rapidly with
increasing $\omega$, making cranking calculations at lower temperatures
difficult. Moments of inertia were calculated from ${\cal I}_2=d\langle
\hat J_z\rangle/d\omega=\beta[\langle \hat J^2_z\rangle-
\langle \hat J_z\rangle^2]$. At high
temperatures, the nucleus is unpaired and the moment of inertia decreases as
the system is cranked. However, for lower temperatures when the nucleons are
paired, the moment of inertia initially increases as we begin to crank, but
then decreases at larger cranking frequencies as pairs break;
Figure~8.19 shows
that the pairing gap also decreases as a function of $\omega$. It is well
known that the moment of inertia depends on the pairing gap \cite{8.7B}, and
that
initially ${\cal I}_2$ should increase with increasing $\omega$. Once the
pairs have been broken, the moment of inertia decreases. These features are
evident in the figure.

It is of particular interest to determine the quadrupole shape of a nucleus
as function of temperature and angular momentum. It is generally expected
that some nuclei may exhibit a sudden phase transition from a prolate to
spherical shape as the temperature increases \cite{8.7C}. In addition, as the
cranking frequency is increased a transition to oblate ellipsoids is also
expected.

In order to obtain a more detailed picture of the deformation, we examine the
components of the quadrupole operator $\hat Q_{\mu} = r^2 Y_{2\mu}^*$. Note,
however, that due to rotational invariance of the uncranked Hamiltonian, the
expectation value of any component $\hat Q_{\mu}$ is expected to vanish. On the
other hand,
there is an intrinsic frame
for each Monte Carlo sample,
in which it is possible to compute the three non-zero components $Q_0'$,
$Q_{2}'$, and $Q_{-2}'$ (the prime is used to denote the intrinsic frame).
The intrinsic quadrupole moments can then be related to the standard
deformation coordinates $\beta$ and $\gamma$ \cite{Bohr,Ormand}.
The task remains to compute the quadrupole moments in the intrinsic frame for
each Monte Carlo sample. This is accomplished by computing and diagonalizing
the expectation value of the cartesian quadrupole tensor $Q_{ij} = 3x_ix_j -
\delta_{ij}r^2$ for each Monte Carlo sample. From the three eigenvalues, it
is straightforward to determine the deformation parameters as in \cite{8.7E}.

Figure~{8.20}
shows the evolution of the shape distribution for ${}^{170}$Dy at
inverse temperatures $T^{-1}=0.5$, 1.0, 2.0, and 3.0~MeV$^{-1}$. These
contour plots show the free energy $F(\beta,\gamma)$, obtained from the shape
probability distribution, $P(\beta,\gamma)$, by
\begin{equation}
F(\beta,\gamma)=-T\ln {{\cal P}(\beta,\gamma)\over\beta^3\sin 3\gamma}
\end{equation}
where the $\beta^3\sin3\gamma$ is the metric in the usual deformation
coordinates; ${\cal P}$ was obtained simply by
binning the Monte Carlo samples in the $\beta-\gamma$ plane.
As is seen from the plots, deformation clearly sets in with
decreasing temperature. At high temperatures, the system is nearly spherical,
whereas at lower temperatures, especially at $T^{-1}=3.0$ MeV$^{-1}$,
there is a prolate
minimum on the $\gamma=0$ axis.

\begin{figure}
$$\epsfxsize=4truein\epsffile{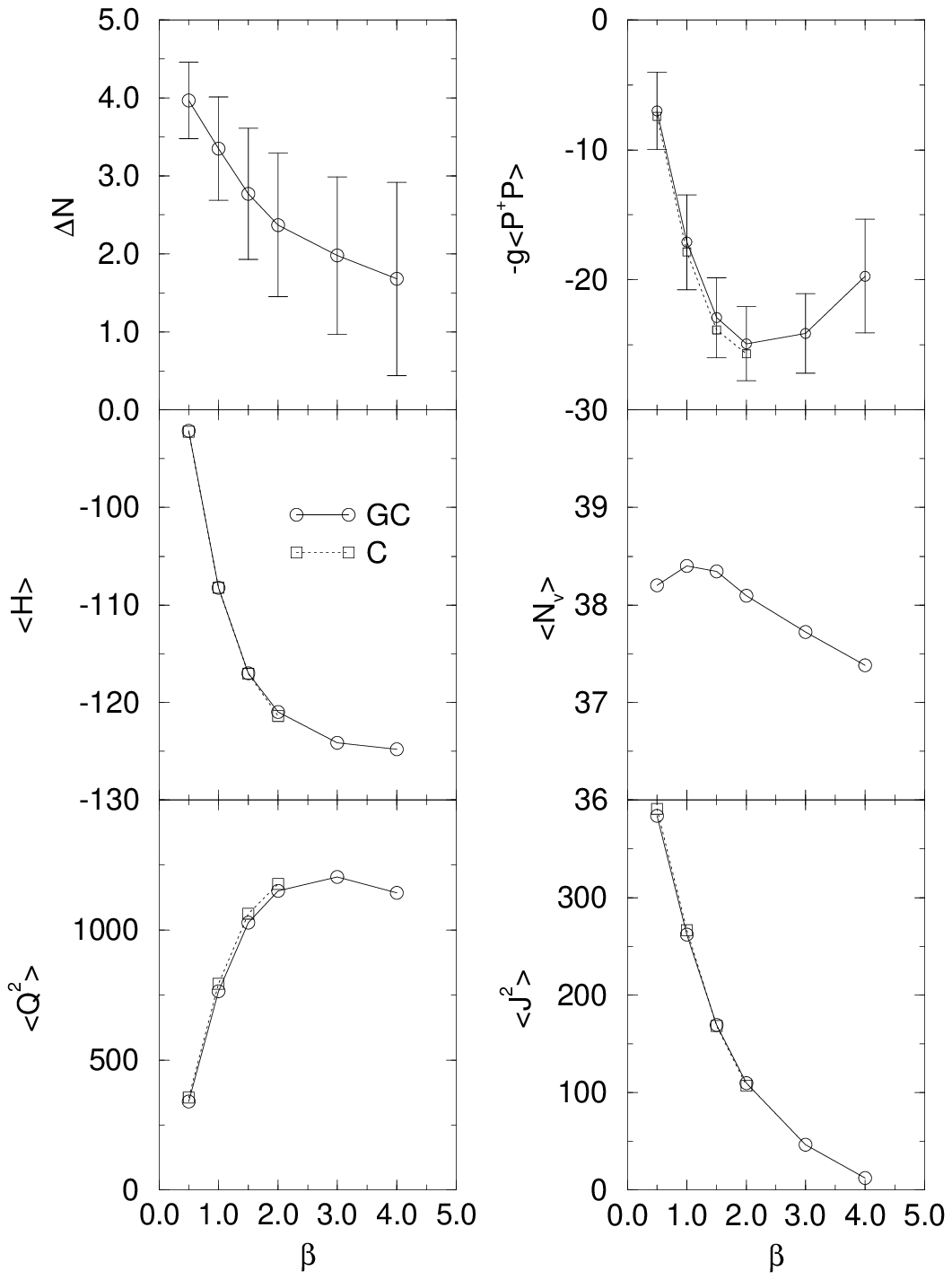}$$
{FIG~8.18  Canonical and grand-canonical static observables are shown for the
uncranked $^{170}$Dy
system as a function of $\beta$.
Circles represent results of canonical projection of the
grand-canonical fields, and squares show the grand-canonical results.
Lines are drawn to guide the eye.
We show expectation values of the
energy $\langle \hat H\rangle$, the isoscalar quadrupole
$\langle \hat Q^2\rangle$,
the valence nucleon number
$\langle \hat N \rangle$, and the number variance
$\langle\Delta N\rangle = \sqrt{\langle \hat N^2\rangle -\langle \hat
N\rangle^2}$
(grand-canonical only), the squared angular momentum $\langle \hat J^2\rangle$
and the pairing energy $-g\langle \hat P^\dagger \hat P\rangle$, which is the
expectation value of the pairing terms in the Hamiltonian (from
\protect\cite{Dean93}).
}
\end{figure}

\begin{figure}
$$\epsfxsize=4truein\epsffile{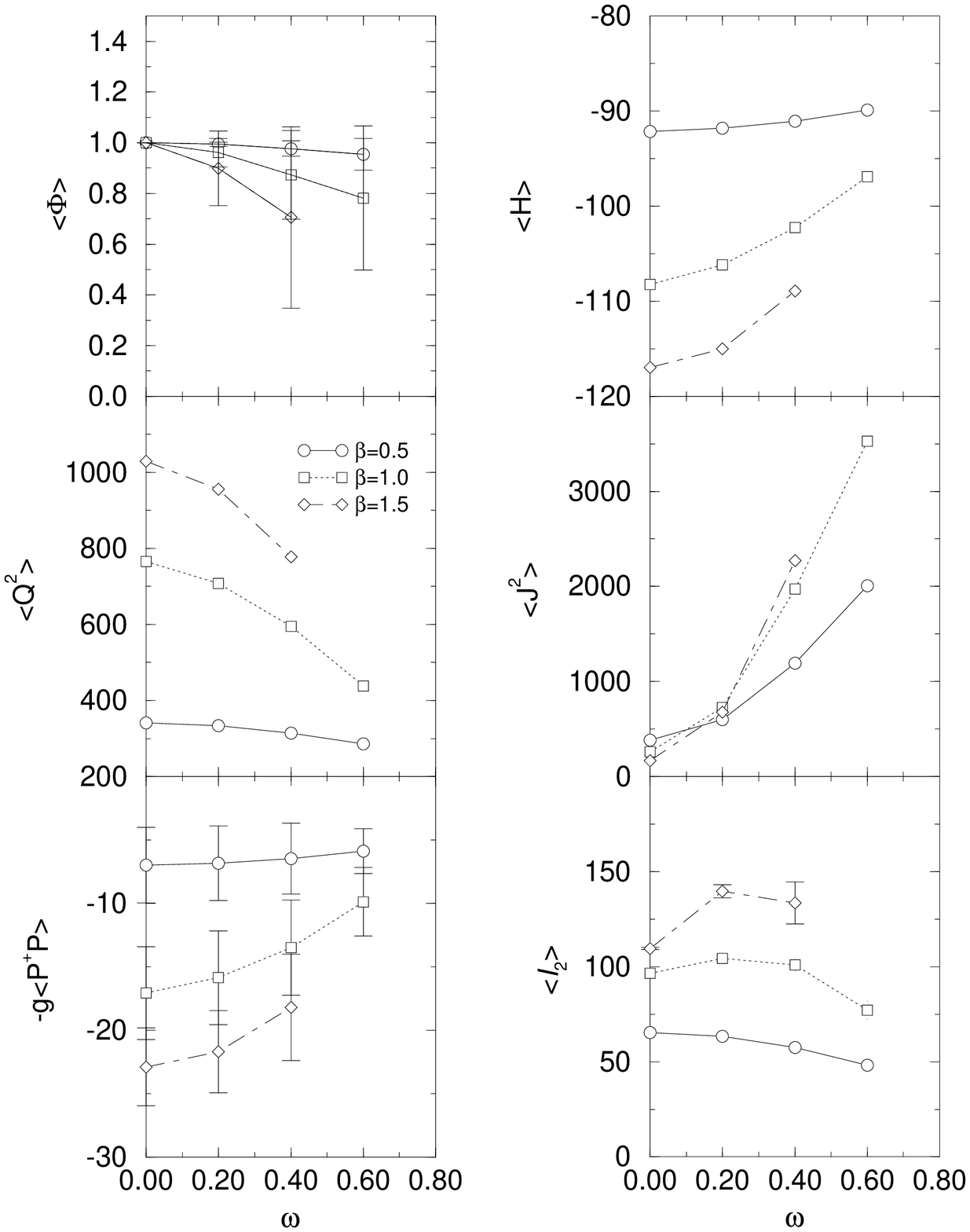}$$
{FIG~8.19. Grand-canonical observables for $^{170}$Dy at
various cranking frequencies and temperatures. We show the
average sign $\langle \Phi\rangle$, the isoscalar quadrupole moment
$\langle \hat Q^2\rangle$,
the energy $\langle \hat H \rangle$, the square of the
angular momentum $\langle \hat J^2\rangle$, the moment of inertia
$\langle{\cal I}_2\rangle$, and the expectation value of the pairing terms
in the Hamiltonian, $-g\langle \hat P^\dagger \hat P\rangle$ (from
\protect\cite{Dean93}).
}
\end{figure}

\begin{figure}
$$\epsfxsize=4truein\epsffile{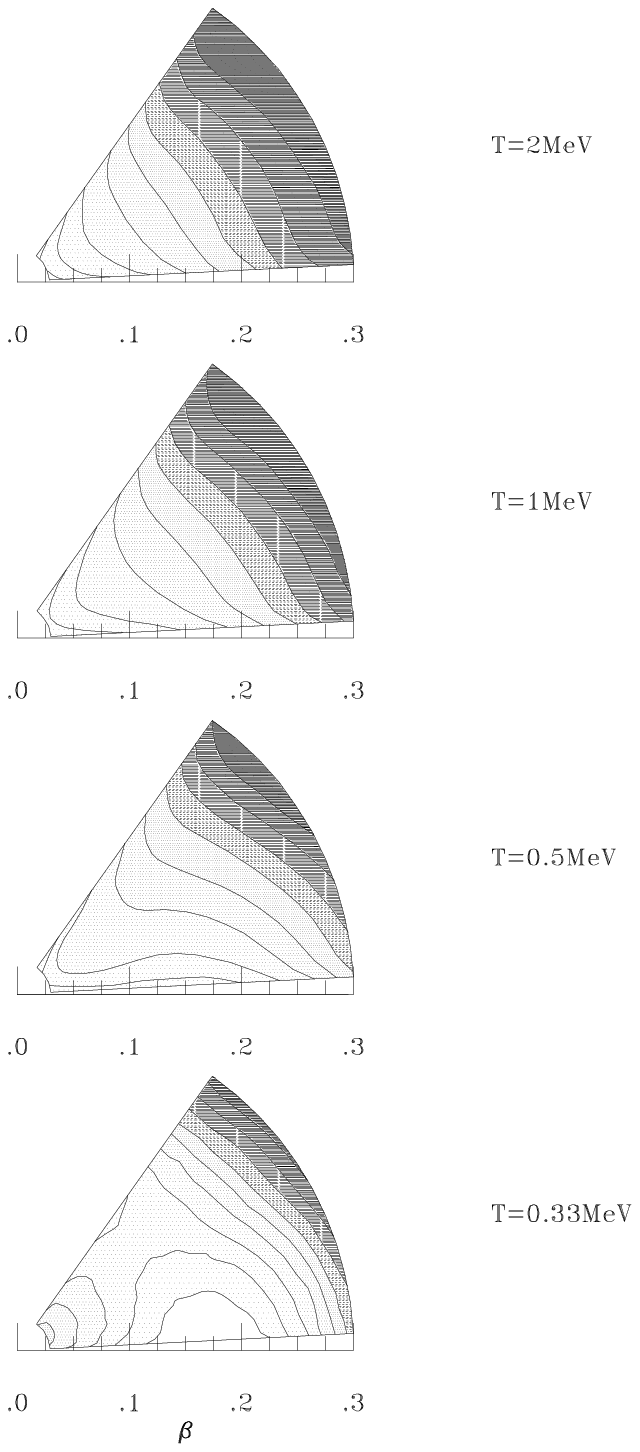}$$
{FIG~8.20.
 Contour plots of the free energy
(as described in the text) in polar coordinates
in the $\beta$-$\gamma$ plane are shown for the $^{170}$Dy system.
Inverse temperatures are from 0.5 (top) 1.0, 2.0,
and 3.0 MeV$^{-1}$
(bottom). Contours are shown at 0.3 MeV intervals. Lighter shades indicate
the more probable nuclear shapes (from \protect\cite{Dean93}).
}
\end{figure}

\noindent
\subsection{$\gamma$-soft nuclei}

Nuclei with mass number $100\leq A \leq140$ are believed to have large shape
fluctuations in their ground states.  Associated with this
softness are spectra with an approximate
$O(5)$ symmetry and bands with energy spacings intermediate
between rotational and vibrational. In the geometrical model  these nuclei
are described by potential energy surfaces with a minimum at $\beta \neq 0$ but
independent of $\gamma$ \cite{8.7F}.  Some of these nuclei have been described
in
terms of a quartic five-dimensional oscillator \cite{8.7G}.  In the Interacting
Boson Model (IBM) they are described by an $O(6)$ dynamical symmetry
\cite{Arima,8.7I,8.7J}.  In the following we review the first fully microscopic
calculations
for soft nuclei with  $100\leq A \leq140$ \cite{8.7K}.

For the two-body interaction we used a
monopole ($J=0)$ plus quadrupole $(J=2)$ force \cite{8.7L}
supplemented by a collective quadrupole interaction:
\begin{equation}
\hat H_2 = - \sum_{\lambda \mu}{{\pi g_\lambda}\over{2\lambda+1}}
 \hat P^\dagger_{\lambda \mu} \hat P_{\lambda \mu}
-{1\over 2} \chi :  \sum_\mu (-)^{\mu} \hat Q_\mu  \hat Q_{-\mu}: \;,
\end{equation}
where $::$ denotes normal ordering.
The single particle energies and the other parameters were determined
as described in Ref. \cite{8.7K}.

We begin discussion of our results with the probability distribution of
the quadrupole moment.  The calculation of the shape distributions
included the quantum-mechanical fluctuations
through the variance of the ${\hat Q}$ operator for each sample,
$\Delta_{\sigma}^2 =
{\rm Tr} ({\hat Q}^2 {\hat U}_\sigma)/
{\rm Tr} ({\hat U}_\sigma) -\langle {\hat Q}\rangle_\sigma^2 $.
The shape distribution $P(\beta,\gamma)$ can be converted to a free
energy surface as described by Eq.~(8.19).

The  shape distributions of  ${}^{128}$Te
and ${}^{124}$Xe are shown in Fig.\  8.21 at different temperatures.
These nuclei are clearly $\gamma$-soft, with energy minima
at $\beta \sim 0.06$ and $\beta \sim 0.15$,
respectively.  Energy surfaces calculated with
Strutinsky-BCS using a deformed Woods-Saxon potential \cite{8.7N}
also indicate $\gamma$-softness with values
of $\beta$ comparable to the SMMC values. These calculations predict
for  ${}^{124}$Xe a prolate minimum with $\beta \approx 0.20$ which is lower
than the spherical configuration by 1.7 MeV but is only 0.3 MeV below the
oblate saddle point, and for
${}^{128}$Te a shallow oblate minimum with $\beta \approx 0.03$.
These $\gamma$-soft
surfaces are similar  to those obtained
in  the $O(6)$ symmetry of the IBM,
or more generally  when the Hamiltonian has mixed
$U(5)$ and  $O(6)$ symmetries but a common $O(5)$ symmetry.
In the Bohr Hamiltonian, an $O(5)$ symmetry occurs when the
collective potential energy depends only on $\beta$ \cite{8.7F}.
The same results are consistent with a potential  energy $V(\beta)$
that has a  quartic anharmonicity \cite{8.7G}, but with a negative quadratic
term
that leads to a minimum at finite $\beta$.

The  total E2 strengths were estimated from $\langle \hat Q^2\rangle$
where ${\hat Q}=e_p \hat Q_p + e_n \hat Q_n$ is the electric quadrupole
operator with
 effective charges of $e_p=1.5 e$ and $e_n=0.5 e$,
and $B(E2; 0^+_1\rightarrow 2^+_1)$ determined by
assuming that most of the strength  is in the $2_1^+$ state.
Values for
$B(E2; 0^+_1\rightarrow 2^+_1)$ of $663\pm 10$, $2106\pm 15$,
and $5491\pm 36$ $e^2 {\rm fm}^4$ were found, to be compared
with the experimental values \cite{8.7O} of 1164, 3910, and 9103  $e^2
{\rm fm}^4$
for  ${}^{124}$Sn, ${}^{128}$Te  and ${}^{124}$Xe, respectively.
Thus, the SMMC calculations reproduce the correct qualitative
trend.
The  $2_1^+$  excitation energies  were also calculated
from the E2 response function. The values of  $1.12 \pm 0.02$, $0.96 \pm
0.02$ and $0.52 \pm 0.01$ MeV
are in close agreement with the experimental values of 1.2, 0.8 and 0.6 MeV
for  ${}^{124}$Sn, ${}^{128}$Te  and ${}^{124}$Xe, respectively.

Another indication of softness is the response of the nucleus to
rotations, probed by adding a cranking field $\omega J_z$ to the Hamiltonian
and examining the
moment of inertia as a function of the cranking frequency $\omega$.
For a soft nucleus one expects a behavior
intermediate between a deformed nucleus, where the
inertia is independent of the cranking frequency, and the
harmonic oscillator, where the inertia becomes singular.
This is confirmed in Fig.\ 8.22 which shows the
moment of inertia ${\cal I}_2$
for ${}^{124}$Xe and ${}^{128}$Te as a function of $\omega$,
and indicates that  $^{128}$Te has a more harmonic character.
The moment of inertia for $\omega=0$ in both nuclei is significantly
lower than the rigid body value ($\approx 43 \hbar^2/$MeV
for $A=124$) due to pairing correlations.

Also shown in Fig.\ 8.22 are
$\langle \hat Q^2 \rangle$ where $\hat Q$ is the mass quadrupole,
the BCS-like
pairing correlation $\langle \hat \Delta^\dagger \hat \Delta\rangle$ for
the protons and $\langle \hat J_z\rangle$ (neutron pairing is less affected,
and therefore not shown).
Notice that the increase in $I_2$ as a function of $\omega$ is
strongly correlated with the rapid decrease of pairing correlations
and that  the peaks in $I_2$ are associated with the onset of
a decrease in collectivity (as measured by  $\langle \hat Q^2 \rangle$).
This suggests band crossing along the yrast line  associated with
pair breaking and alignment of the quasi-particle spins at
$\omega \approx 0.2$ MeV ($\langle \hat J_z \rangle \approx 7 \hbar$)
for  ${}^{128}$Te and $\omega \approx 0.3$ MeV
($\langle \hat J_z \rangle \approx 11 \hbar$) for  ${}^{124}$Xe.
The results are consistent with an experimental evidence of  band
crossing in the yrast sequence of  ${}^{124}$Xe around spin of
10 $\hbar$ \cite{8.7P}.
The alignment effect is clearly seen in the behavior of
$\langle \hat J_z \rangle$ at the lower
temperature, which shows a rapid increase after an
 initial moderate change. Deformation and pairing  decrease
also as a function of temperature.

The total number of
$J$-pairs ($n_J=\sum_\alpha n_{\alpha J}$)  in the various pairing channels
was also calculated;
the results for the number of correlated pairs (after subtracting the
mean-field values) are shown in Fig.~8.23. Since
the number of neutrons in ${}^{124}$Xe is larger than the mid-shell value,
they are treated
as holes.
For $J=0$ and $J=2$ one can compare the largest  $n_{\alpha J}$
with the number of $s$ and $d$ bosons obtained from the $O(6)$ limit of the
IBM.
For ${}^{124}$Xe
the SMMC (IBM) results in the proton-proton pairing channel
are 0.85 (1.22) $s$ ($J=0$) pairs, and 0.76 (0.78)  $d$ ($J=2$)
pairs, while in the neutron-neutron channel we find 1.76 (3.67)  $s$
pairs and 2.14 (2.33)  $d$ pairs.  For the protons the SMMC $d$ to $s$ pair
ratio
0.89 is close to its $O(6)$ value  of 0.64. However, the same ratio for
the neutrons, 1.21,  is intermediate between $O(6)$ and $SU(3)$
(where its value is 1.64) and is consistent with the neutrons filling the
middle of the shell.
The total numbers  of $s$ and $d$ pairs -- 1.61  proton pairs and  3.8
neutron (hole) pairs -- are below the IBM values of 2 and 6, respectively.

\begin{figure}
$$\epsfxsize=4truein\epsffile{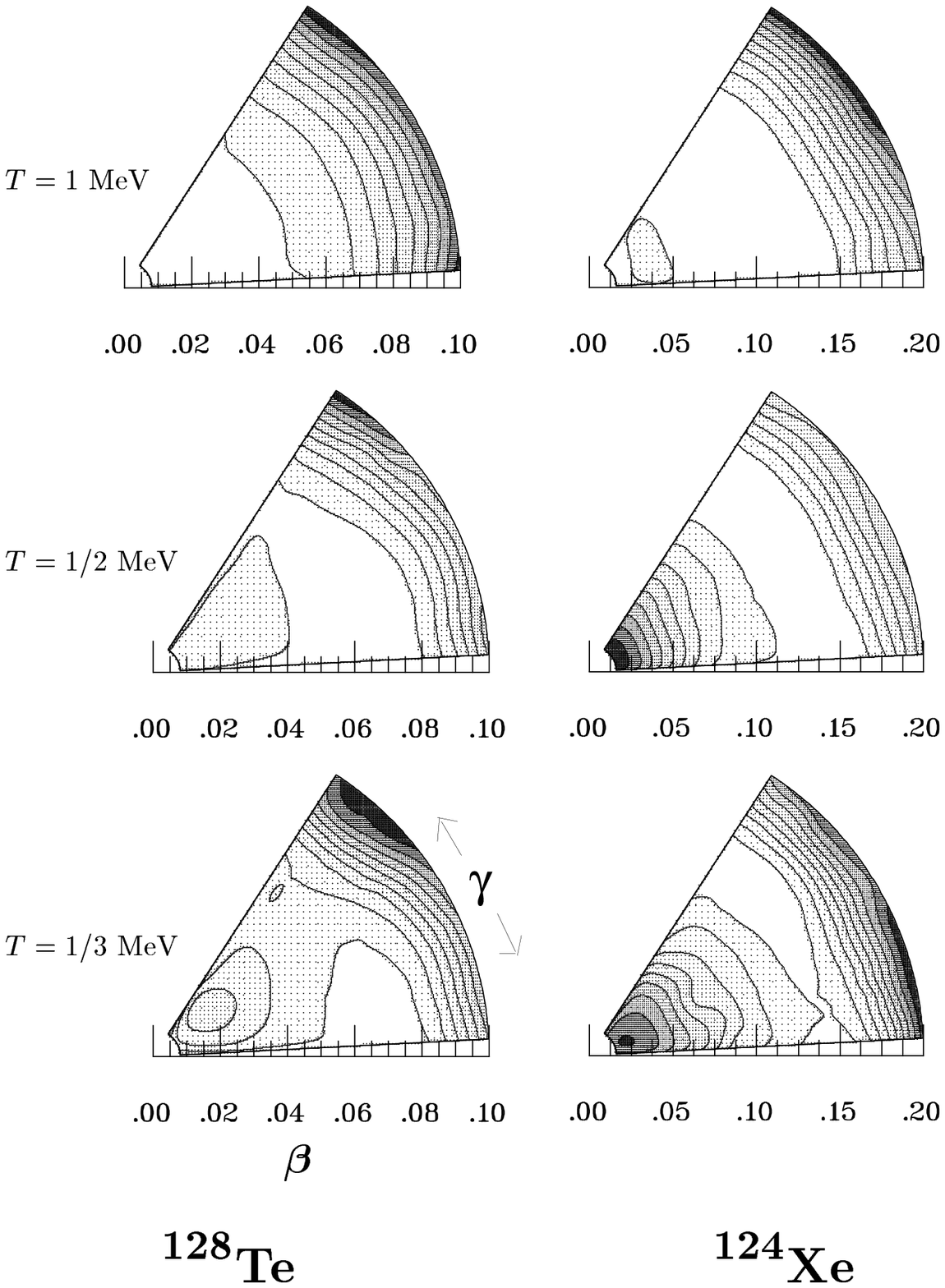}$$
{FIG~8.21.
Contours of the free energy (as described in the text)
in the polar-coordinate $\beta-\gamma$ plane for ${}^{128}$Te and
${}^{124}$Xe.  Contours are shown at 0.3 MeV intervals,
with lighter shades indicating the
more probable nuclear shapes (from \protect\cite{8.7K}).
}
\end{figure}

\begin{figure}
$$\epsfxsize=4truein\epsffile{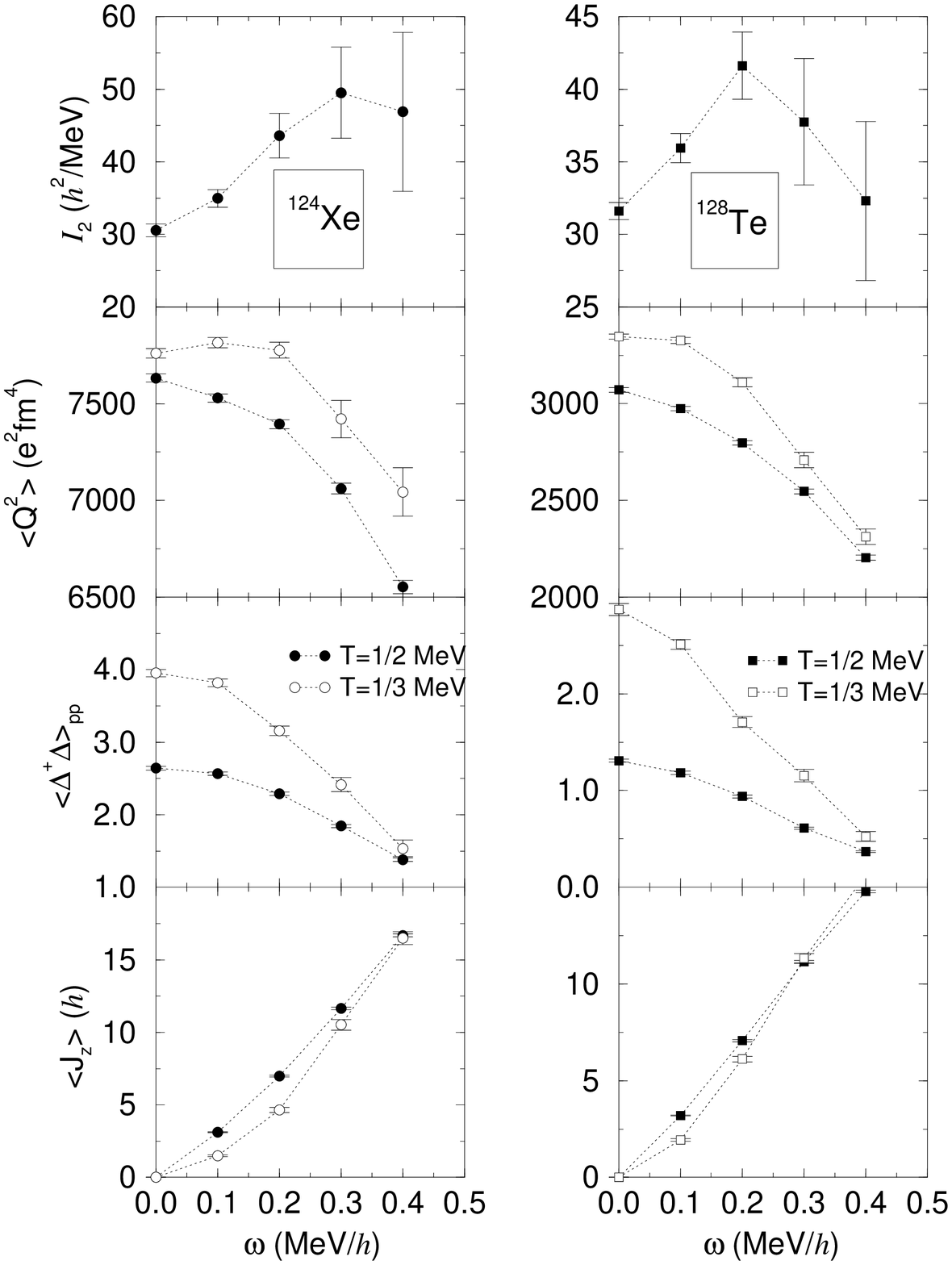}$$
{FIG~8.22.
Observables for ${}^{124}$Xe and ${}^{128}$Te as a function of
cranking frequency $\omega$ and for two temperatures. $I_2$ is the moment
of inertia, $Q$
is the mass quadrupole moment, $\Delta$  is the $J=0$ pairing operator,
and $J_z$ is the angular momentum along the cranking axis (from
\protect\cite{8.7K}).
}
\end{figure}

\begin{figure}
$$\epsfxsize=4truein\epsffile{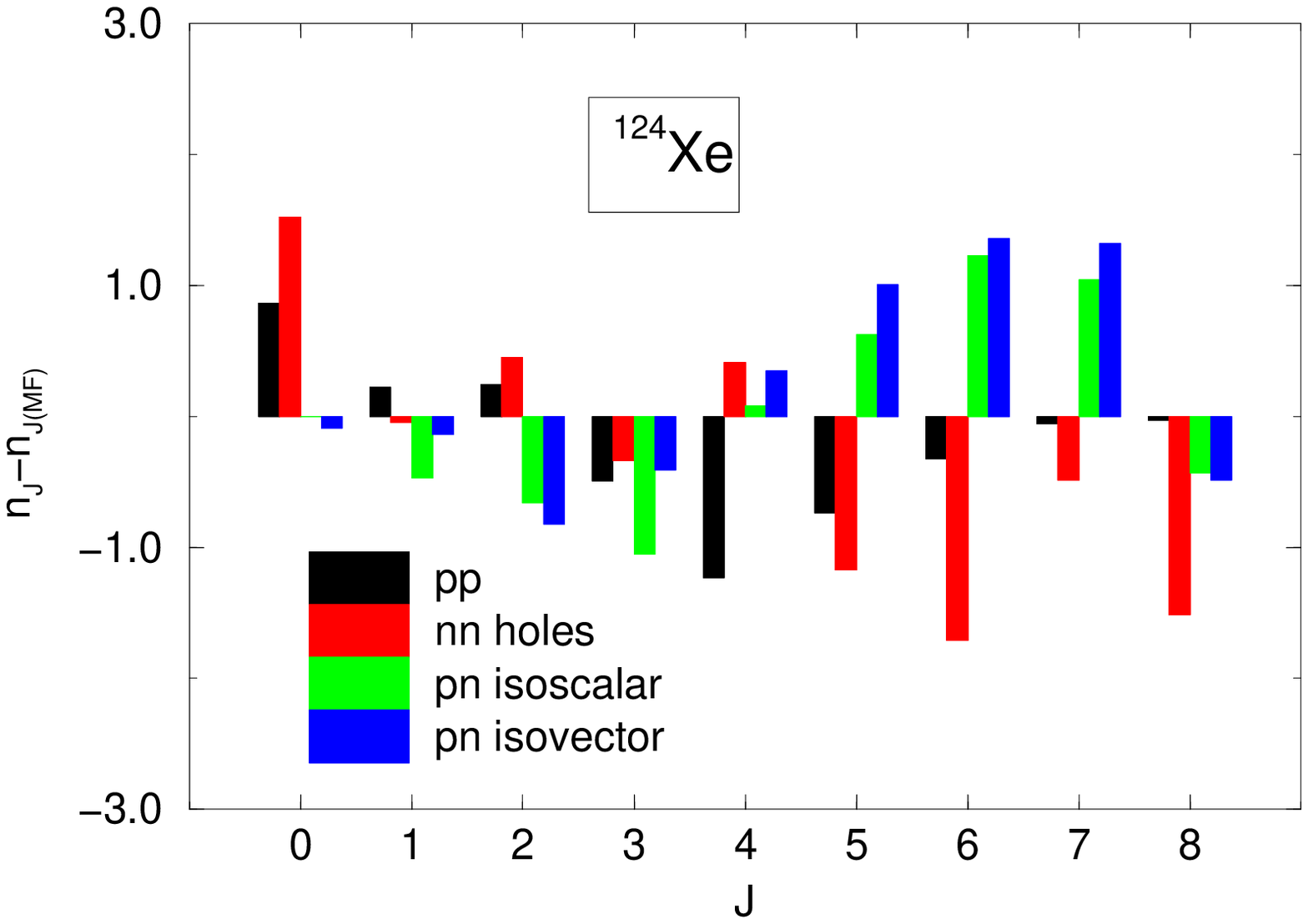}$$
{FIG~8.23.
Number of correlated pairs of angular momentum $J$
in $^{124}$Xe. Different shades correspond to p-p, n-n, isoscalar and
isovector p-n pairs. Neutrons are treated as holes.
}
\end{figure}

\noindent
\subsection{Double-beta decay}

The second-order weak process $(Z,A)\rightarrow(Z+2,A)+2e^-+2\bar\nu_e$ is an
important ``background'' to searches for the lepton-number violating
neutrinoless mode, $(Z,A)\rightarrow(Z+2,A)$. The calculation of the nuclear
matrix element for these two processes is a challenging problem in nuclear
structure, and has been done in a full $pf$ model space for only the lightest
of several candidates, ${}^{48}$Ca.
P.B. Radha {\it et al.} have performed
first Monte Carlo calculations of
the $2\nu\;\beta\beta$ matrix elements in very large model spaces \cite{Radha}.

In two-neutrino double $\beta$-decay, the nuclear matrix element of interest
is
\begin{equation}
M^{2\nu}\equiv \sum_m
{\langle f_0\vert \hat{\bf G}\vert m\rangle \cdot
\langle m\vert \hat{\bf G}\vert i_0\rangle\over
E_m-\Omega}\;,
\end{equation}
where $\vert i_0\rangle$ and $\vert f_0\rangle$ are the $0^+$ ground states
of the initial and final even-even nuclei, and $\vert m\rangle$ is a $1^+$
state of the intermediate odd-odd nucleus; the sum is over all such states.
In this expression, $\hat{\bf G}= \hbox{\bf{$\sigma$}}\tau_-$ is the
Gamow-Teller operator for $\beta^-$-decay (i.e., that which changes a neutron
into a proton) and $\Omega=(E_{i_0}+E_{f_0})/2$.
A common approximation to $M^{2\nu}$
is the closure value, \begin{equation}M^{2\nu}={M_c\over {\bar E}}
\end{equation}
where $\bar E$ is an average energy dominator and
\begin{equation}
M_c\equiv \sum_m
\langle f_0\vert \hat{\bf G}\vert m\rangle\langle m\vert\hat{\bf G}\vert
i_0\rangle=
\langle f_0\vert\hat{\bf G}\cdot\hat{\bf G}\vert i_0\rangle\;.
\end{equation}

SMMC methods can be used to calculate both $M_c$ and $M^{2\nu}$.
To do so,
consider the function
\begin{eqnarray}
\phi (\tau , \tau') & = &
\langle e^{\hat{H}(\tau+\tau')} \hat{\bf G}^\dagger\cdot\hat{\bf
G}^\dagger e^{-\hat{H}\tau}
\hat{\bf G} e^{-\hat{H}\tau'}\hat{\bf G}\rangle \nonumber \\
&=&{1\over Z}{\rm Tr}_A\,
\left[ e^{-(\beta-\tau-\tau')\hat{H}} \hat{\bf G}^\dagger\cdot\hat{\bf
G}^\dagger
e^{-\tau \hat{H}}
\hat{\bf G} e^{-\tau' \hat{H}} \hat{\bf G}\right]\;,
\end{eqnarray}
where $Z={\rm Tr}_A\, e^{-\beta \hat{H}}$ is the partition function for the
initial nucleus,
$\hat H$ is the many-body Hamiltonian, and the trace is
over all states of the initial nucleus.
The quantities $(\beta-\tau-\tau^{\prime})$ and
$\tau$ play the role of the inverse temperature in the parent and
daughter nucleus respectively. A spectral expansion of $\phi$ shows that large
values of these parameters guarantee cooling to the parent and
daughter ground states. In these limits, we note that
$\phi(\tau,\tau^{\prime}=0)$
approaches $e^{-\tau Q} |M_c|^2$,  where $Q=E_i^0-E_f^0$
is the energy release, so that a calculation of $\phi(\tau,0)$
leads directly to the closure matrix
element. If we then define
\begin{equation}
\eta(T,\tau)\equiv \int_0^T d\tau^{\prime} \phi(\tau,\tau^{\prime})
e^{-\tau^{\prime} Q/2},
\end{equation}
and
\begin{equation}
M^{2\nu}(T,\tau)\equiv{\eta(T,\tau) M^{*}_{c} \over \phi(\tau,0)},
\end{equation}
it is easy to see that in the limit of large
$\tau$, $(\beta-\tau-\tau^{\prime})$, and $T$,
$M^{2\nu}(T,\tau)$ becomes
independent of these parameters and is equal to the matrix element
in Eq. (8.23).

In the first applications, Radha {\it et al.} calculated the $2\nu$
matrix elements for $^{48}$Ca and $^{76}$Ge \cite{Radha}.
The first nucleus allowed
a benchmarking of the SMMC method against direct diagonalization. A large-basis
shell model calculation for $^{76}$Ge has long been waited for, as
$^{76}$Ge is one of the few nuclei where the $2\nu\beta\beta$ decay
has been precisely measured and the best limits on the $0\nu$ decay
mode have been established \cite{Ge1,Ge2,Klapdor}.

The SMMC calculation for $^{48}$Ca was performed for the complete
$pf$ shell using the KB3 interaction and compared
to a direct diagonalization using an implementation of the Lanczos
algorithm.
To avoid the sign problem (section 7), the SMMC result was obtained
by extrapolation from a family of sign-problem-free Hamiltonian ${\hat H}_g$
with $g\leq0$ and $\chi=4$.
For both the exact and the closure matrix elements,
the SMMC and direct diagonalization results
agree well for all values of $g\leq0$ (Fig. 8.21).
Upon extrapolation to the physical value $g=1$
the SMMC study yields $M^{2\nu}=0.15\pm0.07$ MeV, consistent
with the matrix element obtained by direct diagonalization
($M^{2\nu}=0.08$ \cite{Caurier90}, with the erratum in \cite{Caurier94}).
Note that
the $g$-dependence was found to be linear
at $g\leq0$. However, when monitoring the $g$-dependence for $g>0$ within the
direct diagonalization, a small curvature in the extrapolation was found at
$g \approx 1$ which could not be detected in the SMMC results at $g\leq0$
and so leads to some uncertainty in the extrapolated SMMC result.

We note that the shell model estimate for the $^{48}$Ca
lifetime \cite{Caurier94} appears to be shorter
than the recently established experimental value
\cite{Vogel95}. While this experimental
lifetime might still be compatible with complete $pf$-shell model calculation
after slight modifications of the important $J=0,T=1$ and $J=1,T=0$ matrix
elements \cite{Radha95}, it appears desirable to calculate the
$2\nu$ matrix element for $^{48}$Ca in the complete $(0d1s)(1p0f)$ shells
taking possible polarization effects into account. Such a challenging
calculation is computationally feasible within the SMMC.

To monitor the possible uncertainty related to the $g$-extrapolation
in the calculation of the $2\nu$ matrix
element for $^{76}$Ge, SMMC studies have been performed for two quite
different families of sign-problem-free Hamiltonians ($\chi=\infty$
and $\chi=4$). The calculation comprises the complete $(0f_{5/2},1p,0g_{9/2})$
model space, which is significantly larger than in previous shell model
studies \cite{Haxton}. The adopted effective interaction is based
on the Paris potential and has been constructed for this model space
using the Q-box method developed by Kuo \cite{Kuo95}.

As is shown in Fig. 8.25, upon linear extrapolation both families of
Hamiltonians predict a consistent value for the $2\nu$ matrix element
of $^{76}$Ge. The results $M^{2\nu}=0.12\pm0.07$ and
$M^{2\nu}=0.12\pm0.06$ are only slightly
lower than the experimental values ($M^{2\nu}=0.22\pm0.01$, \cite{Klapdor}).
This comparison, however, should not be overinterpreted yet,
as the detailed reliability of the effective interaction is still to be
checked.

It is interesting that the closure matrix element found in the SMMC
calculation and the average energy denominator
($M_c=-0.36\pm0.37$, ${\bar E}=-3.0\pm 3.3$ MeV and $M_c=0.08\pm0.17$,
${\bar E}=0.57\pm 1.26$ MeV for the two
families of Hamiltonians with ${\chi}=\infty$ and ${\chi}=4$,
respecitvely) are both
significantly smaller than had been assumed previously. This is confirmed
by a recent truncated diagonalization study \cite{Caurier96}.

\begin{figure}
$$\epsfxsize=3truein\epsffile{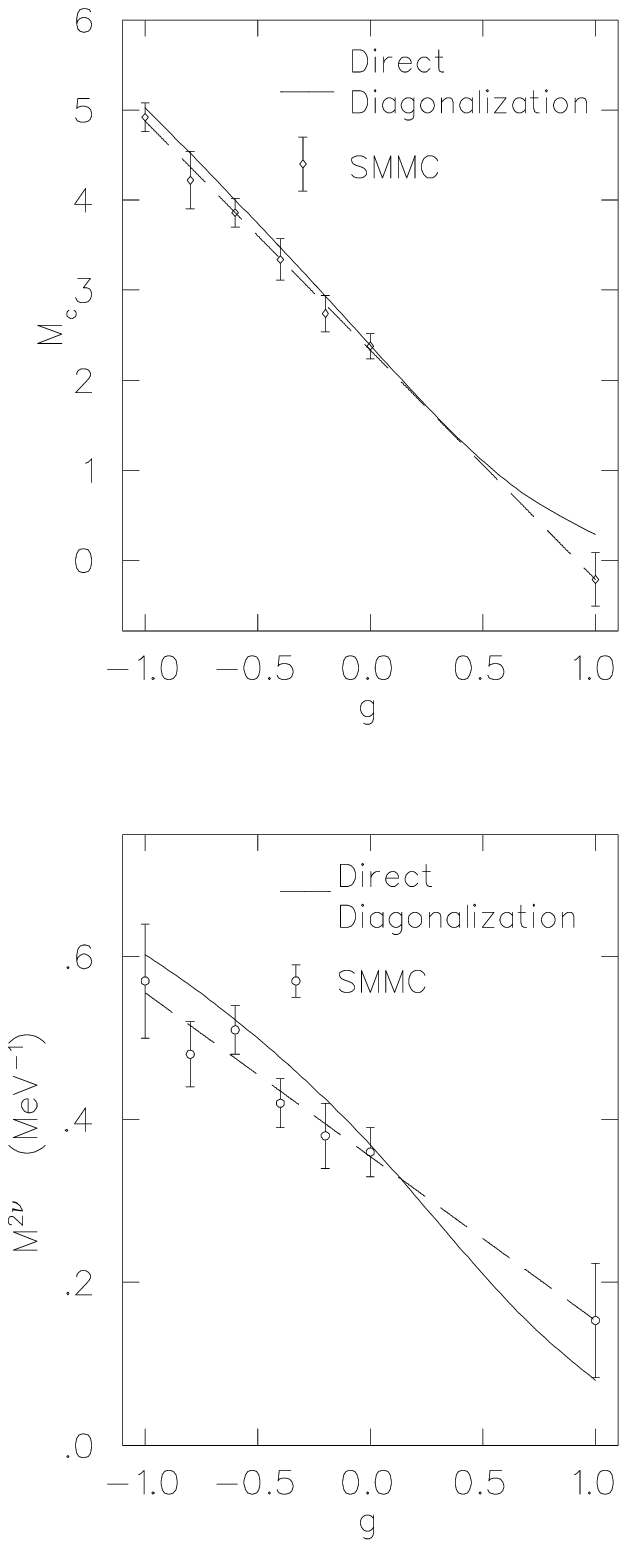}$$
{FIG~8.24.
The closure matrix element (upper panel) and the exact
matrix element (lower panel) for the double-$\beta$ decay of $^{48}$Ca
as a function of $g$-values in the adopted family of Hamiltonian (see text).
The SMMC results (open circles) are compared to the direct
diagonalization (solid circles). The SMMC result at $g=1$ has been obtained
by linear extrapolation (from \protect\cite{Radha}).
}
\end{figure}
\begin{figure}
$$\epsfxsize=4truein\epsffile{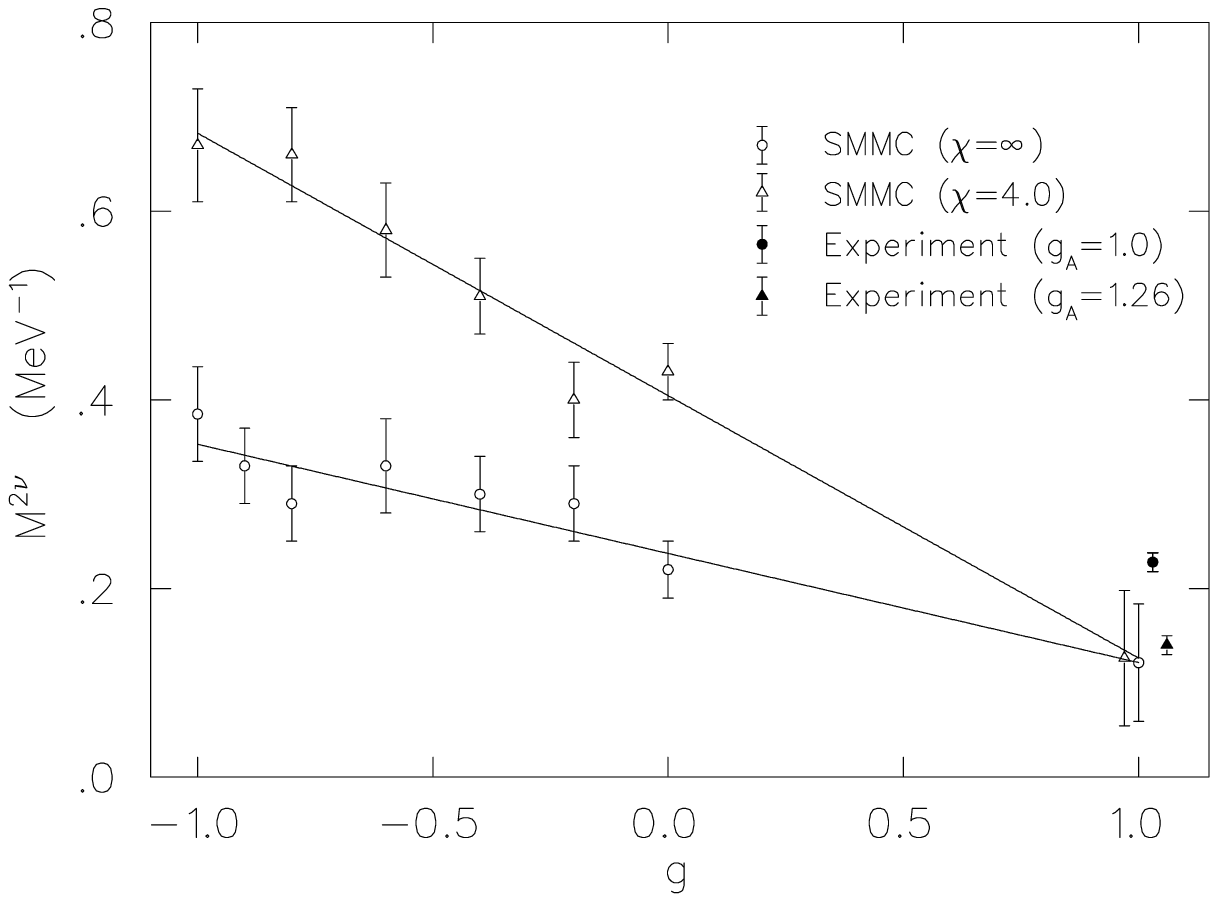}$$
{FIG~8.25.
The $2\nu$ matrix element for $^{76}$Ge calculated within SMMC studies
based on two families of Hamiltonians which are free of sign problems.
The physical values are obtained by linear extrapolation to $g=1$.
The experimental value for this matrix element \protect\cite{Klapdor}
is indicated by the diamond (from \protect\cite{Radha}.
}
\end{figure}

\noindent
\section{Prospects}

We have presented in section 8 a
sampling of results from SMMC calculations.  These
demonstrate both the power and limitations of the methods and the physical
insights they offer.  SMMC calculations, while not a panacea, clearly have
certain advantages over conventional shell model approaches, particularly for
properties of ground states or thermal ensembles.  Of the results
discussed in this review,
the most significant bear on the quenching of GT strength, the
pairing structure, and nuclear shapes. Many of the first applications
of the SMMC approach have significant bearing on astrophysical questions.

With respect to the technical aspects of these calculations we note
the following.
\begin{itemize}
\item SMMC methods are computationally intensive.  However,
computing power is becoming both
less expensive and more widely available at an astonishing rate.  It is
a great advantage that these calculations can efficiently exploit loosely
connected "farms" of work-station-class machines.

\item A number of interesting physics questions require multi-shell
model spaces. Among these are the origin of the apparent renormalization
of $g_A=1$ in Gamow-Teller transitions, strengths of first-forbidden
weak operators, the properties of the giant dipole resonance, and
nuclear behavior at temperatures above 1.5 MeV. Preliminary SMMC studies
offer circumstantial evidence that center-of-mass (CM)
motion  is not a significant concern for some of the operators of interest.
This is not too surprising at finite temperature, since the CM is only three
degrees of freedom (far fewer than the internal dynamics).

\item We lack the capability to treat odd-$A$ or odd-odd
$N \neq Z$ systems at temperatures below $\sim$800 keV,
because of "sign" problems in the Monte Carlo sampling.
However, meaningful results are possible at modest temperatures for such
nuclei, as examplified by the $^{51}$V, $^{55}$Mn, and $^{59}$Co calculations
in section 8.1.
Similar problems prevent spin projection, which would enable yrast
spectroscopy.
\end{itemize}

Otsuka and collaborators \cite{Otsuka}
have recently proposed a hybrid scheme
whereby SMMC methods are used to select a many-body basis, which is then
employed in a conventional diagonalization.  The sign problems alluded to
above are absent, and detailed spectroscopy is possible.  Test applications
to
boson problems have shown some promise, although the utility for realistic
fermion systems remains to be demonstrated.
It is unfortunate that explicit angular momentum projection apparently
plays an important role in these methods, as the numerical effort
of that procedure increases strongly with the size of the model space.

Additional physics results in finite nuclei
that should emerge in the future
include:
more realistic electron capture rates in pre-supernova conditions, the
double-beta decay matrix elements for candidates other than
${}^{76}$Ge, systematic
studies of rare earth nuclei at finite temperature and spin, studies to
improve
the effective interactions used, tests of such models as the IBM and RPA, and
predictions of nuclear properties far from $\beta$-stability.

The ground state and thermal properties of nuclear matter are another
intriguing
application of SMMC methods. One approach is to use single particle states
that are plane waves with periodic boundary conditions and a G-matrix
derived from a realistic inter-nucleon interaction; the formalism and
algorithms we have presented here are then directly applicable.
An alternative approach is to work on a regular lattice of sites in
coordinate space and employ Skyrme-like effective interactions that couple
neighboring sites; the calculation is then similar to that for the Hubbard
model for which special techniques must be used to handle the large,
sparse matrices involved \cite{Linden}. Both approaches are currently being
pursued.

\newpage

\noindent
{\bf Acknowledgements}

Many people have contributed to the development and initial application
of SMMC methods. These include
Y. Alhassid, C. Johnson, G. Lang, W.E. Ormand, P.B. Radha,
and, more recently, M.T. Ressell, P. Vogel, J. White, and D.C. Zheng.
T. Kuo supplied us with effective interactions for those large model
spaces that had never been treated before.
We also wish to thank A. Arima, G. Bertsch, A. Brown, B. Mottelson,
W. Nazarewicz, and A. Poves for useful discussions.
Finally, we would like to thank P. Vogel and G.E. Brown for careful
and critical reading of the final manuscript.

This work was supported by NSF grants PHY91-15574, PHY94-12818, and
PHY94-20470.  Computational cycles on the Intel DELTA and PARAGON
were provided by the Concurrent Supercomputing
Consortium at Caltech. Cycles were also provided on
the VPP500, a vector parallel processor at the RIKEN
supercomputing facility, and on the IBM SP-2 at the Maui HPCC facility.
We thank Drs. I. Tanihata and S. Ohta for their
assistance with the former. The Maui HPCC facility is
sponsored in part by the Phillips Laboratory, Air
Force Materiel Command, USAF, under cooperative agreement
number F29601-93-2-0001.

\vfill
\eject

\noindent
\appendix
\section{Appendix: Properties of one-fermion operators}

In this appendix, we review some of the properties of one-fermion operators
relevant to Monte Carlo methods for the shell model. A more complete
discussion can be found in Ref.~\cite{Lang}.

Consider a set of $N_s$ single-particle states labeled by ${\alpha=1},
\ldots, N_s$. Corresponding to each state $\alpha$ are anticommuting fermion
creation $(a^\dagger_\alpha)$ and annihilation $(a_\alpha)$ operators that
satisfy
\begin{equation}\{ a_\alpha, a_\beta\}=
\{a^\dagger_\alpha, a^\dagger_\beta\}=0;\qquad
\{a^\dagger_\alpha, a_\beta\}=
\delta_{\alpha\beta}\;.
\end{equation}
The associated hermitian number operators, $\hat n_\alpha\equiv
a^\dagger_\alpha a_\alpha$, are mutually commuting, and satisfy the operator
identity
\begin{equation}\hat n^2_\alpha = \hat n_\alpha\;,
\end{equation}
implying that their eigenvalues are either 0 or 1; i.e., a single-particle
state can either be empty or occupied. The total number operator is $\hat
N=\displaystyle\sum_\alpha \hat n_\alpha$.
The fermion vacuum, $\vert {\bf 0}\rangle$, is annihilated by all operators,
$a_\alpha\vert{\bf 0}\rangle=0$, and so satisfies $\hat n_\alpha \vert{\bf
0}\rangle=0$; i.e., all single-particle states are empty.

A complete orthonormal basis of $A$-fermion states can be constructed by
choosing $A$ different single-particle states to be occupied (let them be
labeled by $i=1,\ldots, A$), and then constructing the Slater determinant
\begin{equation}\vert\Phi\rangle=
a^\dagger_1 a^\dagger_2 \ldots a^\dagger_A\vert{\bf 0}\rangle\;,
\end{equation}
where the product is over all occupied states and the order of the creation
operators follows a fixed, predetermined sequence. This state is an
eigenvalue of each $\hat n_i$ with eigenvalue 1 or 0, depending upon
whether the state $i$ is occupied or not.

One-body operators are defined as
\begin{equation}\hat O= \sum_{\alpha\beta} {\bf O}_{\alpha\beta}
a^\dagger_\alpha
a_\beta\;,
\end{equation}
where the ${\bf O}_{\alpha\beta}\equiv \langle\alpha\vert O\vert
\beta\rangle$ are the one-body matrix elements. (Note that we must be careful
to distinguish between the second-quantized operator $\hat O$ and the
$c$-number matrix of its one-body matrix elements, ${\bf O}$.) When acting on
a determinant, a one-body operator produces a linear combination of other
determinants that differ from the original by at most the occupation number
of one state.

Operators and determinants can be defined in any orthonormal single-particle
basis. Suppose that there is a second basis $\lambda\mu\ldots$ distinct from
the $\alpha\beta\ldots$ basis, and let $T_{\lambda\alpha}=
\langle\lambda\vert\alpha\rangle$ be the $N_s\times N_s$ unitary matrix
effecting transformations between them. Then the creation operators transform
as
\begin{equation}a^\dagger_\lambda = \sum_\alpha
T_{\lambda\alpha}a^\dagger_\alpha
\end{equation}
and one-body matrix elements transform as
\begin{equation}{\bf O}_{\lambda\mu}= \sum_{\alpha\beta} T_{\lambda\alpha} {\bf
O}_{\alpha\beta} T^\ast_{\mu\beta}\;.
\end{equation}

Of particular relevance for SMMC are the exponentials of one-body operators.
Consider the action of $\hat u=e^{-\Delta\beta\hat h}$ on a determinant of
the form (A.3), where $\hat h$ is a one-body hamiltonian. Since the
Baker-Hausdorff identity implies that
\begin{equation}e^{-\Delta\beta\hat h} a^\dagger_1 e^{\Delta\beta\hat h}=
\sum_\beta \left[ e^{-\Delta\beta{\bf h}}\right]_{\alpha1}
a^\dagger_\alpha\equiv \sum_\beta {\bf u}_{\alpha1} a^\dagger_\alpha
\end{equation}
and $\hat u\vert{\bf 0}\rangle= \vert{\bf 0}\rangle$, then
\begin{equation}\hat u\vert 1,2,\ldots,A\rangle=
\left(\sum_\alpha {\bf u}_{\alpha1}a^\dagger_\alpha\right)
\left(\sum_\beta {\bf u}_{\beta2}a^\dagger_\beta\right)\ldots
\left(\sum_\zeta {\bf u}_{\zeta A}a^\dagger_\zeta\right)
\vert {\bf 0}\rangle\;.
\end{equation}
Thus, the action of the exponential of a one-body operator on a determinant
is to redefine the occupied states by the linear transformation associated
with ${\bf u}$. For evolution at a finite $\beta$, we deal with products of
exponentials of one-body operators, $\hat U=\ldots \hat u_2\hat u_1$; the net
result of this operator on a determinant is clearly a linear transformation
of the occupied states by the matrix ${\bf U}=\ldots {\bf u}_2{\bf u}_1$. It
then follows that the expectation value of $\hat U$ in a determinental trial
function, as is required in the zero-temperature formalism, is
\begin{equation}\langle\Phi\vert\hat U\vert \Phi\rangle=\det U_{ij}\;,
\end{equation}
where $U_{ij}$ is the $A\times A$ matrix of one-body quantities, $\langle
i\vert U\vert j\rangle$.

The grand-canonical trace of $\hat U$ is the sum of the expectation values in
all possible many-particle states:
\begin{equation}{\rm Tr}\,\hat U=
\sum^{N_s}_{A=1}\,{\rm Tr}_A\,\hat U\;,
\end{equation}
where ${\rm Tr}_A$ is the canonical trace (sum over all $A$-particle states).
The enumeration of the canonical traces is straightforward:
\begin{eqnarray}
{\rm Tr}_0\,\hat U&=&
	\langle{\bf 0}\vert\hat U\vert {\bf 0}\rangle=1 \nonumber\\
{\rm Tr}_1\,\hat U&=&
	\sum_1 \langle 1\vert\hat U\vert 1\rangle={\rm tr}{\bf U} \nonumber\\
{\rm Tr}_2\,\hat U&=&
	{1\over2}\sum_{12} \langle 12\vert\hat U\vert 12\rangle=
	{1\over2}\left[({\rm tr}{\bf U})^2-{\rm tr}{\bf U}^2\right] \nonumber\\
&\vdots& \cr \nonumber\\
{\rm Tr}_{N_s}\,\hat U&=&
	\det {\bf U}\;.
\end{eqnarray}
These can be summed into the single expression
\begin{equation}{\rm Tr}\,\hat U= \det ({\bf 1}+{\bf U})\;,
\end{equation}
which can be verified by a direct expansion in powers of ${\bf U}$.
Here, the symbol tr denotes the matrix trace, while Tr is reserved
for many-body traces.

We will also be interested in the thermodynamic expectation values of
few-body operators. For a one-body operator $\hat \Omega$, we require
\begin{equation}\langle \hat \Omega\rangle\equiv
{{\rm Tr}\,\hat U\hat \Omega\over {\rm Tr}\,\hat U}\;.
\end{equation}
This expression is most conveniently evaluated by considering the operator
$\hat U_\varepsilon= \hat U e^{\varepsilon\hat \Omega}$, so that
\begin{equation}\langle \hat \Omega\rangle =
{d\over d\varepsilon}\ln\,{\rm Tr}\,\hat
U_\varepsilon\bigl\vert_{\varepsilon=0}\;.
\end{equation}
Since ${\rm Tr}\,\hat U_\varepsilon=\det (1+{\bf U}e^{\varepsilon
{\bf \Omega}})$
and $\det ({\bf A}+\varepsilon{\bf B})\sim (\det{\bf A})(1+\varepsilon{\rm
Tr}\,{\bf A}^{-1}{\bf B})$ to linear order in $\varepsilon$, we have
\begin{equation}\langle\hat \Omega\rangle=
{\rm tr}\,{{\bf 1}\over{\bf 1}+{\bf U}} {\bf U \Omega}\;.
\end{equation}
Similarly, the expectation value of the product of two one-body operators can
be found to be
\begin{equation}\langle \hat \Omega'\hat \Omega\rangle=
\langle\hat \Omega'\rangle\langle\hat \Omega\rangle+
{\rm tr}\,{{\bf 1}\over {\bf 1}+{\bf U}}
{\bf U\Omega'\Omega}-{\rm tr}\,{{\bf 1}\over {\bf 1}+{\bf U}}
{\bf U\Omega'}{{\bf 1}\over {\bf 1}+{\bf U}}{\bf U\Omega}\;.
\end{equation}

\vfill
\eject


\begin{thebibliography}{999}
\bibitem{7D} If all single-particle orbits in the model space have the same
parity, $\pi=+1$ and the sign rule of Ref.~\cite{Lang}
is reproduced. The sign rule
of Ref.~\cite{Lang}
holds in the Biedenharn-Rose convention, but in the calculation
of the two-body matrix elements the spherical harmonics $Y_{lm}$ should be
multiplied by~$i^l$.
\end{thebibliography}
\end{document}